% spell -b rub +list6.txt > rubby
%%%%%%%%%%%%%%%%%%%%%
% Input macro files %
%%%%%%%%%%%%%%%%%%%%%

%\input psfig
%\input texMNRAS_new/mn
% MN.TEX (Computer Modern version)
%
% plain TeX single / double column macros for the
% Monthly Notices of Royal Astronomical Society
%
% v1.4  (mn.tex)  --- released 22nd February 1994 (M. Reed)
% v1.3  (mnd.tex) --- released 28th November 1992 (M. Reed)
% v1.26     "     --- released  1st August 1992 (M. Reed)
% v1.25     "     --- released 25th February 1992 (M. Reed)
%
% Copyright Cambridge University Press
%
% > Incorporating special symbol code from laa.sty v1.1 (25th Feb 1991)
%   used with the permission of Springer Verlag.
% > Incorporating parts of mssymb.tex (8th July 1987).

\catcode `\@=11 % @ signs are letters

\def\@version{1.4}
\def\@verdate{22nd Feb 1994}

% Fonts: Computer Modern / Monotype Times (CUP only)
%
% Font family sizes available:
%   8pt, 9pt, 10pt, 11pt, 14pt and 17pt.
%
% Faces available:
%   \rm, math italic, symbol, \it, \bf, \sl, \tt, \sc, \sf, \cal, \em,
%   \mit, \bmit, \cal, \bcal, \oldstyle and \boldstyle.
%
% Times version only:
%   AMS symbol fonts, \Bbb (blackboard bold)
%
% Odd sizes:
%   none

% define the typeface in use

\newif\ifprod@font

\ifx\@typeface\undefined
  \def\@typeface{Comp. Modern}\prod@fontfalse
\else
  \prod@fonttrue % We want Times
\fi

\def\newfam{\alloc@8\fam\chardef\sixt@@n} % made not outer

\ifprod@font
\font\fiverm=mtr10 at 5pt
\font\fivebf=mtbx10 at 5pt
\font\fiveit=mtti10 at 5pt
\font\fivesl=mtsl10 at 5pt
\font\fivett=mttt10 at 5pt     \hyphenchar\fivett=-1
\font\fivecsc=mtcsc10 at 5pt
\font\fivesf=mtss10 at 5pt
\font\fivei=mtmi10 at 5pt      \skewchar\fivei='177
\font\fivemib=mtmib10 at 5pt   \skewchar\fivemib='177
\font\fivesy=mtsy10 at 5pt     \skewchar\fivesy='60
\font\fivesyb=mtbsy10 at 5pt   \skewchar\fivesyb='60

\font\sixrm=mtr10 at 6pt
\font\sixbf=mtbx10 at 6pt
\font\sixit=mtti10 at 6pt
\font\sixsl=mtsl10 at 6pt
\font\sixtt=mttt10 at 6pt      \hyphenchar\sixtt=-1
\font\sixcsc=mtcsc10 at 6pt
\font\sixsf=mtss10 at 6pt
\font\sixi=mtmi10 at 6pt       \skewchar\sixi='177
\font\sixmib=mtmib10 at 6pt    \skewchar\sixmib='177
\font\sixsy=mtsy10 at 6pt      \skewchar\sixsy='60
\font\sixsyb=mtbsy10 at 6pt    \skewchar\sixsyb='60

\font\sevenrm=mtr10 at 7pt
\font\sevenbf=mtbx10 at 7pt
\font\sevenit=mtti10 at 7pt
\font\sevensl=mtsl10 at 7pt
\font\seventt=mttt10 at 7pt     \hyphenchar\seventt=-1
\font\sevencsc=mtcsc10 at 7pt
\font\sevensf=mtss10 at 7pt
\font\seveni=mtmi10 at 7pt      \skewchar\seveni='177
\font\sevenmib=mtmib10 at 7pt   \skewchar\sevenmib='177
\font\sevensy=mtsy10 at 7pt     \skewchar\sevensy='60
\font\sevensyb=mtbsy10 at 7pt   \skewchar\sevensyb='60

\font\eightrm=mtr10 at 8pt
\font\eightbf=mtbx10 at 8pt
\font\eightit=mtti10 at 8pt
\font\eighti=mtmi10 at 8pt      \skewchar\eighti='177
\font\eightmib=mtmib10 at 8pt   \skewchar\eightmib='177
\font\eightsy=mtsy10 at 8pt     \skewchar\eightsy='60
\font\eightsyb=mtbsy10 at 8pt   \skewchar\eightsyb='60
\font\eightsl=mtsl10 at 8pt
\font\eighttt=mttt10 at 8pt     \hyphenchar\eighttt=-1
\font\eightcsc=mtcsc10 at 8pt
\font\eightsf=mtss10 at 8pt

\font\ninerm=mtr10 at 9pt
\font\ninebf=mtbx10 at 9pt
\font\nineit=mtti10 at 9pt
\font\ninei=mtmi10 at 9pt      \skewchar\ninei='177
\font\ninemib=mtmib10 at 9pt   \skewchar\ninemib='177
\font\ninesy=mtsy10 at 9pt     \skewchar\ninesy='60
\font\ninesyb=mtbsy10 at 9pt   \skewchar\ninesyb='60
\font\ninesl=mtsl10 at 9pt
\font\ninett=mttt10 at 9pt     \hyphenchar\ninett=-1
\font\ninecsc=mtcsc10 at 9pt
\font\ninesf=mtss10 at 9pt

\font\tenrm=mtr10
\font\tenbf=mtbx10
\font\tenit=mtti10
\font\teni=mtmi10		\skewchar\teni='177
\font\tenmib=mtmib10	\skewchar\tenmib='177
\font\tensy=mtsy10		\skewchar\tensy='60
\font\tensyb=mtbsy10	\skewchar\tensyb='60
\font\tenex=cmex10
\font\tensl=mtsl10
\font\tentt=mttt10		\hyphenchar\tentt=-1
\font\tencsc=mtcsc10
\font\tensf=mtss10

\font\elevenrm=mtr10 at 11pt
\font\elevenbf=mtbx10 at 11pt
\font\elevenit=mtti10 at 11pt
\font\eleveni=mtmi10 at 11pt      \skewchar\eleveni='177
\font\elevenmib=mtmib10 at 11pt   \skewchar\elevenmib='177
\font\elevensy=mtsy10 at 11pt     \skewchar\elevensy='60
\font\elevensyb=mtbsy10 at 11pt   \skewchar\elevensyb='60
\font\elevensl=mtsl10 at 11pt
\font\eleventt=mttt10 at 11pt     \hyphenchar\eleventt=-1
\font\elevencsc=mtcsc10 at 11pt
\font\elevensf=mtss10 at 11pt

\font\twelverm=mtr10 at 12pt
\font\twelvebf=mtbx10 at 12pt
\font\twelveit=mtti10 at 12pt
\font\twelvesl=mtsl10 at 12pt
\font\twelvett=mttt10 at 12pt     \hyphenchar\twelvett=-1
\font\twelvecsc=mtcsc10 at 12pt
\font\twelvesf=mtss10 at 12pt
\font\twelvei=mtmi10 at 12pt      \skewchar\twelvei='177
\font\twelvemib=mtmib10 at 12pt   \skewchar\twelvemib='177
\font\twelvesy=mtsy10 at 12pt     \skewchar\twelvesy='60
\font\twelvesyb=mtbsy10 at 12pt   \skewchar\twelvesyb='60

\font\fourteenrm=mtr10 at 14pt
\font\fourteenbf=mtbx10 at 14pt
\font\fourteenit=mtti10 at 14pt
\font\fourteeni=mtmi10 at 14pt      \skewchar\fourteeni='177
\font\fourteenmib=mtmib10 at 14pt   \skewchar\fourteenmib='177
\font\fourteensy=mtsy10 at 14pt     \skewchar\fourteensy='60
\font\fourteensyb=mtbsy10 at 14pt   \skewchar\fourteensyb='60
\font\fourteensl=mtsl10 at 14pt
\font\fourteentt=mttt10 at 14pt     \hyphenchar\fourteentt=-1
\font\fourteencsc=mtcsc10 at 14pt
\font\fourteensf=mtss10 at 14pt

\font\seventeenrm=mtr10 at 17pt
\font\seventeenbf=mtbx10 at 17pt
\font\seventeenit=mtti10 at 17pt
\font\seventeeni=mtmi10 at 17pt      \skewchar\seventeeni='177
\font\seventeenmib=mtmib10 at 17pt   \skewchar\seventeenmib='177
\font\seventeensy=mtsy10 at 17pt     \skewchar\seventeensy='60
\font\seventeensyb=mtbsy10 at 17pt   \skewchar\seventeensyb='60
\font\seventeensl=mtsl10 at 17pt
\font\seventeentt=mttt10 at 17pt     \hyphenchar\seventeentt=-1
\font\seventeencsc=mtcsc10 at 17pt
\font\seventeensf=mtss10 at 17pt

% load fonts for AMS special symbols

\newfam\xmfam
\newfam\ymfam

\font\fivexm=mtxm10 at 5pt
\font\sixxm=mtxm10 at 6pt
\font\sevenxm=mtxm10 at 7pt
\font\eightxm=mtxm10 at 8pt
\font\ninexm=mtxm10 at 9pt
\font\tenxm=mtxm10
\font\elevenxm=mtxm10 at 11pt
\font\twelvexm=mtxm10 at 12pt
\font\fourteenxm=mtxm10 at 14pt
\font\seventeenxm=mtxm10 at 17pt

\font\fiveym=mtym10 at 5pt
\font\sixym=mtym10 at 6pt
\font\sevenym=mtym10 at 7pt
\font\eightym=mtym10 at 8pt
\font\nineym=mtym10 at 9pt
\font\tenym=mtym10
\font\elevenym=mtym10 at 11pt
\font\twelveym=mtym10 at 12pt
\font\fourteenym=mtym10 at 14pt
\font\seventeenym=mtym10 at 17pt
\else
\font\fiverm=cmr5
\font\fivei=cmmi5             \skewchar\fivei='177
\font\fivemib=cmmib10 at 5pt  \skewchar\fivemib='177
\font\fivesy=cmsy5            \skewchar\fivesy='60
\font\fivesyb=cmbsy10 at 5pt  \skewchar\fivesyb='60
\font\fivebf=cmbx5

\font\sixrm=cmr6
\font\sixi=cmmi6             \skewchar\sixi='177
\font\sixmib=cmmib10 at 6pt  \skewchar\sixmib='177
\font\sixsy=cmsy6            \skewchar\sixsy='60
\font\sixsyb=cmbsy10 at 6pt  \skewchar\sixsyb='60
\font\sixbf=cmbx6

\font\sevenrm=cmr7
\font\seveni=cmmi7             \skewchar\seveni='177
\font\sevenmib=cmmib10 at 7pt  \skewchar\sevenmib='177
\font\sevensy=cmsy7            \skewchar\sevensy='60
\font\sevensyb=cmbsy10 at 7pt  \skewchar\sevensyb='60
\font\sevenbf=cmbx7

\font\eightrm=cmr8
\font\eightbf=cmbx8
\font\eightit=cmti8
\font\eighti=cmmi8			\skewchar\eighti='177
\font\eightmib=cmmib10 at 8pt	\skewchar\eightmib='177
\font\eightsy=cmsy8			\skewchar\eightsy='60
\font\eightsyb=cmbsy10 at 8pt	\skewchar\eightsyb='60
\font\eightsl=cmsl8
\font\eighttt=cmtt8			\hyphenchar\eighttt=-1
\font\eightcsc=cmcsc10 at 8pt
\font\eightsf=cmss8

\font\ninerm=cmr9
\font\ninebf=cmbx9
\font\nineit=cmti9
\font\ninei=cmmi9			\skewchar\ninei='177
\font\ninemib=cmmib10 at 9pt	\skewchar\ninemib='177
\font\ninesy=cmsy9			\skewchar\ninesy='60
\font\ninesyb=cmbsy10 at 9pt	\skewchar\ninesyb='60
\font\ninesl=cmsl9
\font\ninett=cmtt9			\hyphenchar\ninett=-1
\font\ninecsc=cmcsc10 at 9pt
\font\ninesf=cmss9

\font\tenrm=cmr10
\font\tenbf=cmbx10
\font\tenit=cmti10
\font\teni=cmmi10		\skewchar\teni='177
\font\tenmib=cmmib10	\skewchar\tenmib='177
\font\tensy=cmsy10		\skewchar\tensy='60
\font\tensyb=cmbsy10	\skewchar\tensyb='60
\font\tenex=cmex10
\font\tensl=cmsl10
\font\tentt=cmtt10		\hyphenchar\tentt=-1
\font\tencsc=cmcsc10
\font\tensf=cmss10

\font\elevenrm=cmr10 scaled \magstephalf
\font\elevenbf=cmbx10 scaled \magstephalf
\font\elevenit=cmti10 scaled \magstephalf
\font\eleveni=cmmi10 scaled \magstephalf	\skewchar\eleveni='177
\font\elevenmib=cmmib10 scaled \magstephalf	\skewchar\elevenmib='177
\font\elevensy=cmsy10 scaled \magstephalf	\skewchar\elevensy='60
\font\elevensyb=cmbsy10 scaled \magstephalf	\skewchar\elevensyb='60
\font\elevensl=cmsl10 scaled \magstephalf
\font\eleventt=cmtt10 scaled \magstephalf	\hyphenchar\eleventt=-1
\font\elevencsc=cmcsc10 scaled \magstephalf
\font\elevensf=cmss10 scaled \magstephalf

\font\twelverm=cmr10 scaled \magstep1
\font\twelvebf=cmbx10 scaled \magstep1
\font\twelvei=cmmi10 scaled \magstep1      \skewchar\twelvei='177
\font\twelvemib=cmmib10 scaled \magstep1   \skewchar\twelvemib='177
\font\twelvesy=cmsy10 scaled \magstep1     \skewchar\twelvesy='60
\font\twelvesyb=cmbsy10 scaled \magstep1   \skewchar\twelvesyb='60

\font\fourteenrm=cmr10 scaled \magstep2
\font\fourteenbf=cmbx10 scaled \magstep2
\font\fourteenit=cmti10 scaled \magstep2
\font\fourteeni=cmmi10 scaled \magstep2		\skewchar\fourteeni='177
\font\fourteenmib=cmmib10 scaled \magstep2	\skewchar\fourteenmib='177
\font\fourteensy=cmsy10 scaled \magstep2	\skewchar\fourteensy='60
\font\fourteensyb=cmbsy10 scaled \magstep2	\skewchar\fourteensyb='60
\font\fourteensl=cmsl10 scaled \magstep2
\font\fourteentt=cmtt10 scaled \magstep2	\hyphenchar\fourteentt=-1
\font\fourteencsc=cmcsc10 scaled \magstep2
\font\fourteensf=cmss10 scaled \magstep2

\font\seventeenrm=cmr10 scaled \magstep3
\font\seventeenbf=cmbx10 scaled \magstep3
\font\seventeenit=cmti10 scaled \magstep3
\font\seventeeni=cmmi10 scaled \magstep3	\skewchar\seventeeni='177
\font\seventeenmib=cmmib10 scaled \magstep3	\skewchar\seventeenmib='177
\font\seventeensy=cmsy10 scaled \magstep3	\skewchar\seventeensy='60
\font\seventeensyb=cmbsy10 scaled \magstep3	\skewchar\seventeensyb='60
\font\seventeensl=cmsl10 scaled \magstep3
\font\seventeentt=cmtt10 scaled \magstep3	\hyphenchar\seventeentt=-1
\font\seventeencsc=cmcsc10 scaled \magstep3
\font\seventeensf=cmss10 scaled \magstep3
\fi

\def\hexnumber#1{\ifcase#1 0\or1\or2\or3\or4\or5\or6\or7\or8\or9\or
  A\or B\or C\or D\or E\or F\fi}

\ifprod@font
  \edef\@xm{\hexnumber\xmfam}
  \edef\@ym{\hexnumber\ymfam}
\fi

\def\makestrut{%
  \setbox\strutbox=\hbox{%
    \vrule height.7\baselineskip depth.3\baselineskip width \z@}%
}

\def\baselinestretch{1}
\newskip\tmp@bls

\def\b@ls#1{% set baseline using \baselinestretch as a scale factor
  \tmp@bls=#1\relax
  \baselineskip=#1\relax\makestrut
  \normalbaselineskip=\baselinestretch\tmp@bls
  \normalbaselines
}

\def\nostb@ls#1{% set baseline skip ignoring \baselinestretch
  \normalbaselineskip=#1\relax
  \normalbaselines
  \makestrut
}

% families \itfam, \slfam, \bffam, \ttfam defined in PLAIN.
%
% \itfam is \fam4
% \slfam is \fam5
% \bffam is \fam6
% \ttfam is \fam7

\newfam\mibfam % \fam8
\newfam\sybfam % \fam9
\newfam\scfam  % \fam10
\newfam\sffam  % \fam11

\def\mit{\fam\@ne}

\def\cal{\fam\tw@}

\def\em{\ifdim\fontdimen1\font>\z@ \rm\else\it\fi}

\textfont3=\tenex
\scriptfont3=\tenex
\scriptscriptfont3=\tenex

\setbox0=\hbox{\tenex B} \p@renwd=\wd0 % width of the big left (

\def\eightpoint{% 8^6^5 on 10pt
  \def\rm{\fam0\eightrm}%
  \textfont0=\eightrm \scriptfont0=\sixrm \scriptscriptfont0=\fiverm%
  \textfont1=\eighti  \scriptfont1=\sixi  \scriptscriptfont1=\fivei%
  \textfont2=\eightsy \scriptfont2=\sixsy \scriptscriptfont2=\fivesy%
  \textfont\itfam=\eightit\def\it{\fam\itfam\eightit}%
  \ifprod@font
    \scriptfont\itfam=\sixit
      \scriptscriptfont\itfam=\fiveit
  \else
    \scriptfont\itfam=\eightit
      \scriptscriptfont\itfam=\eightit
  \fi
  \textfont\bffam=\eightbf%
    \scriptfont\bffam=\sixbf%
      \scriptscriptfont\bffam=\fivebf%
  \def\bf{\fam\bffam\eightbf}%
  \textfont\slfam=\eightsl\def\sl{\fam\slfam\eightsl}%
  \ifprod@font
    \scriptfont\slfam=\sixsl
      \scriptscriptfont\slfam=\fivesl
  \else
    \scriptfont\slfam=\eightsl
      \scriptscriptfont\slfam=\eightsl
  \fi
  \textfont\ttfam=\eighttt\def\tt{\fam\ttfam\eighttt}%
  \ifprod@font
    \scriptfont\ttfam=\sixtt
      \scriptscriptfont\ttfam=\fivett
  \else
    \scriptfont\ttfam=\eighttt
      \scriptscriptfont\ttfam=\eighttt
  \fi
  \textfont\scfam=\eightcsc\def\sc{\fam\scfam\eightcsc}%
  \ifprod@font
    \scriptfont\scfam=\sixcsc
      \scriptscriptfont\scfam=\fivecsc
  \else
    \scriptfont\scfam=\eightcsc
      \scriptscriptfont\scfam=\eightcsc
  \fi
  \textfont\sffam=\eightsf\def\sf{\fam\sffam\eightsf}%
  \ifprod@font
    \scriptfont\sffam=\sixsf
      \scriptscriptfont\sffam=\fivesf
  \else
    \scriptfont\sffam=\eightsf
      \scriptscriptfont\sffam=\eightsf
  \fi
  \textfont\mibfam=\eightmib
    \scriptfont\mibfam=\sixmib
      \scriptscriptfont\mibfam=\fivemib
  \textfont\sybfam=\eightsyb
    \scriptfont\sybfam=\sixsyb
      \scriptscriptfont\sybfam=\fivesyb
  \ifprod@font
    \textfont\xmfam=\eightxm
      \scriptfont\xmfam=\sixxm
        \scriptscriptfont\xmfam=\fivexm
    \textfont\ymfam=\eightym
      \scriptfont\ymfam=\sixym
        \scriptscriptfont\ymfam=\fiveym
  \fi
  \def\oldstyle{\fam\@ne\eighti}%
  \def\boldstyle{\fam\mibfam\eightmib}%
  \b@ls{10pt}\rm%
}

\def\ninepoint{% 9^6^5 on 11pt (two col) / 12 (single col)
  \def\rm{\fam0\ninerm}%
  \textfont0=\ninerm \scriptfont0=\sixrm \scriptscriptfont0=\fiverm%
  \textfont1=\ninei  \scriptfont1=\sixi  \scriptscriptfont1=\fivei%
  \textfont2=\ninesy \scriptfont2=\sixsy \scriptscriptfont2=\fivesy%
  \textfont\itfam=\nineit\def\it{\fam\itfam\nineit}%
  \ifprod@font
    \scriptfont\itfam=\sixit
      \scriptscriptfont\itfam=\fiveit
  \else
    \scriptfont\itfam=\nineit
      \scriptscriptfont\itfam=\nineit
  \fi
  \textfont\bffam=\ninebf%
    \scriptfont\bffam=\sixbf%
      \scriptscriptfont\bffam=\fivebf%
  \def\bf{\fam\bffam\ninebf}%
  \textfont\slfam=\ninesl\def\sl{\fam\slfam\ninesl}%
  \ifprod@font
    \scriptfont\slfam=\sixsl
      \scriptscriptfont\slfam=\fivesl
  \else
    \scriptfont\slfam=\ninesl
      \scriptscriptfont\slfam=\ninesl
  \fi
  \textfont\ttfam=\ninett\def\tt{\fam\ttfam\ninett}%
  \ifprod@font
    \scriptfont\ttfam=\sixtt
      \scriptscriptfont\ttfam=\fivett
  \else
    \scriptfont\ttfam=\ninett
      \scriptscriptfont\ttfam=\ninett
  \fi
  \textfont\scfam=\ninecsc\def\sc{\fam\scfam\ninecsc}%
  \ifprod@font
    \scriptfont\scfam=\sixcsc
      \scriptscriptfont\scfam=\fivecsc
  \else
    \scriptfont\scfam=\ninecsc
      \scriptscriptfont\scfam=\ninecsc
  \fi
  \textfont\sffam=\ninesf\def\sf{\fam\sffam\ninesf}%
  \ifprod@font
    \scriptfont\sffam=\sixsf
      \scriptscriptfont\sffam=\fivesf
  \else
    \scriptfont\sffam=\ninesf
      \scriptscriptfont\sffam=\ninesf
  \fi
  \textfont\mibfam=\ninemib
    \scriptfont\mibfam=\sixmib
      \scriptscriptfont\mibfam=\fivemib
  \textfont\sybfam=\ninesyb
    \scriptfont\sybfam=\sixsyb
      \scriptscriptfont\sybfam=\fivesyb
  \ifprod@font
    \textfont\xmfam=\ninexm
      \scriptfont\xmfam=\sixxm
        \scriptscriptfont\xmfam=\fivexm
    \textfont\ymfam=\nineym
      \scriptfont\ymfam=\sixym
        \scriptscriptfont\ymfam=\fiveym
  \fi
  \def\oldstyle{\fam\@ne\ninei}%
  \def\boldstyle{\fam\mibfam\ninemib}%
  \b@ls{\TextLeading plus \Feathering}\rm%
}

\def\tenpoint{% 10^7^5 on 11pt
  \def\rm{\fam0\tenrm}%
  \textfont0=\tenrm \scriptfont0=\sevenrm \scriptscriptfont0=\fiverm%
  \textfont1=\teni  \scriptfont1=\seveni  \scriptscriptfont1=\fivei%
  \textfont2=\tensy \scriptfont2=\sevensy \scriptscriptfont2=\fivesy%
  \textfont\itfam=\tenit\def\it{\fam\itfam\tenit}%
  \ifprod@font
    \scriptfont\itfam=\sevenit
      \scriptscriptfont\itfam=\fiveit
  \else
    \scriptfont\itfam=\tenit
      \scriptscriptfont\itfam=\tenit
  \fi
  \textfont\bffam=\tenbf%
    \scriptfont\bffam=\sevenbf%
      \scriptscriptfont\bffam=\fivebf%
  \def\bf{\fam\bffam\tenbf}%
  \textfont\slfam=\tensl\def\sl{\fam\slfam\tensl}%
  \ifprod@font
    \scriptfont\slfam=\sevensl
      \scriptscriptfont\slfam=\fivesl
  \else
    \scriptfont\slfam=\tensl
      \scriptscriptfont\slfam=\tensl
  \fi
  \textfont\ttfam=\tentt\def\tt{\fam\ttfam\tentt}%
  \ifprod@font
    \scriptfont\ttfam=\seventt
      \scriptscriptfont\ttfam=\fivett
  \else
    \scriptfont\ttfam=\tentt
      \scriptscriptfont\ttfam=\tentt
  \fi
  \textfont\scfam=\tencsc\def\sc{\fam\scfam\tencsc}%
  \ifprod@font
    \scriptfont\scfam=\sevencsc
      \scriptscriptfont\scfam=\fivecsc
  \else
    \scriptfont\scfam=\tencsc
      \scriptscriptfont\scfam=\tencsc
  \fi
  \textfont\sffam=\tensf\def\sf{\fam\sffam\tensf}%
  \ifprod@font
    \scriptfont\sffam=\sevensf
      \scriptscriptfont\sffam=\fivesf
  \else
    \scriptfont\sffam=\tensf
      \scriptscriptfont\sffam=\tensf
  \fi
  \textfont\mibfam=\tenmib
    \scriptfont\mibfam=\sevenmib
      \scriptscriptfont\mibfam=\fivemib
  \textfont\sybfam=\tensyb
    \scriptfont\sybfam=\sevensyb
      \scriptscriptfont\sybfam=\fivesyb
  \ifprod@font
    \textfont\xmfam=\tenxm
      \scriptfont\xmfam=\sevenxm
        \scriptscriptfont\xmfam=\fivexm
    \textfont\ymfam=\tenym
      \scriptfont\ymfam=\sevenym
        \scriptscriptfont\ymfam=\fiveym
  \fi
  \def\oldstyle{\fam\@ne\teni}%
  \def\boldstyle{\fam\mibfam\tenmib}%
  \b@ls{11pt}\rm%
}

\def\elevenpoint{% 11^8^6 on 13pt
  \def\rm{\fam0\elevenrm}%
  \textfont0=\elevenrm \scriptfont0=\eightrm \scriptscriptfont0=\sixrm%
  \textfont1=\eleveni  \scriptfont1=\eighti  \scriptscriptfont1=\sixi%
  \textfont2=\elevensy \scriptfont2=\eightsy \scriptscriptfont2=\sixsy%
  \textfont\itfam=\elevenit\def\it{\fam\itfam\elevenit}%
  \ifprod@font
    \scriptfont\itfam=\eightit
      \scriptscriptfont\itfam=\sixit
  \else
    \scriptfont\itfam=\elevenit
      \scriptscriptfont\itfam=\elevenit
  \fi
  \textfont\bffam=\elevenbf%
    \scriptfont\bffam=\eightbf%
      \scriptscriptfont\bffam=\sixbf%
  \def\bf{\fam\bffam\elevenbf}%
  \textfont\slfam=\elevensl\def\sl{\fam\slfam\elevensl}%
  \ifprod@font
    \scriptfont\slfam=\eightsl
      \scriptscriptfont\slfam=\sixsl
  \else
    \scriptfont\slfam=\elevensl
      \scriptscriptfont\slfam=\elevensl
  \fi
  \textfont\ttfam=\eleventt\def\tt{\fam\ttfam\eleventt}%
  \ifprod@font
    \scriptfont\ttfam=\eighttt
      \scriptscriptfont\ttfam=\sixtt
  \else
    \scriptfont\ttfam=\eleventt
      \scriptscriptfont\ttfam=\eleventt
  \fi
  \textfont\scfam=\elevencsc\def\sc{\fam\scfam\elevencsc}%
  \ifprod@font
    \scriptfont\scfam=\eightcsc
      \scriptscriptfont\scfam=\sixcsc
  \else
    \scriptfont\scfam=\elevencsc
      \scriptscriptfont\scfam=\elevencsc
  \fi
  \textfont\sffam=\elevensf\def\sf{\fam\sffam\elevensf}%
  \ifprod@font
    \scriptfont\sffam=\eightsf
      \scriptscriptfont\sffam=\sixsf
  \else
    \scriptfont\sffam=\elevensf
      \scriptscriptfont\sffam=\elevensf
  \fi
  \textfont\mibfam=\elevenmib
    \scriptfont\mibfam=\eightmib
      \scriptscriptfont\mibfam=\sixmib
  \textfont\sybfam=\elevensyb
    \scriptfont\sybfam=\eightsyb
      \scriptscriptfont\sybfam=\sixsyb
  \ifprod@font
    \textfont\xmfam=\elevenxm
      \scriptfont\xmfam=\eightxm
       \scriptscriptfont\xmfam=\sixxm
    \textfont\ymfam=\elevenym
      \scriptfont\ymfam=\eightym
        \scriptscriptfont\ymfam=\sixym
   \fi
  \def\oldstyle{\fam\@ne\eleveni}%
  \def\boldstyle{\fam\mibfam\elevenmib}%
  \b@ls{13pt}\rm%
}

\def\fourteenpoint{% 14^10^7 on 17pt
  \def\rm{\fam0\fourteenrm}%
  \textfont0\fourteenrm  \scriptfont0\tenrm  \scriptscriptfont0\sevenrm%
  \textfont1\fourteeni   \scriptfont1\teni   \scriptscriptfont1\seveni%
  \textfont2\fourteensy  \scriptfont2\tensy  \scriptscriptfont2\sevensy%
  \textfont\itfam=\fourteenit\def\it{\fam\itfam\fourteenit}%
  \ifprod@font
    \scriptfont\itfam=\tenit
      \scriptscriptfont\itfam=\sevenit
  \else
    \scriptfont\itfam=\fourteenit
      \scriptscriptfont\itfam=\fourteenit
  \fi
  \textfont\bffam=\fourteenbf%
    \scriptfont\bffam=\tenbf%
      \scriptscriptfont\bffam=\sevenbf%
  \def\bf{\fam\bffam\fourteenbf}%
  \textfont\slfam=\fourteensl\def\sl{\fam\slfam\fourteensl}%
  \ifprod@font
    \scriptfont\slfam=\tensl
      \scriptscriptfont\slfam=\sevensl
  \else
    \scriptfont\slfam=\fourteensl
      \scriptscriptfont\slfam=\fourteensl
  \fi
  \textfont\ttfam=\fourteentt\def\tt{\fam\ttfam\fourteentt}%
  \ifprod@font
    \scriptfont\ttfam=\tentt
      \scriptscriptfont\ttfam=\seventt
  \else
    \scriptfont\ttfam=\fourteentt
      \scriptscriptfont\ttfam=\fourteentt
  \fi
  \textfont\scfam=\fourteencsc\def\sc{\fam\scfam\fourteencsc}%
  \ifprod@font
    \scriptfont\scfam=\tencsc
      \scriptscriptfont\scfam=\sevencsc
  \else
    \scriptfont\scfam=\fourteencsc
      \scriptscriptfont\scfam=\fourteencsc
  \fi
  \textfont\sffam=\fourteensf\def\sf{\fam\sffam\fourteensf}%
  \ifprod@font
    \scriptfont\sffam=\tensf
      \scriptscriptfont\sffam=\sevensf
  \else
    \scriptfont\sffam=\fourteensf
      \scriptscriptfont\sffam=\fourteensf
  \fi
  \textfont\mibfam=\fourteenmib
    \scriptfont\mibfam=\tenmib
      \scriptscriptfont\mibfam=\sevenmib
  \textfont\sybfam=\fourteensyb
    \scriptfont\sybfam=\tensyb
      \scriptscriptfont\sybfam=\sevensyb
  \ifprod@font
    \textfont\xmfam=\fourteenxm
      \scriptfont\xmfam=\tenxm
        \scriptscriptfont\xmfam=\sevenxm
   \textfont\ymfam=\fourteenym
      \scriptfont\ymfam=\tenym
        \scriptscriptfont\ymfam=\sevenym
  \fi
  \def\oldstyle{\fam\@ne\fourteeni}%
  \def\boldstyle{\fam\mibfam\fourteenmib}%
  \b@ls{17pt}\rm%
}

\def\seventeenpoint{% 17^12^10 on 20pt
  \def\rm{\fam0\seventeenrm}%
  \textfont0\seventeenrm  \scriptfont0\twelverm  \scriptscriptfont0\tenrm%
  \textfont1\seventeeni   \scriptfont1\twelvei   \scriptscriptfont1\teni%
  \textfont2\seventeensy  \scriptfont2\twelvesy  \scriptscriptfont2\tensy%
  \textfont\itfam=\seventeenit\def\it{\fam\itfam\seventeenit}%
  \ifprod@font
    \scriptfont\itfam=\twelveit
      \scriptscriptfont\itfam=\tenit
  \else
    \scriptfont\itfam=\seventeenit
      \scriptscriptfont\itfam=\seventeenit
  \fi
  \textfont\bffam=\seventeenbf%
    \scriptfont\bffam=\twelvebf%
      \scriptscriptfont\bffam=\tenbf%
  \def\bf{\fam\bffam\seventeenbf}%
  \textfont\slfam=\seventeensl\def\sl{\fam\slfam\seventeensl}%
  \ifprod@font
    \scriptfont\slfam=\twelvesl
      \scriptscriptfont\slfam=\tensl
  \else
    \scriptfont\slfam=\seventeensl
      \scriptscriptfont\slfam=\seventeensl
  \fi
  \textfont\ttfam=\seventeentt\def\tt{\fam\ttfam\seventeentt}%
  \ifprod@font
    \scriptfont\ttfam=\twelvett
      \scriptscriptfont\ttfam=\tentt
  \else
    \scriptfont\ttfam=\seventeentt
      \scriptscriptfont\ttfam=\seventeentt
  \fi
  \textfont\scfam=\seventeencsc\def\sc{\fam\scfam\seventeencsc}%
  \ifprod@font
    \scriptfont\scfam=\twelvecsc
      \scriptscriptfont\scfam=\tencsc
  \else
    \scriptfont\scfam=\seventeencsc
      \scriptscriptfont\scfam=\seventeencsc
  \fi
  \textfont\sffam=\seventeensf\def\sf{\fam\sffam\seventeensf}%
  \ifprod@font
    \scriptfont\sffam=\twelvesf
      \scriptscriptfont\sffam=\tensf
  \else
    \scriptfont\sffam=\seventeensf
      \scriptscriptfont\sffam=\seventeensf
  \fi
  \textfont\mibfam=\seventeenmib
    \scriptfont\mibfam=\twelvemib
      \scriptscriptfont\mibfam=\tenmib
  \textfont\sybfam=\seventeensyb
    \scriptfont\sybfam=\twelvesyb
      \scriptscriptfont\sybfam=\tensyb
  \ifprod@font
    \textfont\xmfam=\seventeenxm
      \scriptfont\xmfam=\twelvexm
        \scriptscriptfont\xmfam=\tenxm
    \textfont\ymfam=\seventeenym
      \scriptfont\ymfam=\twelveym
        \scriptscriptfont\ymfam=\tenym
  \fi
  \def\oldstyle{\fam\@ne\seventeeni}%
  \def\boldstyle{\fam\mibfam\seventeenmib}%
  \b@ls{20pt}\rm%
}

\lineskip=1pt      \normallineskip=\lineskip
\lineskiplimit=\z@ \normallineskiplimit=\lineskiplimit

% BOLD MATH SYMBOLS

% Astronomy and Astrophysics symbol macros

\def\la{\mathrel{\mathchoice {\vcenter{\offinterlineskip\halign{\hfil
$\displaystyle##$\hfil\cr<\cr\sim\cr}}}
{\vcenter{\offinterlineskip\halign{\hfil$\textstyle##$\hfil\cr
<\cr\sim\cr}}}
{\vcenter{\offinterlineskip\halign{\hfil$\scriptstyle##$\hfil\cr
<\cr\sim\cr}}}
{\vcenter{\offinterlineskip\halign{\hfil$\scriptscriptstyle##$\hfil\cr
<\cr\sim\cr}}}}}

\def\ga{\mathrel{\mathchoice {\vcenter{\offinterlineskip\halign{\hfil
$\displaystyle##$\hfil\cr>\cr\sim\cr}}}
{\vcenter{\offinterlineskip\halign{\hfil$\textstyle##$\hfil\cr
>\cr\sim\cr}}}
{\vcenter{\offinterlineskip\halign{\hfil$\scriptstyle##$\hfil\cr
>\cr\sim\cr}}}
{\vcenter{\offinterlineskip\halign{\hfil$\scriptscriptstyle##$\hfil\cr
>\cr\sim\cr}}}}}

\def\getsto{\mathrel{\mathchoice {\vcenter{\offinterlineskip
\halign{\hfil
$\displaystyle##$\hfil\cr\gets\cr\to\cr}}}
{\vcenter{\offinterlineskip\halign{\hfil$\textstyle##$\hfil\cr\gets
\cr\to\cr}}}
{\vcenter{\offinterlineskip\halign{\hfil$\scriptstyle##$\hfil\cr\gets
\cr\to\cr}}}
{\vcenter{\offinterlineskip\halign{\hfil$\scriptscriptstyle##$\hfil\cr
\gets\cr\to\cr}}}}}

\def\lid{\mathrel{\mathchoice {\vcenter{\offinterlineskip\halign{\hfil
$\displaystyle##$\hfil\cr<\cr\noalign{\vskip1.2pt}=\cr}}}
{\vcenter{\offinterlineskip\halign{\hfil$\textstyle##$\hfil\cr<\cr
\noalign{\vskip1.2pt}=\cr}}}
{\vcenter{\offinterlineskip\halign{\hfil$\scriptstyle##$\hfil\cr<\cr
\noalign{\vskip1pt}=\cr}}}
{\vcenter{\offinterlineskip\halign{\hfil$\scriptscriptstyle##$\hfil\cr
<\cr
\noalign{\vskip0.9pt}=\cr}}}}}

\def\gid{\mathrel{\mathchoice {\vcenter{\offinterlineskip\halign{\hfil
$\displaystyle##$\hfil\cr>\cr\noalign{\vskip1.2pt}=\cr}}}
{\vcenter{\offinterlineskip\halign{\hfil$\textstyle##$\hfil\cr>\cr
\noalign{\vskip1.2pt}=\cr}}}
{\vcenter{\offinterlineskip\halign{\hfil$\scriptstyle##$\hfil\cr>\cr
\noalign{\vskip1pt}=\cr}}}
{\vcenter{\offinterlineskip\halign{\hfil$\scriptscriptstyle##$\hfil\cr
>\cr
\noalign{\vskip0.9pt}=\cr}}}}}

\def\grole{\mathrel{\mathchoice {\vcenter{\offinterlineskip\halign{\hfil
$\displaystyle##$\hfil\cr>\cr\noalign{\vskip-1.5pt}<\cr}}}
{\vcenter{\offinterlineskip\halign{\hfil$\textstyle##$\hfil\cr
>\cr\noalign{\vskip-1.5pt}<\cr}}}
{\vcenter{\offinterlineskip\halign{\hfil$\scriptstyle##$\hfil\cr
>\cr\noalign{\vskip-1pt}<\cr}}}
{\vcenter{\offinterlineskip\halign{\hfil$\scriptscriptstyle##$\hfil\cr
>\cr\noalign{\vskip-0.5pt}<\cr}}}}}

\def\leogr{\mathrel{\mathchoice {\vcenter{\offinterlineskip\halign{\hfil
$\displaystyle##$\hfil\cr<\cr\noalign{\vskip-1.5pt}>\cr}}}
{\vcenter{\offinterlineskip\halign{\hfil$\textstyle##$\hfil\cr
<\cr\noalign{\vskip-1.5pt}>\cr}}}
{\vcenter{\offinterlineskip\halign{\hfil$\scriptstyle##$\hfil\cr
<\cr\noalign{\vskip-1pt}>\cr}}}
{\vcenter{\offinterlineskip\halign{\hfil$\scriptscriptstyle##$\hfil\cr
<\cr\noalign{\vskip-0.5pt}>\cr}}}}}

\def\loa{\mathrel{\mathchoice {\vcenter{\offinterlineskip\halign{\hfil
$\displaystyle##$\hfil\cr<\cr\approx\cr}}}
{\vcenter{\offinterlineskip\halign{\hfil$\textstyle##$\hfil\cr
<\cr\approx\cr}}}
{\vcenter{\offinterlineskip\halign{\hfil$\scriptstyle##$\hfil\cr
<\cr\approx\cr}}}
{\vcenter{\offinterlineskip\halign{\hfil$\scriptscriptstyle##$\hfil\cr
<\cr\approx\cr}}}}}

\def\goa{\mathrel{\mathchoice {\vcenter{\offinterlineskip\halign{\hfil
$\displaystyle##$\hfil\cr>\cr\approx\cr}}}
{\vcenter{\offinterlineskip\halign{\hfil$\textstyle##$\hfil\cr
>\cr\approx\cr}}}
{\vcenter{\offinterlineskip\halign{\hfil$\scriptstyle##$\hfil\cr
>\cr\approx\cr}}}
{\vcenter{\offinterlineskip\halign{\hfil$\scriptscriptstyle##$\hfil\cr
>\cr\approx\cr}}}}}

\def\diameter{{\ifmmode\mathchoice
{\ooalign{\hfil\hbox{$\displaystyle/$}\hfil\crcr
{\hbox{$\displaystyle\mathchar"20D$}}}}
{\ooalign{\hfil\hbox{$\textstyle/$}\hfil\crcr
{\hbox{$\textstyle\mathchar"20D$}}}}
{\ooalign{\hfil\hbox{$\scriptstyle/$}\hfil\crcr
{\hbox{$\scriptstyle\mathchar"20D$}}}}
{\ooalign{\hfil\hbox{$\scriptscriptstyle/$}\hfil\crcr
{\hbox{$\scriptscriptstyle\mathchar"20D$}}}}
\else{\ooalign{\hfil/\hfil\crcr\mathhexbox20D}}%
\fi}}

\def\sq{\ifmmode\squareforqed\else{\unskip\nobreak\hfil
\penalty50\hskip1em\null\nobreak\hfil\squareforqed
\parfillskip=0pt\finalhyphendemerits=0\endgraf}\fi}
\def\squareforqed{\hbox{\rlap{$\sqcap$}$\sqcup$}}

% Simulated Blackboard Bold symbols

\def\bbbc{{\mathchoice {\setbox0=\hbox{$\displaystyle\rm C$}\hbox{\hbox
to0pt{\kern0.4\wd0\vrule height0.9\ht0\hss}\box0}}
{\setbox0=\hbox{$\textstyle\rm C$}\hbox{\hbox
to0pt{\kern0.4\wd0\vrule height0.9\ht0\hss}\box0}}
{\setbox0=\hbox{$\scriptstyle\rm C$}\hbox{\hbox
to0pt{\kern0.4\wd0\vrule height0.9\ht0\hss}\box0}}
{\setbox0=\hbox{$\scriptscriptstyle\rm C$}\hbox{\hbox
to0pt{\kern0.4\wd0\vrule height0.9\ht0\hss}\box0}}}}
\def\bbbq{{\mathchoice {\setbox0=\hbox{$\displaystyle\rm
Q$}\hbox{\raise
0.15\ht0\hbox to0pt{\kern0.4\wd0\vrule height0.8\ht0\hss}\box0}}
{\setbox0=\hbox{$\textstyle\rm Q$}\hbox{\raise
0.15\ht0\hbox to0pt{\kern0.4\wd0\vrule height0.8\ht0\hss}\box0}}
{\setbox0=\hbox{$\scriptstyle\rm Q$}\hbox{\raise
0.15\ht0\hbox to0pt{\kern0.4\wd0\vrule height0.7\ht0\hss}\box0}}
{\setbox0=\hbox{$\scriptscriptstyle\rm Q$}\hbox{\raise
0.15\ht0\hbox to0pt{\kern0.4\wd0\vrule height0.7\ht0\hss}\box0}}}}
\def\bbbt{{\mathchoice {\setbox0=\hbox{$\displaystyle\rm
T$}\hbox{\hbox to0pt{\kern0.3\wd0\vrule height0.9\ht0\hss}\box0}}
{\setbox0=\hbox{$\textstyle\rm T$}\hbox{\hbox
to0pt{\kern0.3\wd0\vrule height0.9\ht0\hss}\box0}}
{\setbox0=\hbox{$\scriptstyle\rm T$}\hbox{\hbox
to0pt{\kern0.3\wd0\vrule height0.9\ht0\hss}\box0}}
{\setbox0=\hbox{$\scriptscriptstyle\rm T$}\hbox{\hbox
to0pt{\kern0.3\wd0\vrule height0.9\ht0\hss}\box0}}}}
\def\bbbs{{\mathchoice
{\setbox0=\hbox{$\displaystyle     \rm S$}\hbox{\raise0.5\ht0\hbox
to0pt{\kern0.35\wd0\vrule height0.45\ht0\hss}\hbox
to0pt{\kern0.55\wd0\vrule height0.5\ht0\hss}\box0}}
{\setbox0=\hbox{$\textstyle        \rm S$}\hbox{\raise0.5\ht0\hbox
to0pt{\kern0.35\wd0\vrule height0.45\ht0\hss}\hbox
to0pt{\kern0.55\wd0\vrule height0.5\ht0\hss}\box0}}
{\setbox0=\hbox{$\scriptstyle      \rm S$}\hbox{\raise0.5\ht0\hbox
to0pt{\kern0.35\wd0\vrule height0.45\ht0\hss}\raise0.05\ht0\hbox
to0pt{\kern0.5\wd0\vrule height0.45\ht0\hss}\box0}}
{\setbox0=\hbox{$\scriptscriptstyle\rm S$}\hbox{\raise0.5\ht0\hbox
to0pt{\kern0.4\wd0\vrule height0.45\ht0\hss}\raise0.05\ht0\hbox
to0pt{\kern0.55\wd0\vrule height0.45\ht0\hss}\box0}}}}
\def\bbbz{{\mathchoice {\hbox{$\sf\textstyle Z\kern-0.4em Z$}}
{\hbox{$\sf\textstyle Z\kern-0.4em Z$}}
{\hbox{$\sf\scriptstyle Z\kern-0.3em Z$}}
{\hbox{$\sf\scriptscriptstyle Z\kern-0.2em Z$}}}}

% replace some of the above with AMS symbols (from MTXM)

\ifprod@font
  \mathchardef\la="3\@xm2E
  \mathchardef\getsto="3\@xm1C
  \mathchardef\lid="3\@xm35
  \mathchardef\grole="3\@xm3F
  \mathchardef\loa="3\@xm2F
  \mathchardef\ga="3\@xm26
  \mathchardef\gid="3\@xm3D
  \mathchardef\leogr="3\@xm37
  \mathchardef\goa="3\@xm27
  \mathchardef\sq="0\@xm03
%
% Redefine \diameter to work better with Times fonts (still not perfect)
%
\def\diameter{{%
  \ifmmode
    \mathchoice
    {\ooalign{\hfil\hbox{$\displaystyle/$}\hfil\crcr
    {\lower.2ex\hbox{$\displaystyle\mathchar"20D$}}}}%
    {\ooalign{\hfil\hbox{$\textstyle/$}\hfil\crcr
    {\lower.2ex\hbox{$\textstyle\mathchar"20D$}}}}%
    {\ooalign{\hfil\hbox{$\scriptstyle/$}\hfil\crcr
    {\lower.1ex\hbox{$\scriptstyle\mathchar"20D$}}}}%
    {\ooalign{\hfil\hbox{$\scriptscriptstyle/$}\hfil\crcr
    {\lower.1ex\hbox{$\scriptscriptstyle\mathchar"20D$}}}}%
  \else
    {\ooalign{\hfil/\hfil\crcr\lower.2ex\hbox{\mathhexbox20D}}}%
  \fi
}}
%
% Redefine the simulated \bbb macros to use real Blackboard bold characters
% from MTYM. Redefine \bbbone to work better with Times fonts.
%

\def\bbbc{{\Bbb{C}}}
\def\bbbq{{\Bbb{Q}}}
\def\bbbt{{\Bbb{T}}}
\def\bbbs{{\Bbb{S}}}
\def\bbbz{{\Bbb{Z}}}
\fi

% The AMS symbol set (taken from mssymb.tex)

\ifprod@font
\mathchardef\boxdot="2\@xm00
\mathchardef\boxplus="2\@xm01
\mathchardef\boxtimes="2\@xm02
\mathchardef\square="0\@xm03
\mathchardef\blacksquare="0\@xm04
\mathchardef\centerdot="2\@xm05
\mathchardef\lozenge="0\@xm06
\mathchardef\blacklozenge="0\@xm07
\mathchardef\circlearrowright="3\@xm08
\mathchardef\circlearrowleft="3\@xm09
\mathchardef\rightleftharpoons="3\@xm0A
\mathchardef\leftrightharpoons="3\@xm0B
\mathchardef\boxminus="2\@xm0C
\mathchardef\Vdash="3\@xm0D
\mathchardef\Vvdash="3\@xm0E
\mathchardef\vDash="3\@xm0F
\mathchardef\twoheadrightarrow="3\@xm10
\mathchardef\twoheadleftarrow="3\@xm11
\mathchardef\leftleftarrows="3\@xm12
\mathchardef\rightrightarrows="3\@xm13
\mathchardef\upuparrows="3\@xm14
\mathchardef\downdownarrows="3\@xm15
\mathchardef\upharpoonright="3\@xm16

\mathchardef\downharpoonright="3\@xm17
\mathchardef\upharpoonleft="3\@xm18
\mathchardef\downharpoonleft="3\@xm19
\mathchardef\rightarrowtail="3\@xm1A
\mathchardef\leftarrowtail="3\@xm1B
\mathchardef\leftrightarrows="3\@xm1C
\mathchardef\rightleftarrows="3\@xm1D
\mathchardef\Lsh="3\@xm1E
\mathchardef\Rsh="3\@xm1F
\mathchardef\rightsquigarrow="3\@xm20
\mathchardef\leftrightsquigarrow="3\@xm21
\mathchardef\looparrowleft="3\@xm22
\mathchardef\looparrowright="3\@xm23
\mathchardef\circeq="3\@xm24
\mathchardef\succsim="3\@xm25
\mathchardef\gtrsim="3\@xm26
\mathchardef\gtrapprox="3\@xm27
\mathchardef\multimap="3\@xm28
\mathchardef\therefore="3\@xm29
\mathchardef\because="3\@xm2A
\mathchardef\doteqdot="3\@xm2B

\mathchardef\triangleq="3\@xm2C
\mathchardef\precsim="3\@xm2D
\mathchardef\lesssim="3\@xm2E
\mathchardef\lessapprox="3\@xm2F
\mathchardef\eqslantless="3\@xm30
\mathchardef\eqslantgtr="3\@xm31
\mathchardef\curlyeqprec="3\@xm32
\mathchardef\curlyeqsucc="3\@xm33
\mathchardef\preccurlyeq="3\@xm34
\mathchardef\leqq="3\@xm35
\mathchardef\leqslant="3\@xm36
\mathchardef\lessgtr="3\@xm37
\mathchardef\backprime="0\@xm38
\mathchardef\risingdotseq="3\@xm3A
\mathchardef\fallingdotseq="3\@xm3B
\mathchardef\succcurlyeq="3\@xm3C
\mathchardef\geqq="3\@xm3D
\mathchardef\geqslant="3\@xm3E
\mathchardef\gtrless="3\@xm3F
\mathchardef\sqsubset="3\@xm40
\mathchardef\sqsupset="3\@xm41
\mathchardef\vartriangleright="3\@xm42
\mathchardef\vartriangleleft="3\@xm43
\mathchardef\trianglerighteq="3\@xm44
\mathchardef\trianglelefteq="3\@xm45
\mathchardef\bigstar="0\@xm46
\mathchardef\between="3\@xm47
\mathchardef\blacktriangledown="0\@xm48
\mathchardef\blacktriangleright="3\@xm49
\mathchardef\blacktriangleleft="3\@xm4A
\mathchardef\vartriangle="0\@xm4D
\mathchardef\blacktriangle="0\@xm4E
\mathchardef\triangledown="0\@xm4F
\mathchardef\eqcirc="3\@xm50
\mathchardef\lesseqgtr="3\@xm51
\mathchardef\gtreqless="3\@xm52
\mathchardef\lesseqqgtr="3\@xm53
\mathchardef\gtreqqless="3\@xm54
\mathchardef\Rrightarrow="3\@xm56
\mathchardef\Lleftarrow="3\@xm57
\mathchardef\veebar="2\@xm59
\mathchardef\barwedge="2\@xm5A
\mathchardef\doublebarwedge="2\@xm5B
\mathchardef\angle="0\@xm5C
\mathchardef\measuredangle="0\@xm5D
\mathchardef\sphericalangle="0\@xm5E
\mathchardef\varpropto="3\@xm5F
\mathchardef\smallsmile="3\@xm60
\mathchardef\smallfrown="3\@xm61
\mathchardef\Subset="3\@xm62
\mathchardef\Supset="3\@xm63
\mathchardef\Cup="2\@xm64

\mathchardef\Cap="2\@xm65

\mathchardef\curlywedge="2\@xm66
\mathchardef\curlyvee="2\@xm67
\mathchardef\leftthreetimes="2\@xm68
\mathchardef\rightthreetimes="2\@xm69
\mathchardef\subseteqq="3\@xm6A
\mathchardef\supseteqq="3\@xm6B
\mathchardef\bumpeq="3\@xm6C
\mathchardef\Bumpeq="3\@xm6D
\mathchardef\lll="3\@xm6E

\mathchardef\ggg="3\@xm6F

\mathchardef\circledS="0\@xm73
\mathchardef\pitchfork="3\@xm74
\mathchardef\dotplus="2\@xm75
\mathchardef\backsim="3\@xm76
\mathchardef\backsimeq="3\@xm77
\mathchardef\complement="0\@xm7B
\mathchardef\intercal="2\@xm7C
\mathchardef\circledcirc="2\@xm7D
\mathchardef\circledast="2\@xm7E
\mathchardef\circleddash="2\@xm7F
\def\ulcorner{\delimiter"4\@xm70\@xm70 }
\def\urcorner{\delimiter"5\@xm71\@xm71 }
\def\llcorner{\delimiter"4\@xm78\@xm78 }
\def\lrcorner{\delimiter"5\@xm79\@xm79 }
\def\yen{\mathhexbox\@xm55 }
\def\checkmark{\mathhexbox\@xm58 }
\def\circledR{\mathhexbox\@xm72 }
\def\maltese{\mathhexbox\@xm7A }
\mathchardef\lvertneqq="3\@ym00
\mathchardef\gvertneqq="3\@ym01
\mathchardef\nleq="3\@ym02
\mathchardef\ngeq="3\@ym03
\mathchardef\nless="3\@ym04
\mathchardef\ngtr="3\@ym05
\mathchardef\nprec="3\@ym06
\mathchardef\nsucc="3\@ym07
\mathchardef\lneqq="3\@ym08
\mathchardef\gneqq="3\@ym09
\mathchardef\nleqslant="3\@ym0A
\mathchardef\ngeqslant="3\@ym0B
\mathchardef\lneq="3\@ym0C
\mathchardef\gneq="3\@ym0D
\mathchardef\npreceq="3\@ym0E
\mathchardef\nsucceq="3\@ym0F
\mathchardef\precnsim="3\@ym10
\mathchardef\succnsim="3\@ym11
\mathchardef\lnsim="3\@ym12
\mathchardef\gnsim="3\@ym13
\mathchardef\nleqq="3\@ym14
\mathchardef\ngeqq="3\@ym15
\mathchardef\precneqq="3\@ym16
\mathchardef\succneqq="3\@ym17
\mathchardef\precnapprox="3\@ym18
\mathchardef\succnapprox="3\@ym19
\mathchardef\lnapprox="3\@ym1A
\mathchardef\gnapprox="3\@ym1B
\mathchardef\nsim="3\@ym1C
\mathchardef\ncong="3\@ym1D

\mathchardef\varsubsetneq="3\@ym20
\mathchardef\varsupsetneq="3\@ym21
\mathchardef\nsubseteqq="3\@ym22
\mathchardef\nsupseteqq="3\@ym23
\mathchardef\subsetneqq="3\@ym24
\mathchardef\supsetneqq="3\@ym25
\mathchardef\varsubsetneqq="3\@ym26
\mathchardef\varsupsetneqq="3\@ym27
\mathchardef\subsetneq="3\@ym28
\mathchardef\supsetneq="3\@ym29
\mathchardef\nsubseteq="3\@ym2A
\mathchardef\nsupseteq="3\@ym2B
\mathchardef\nparallel="3\@ym2C
\mathchardef\nmid="3\@ym2D
\mathchardef\nshortmid="3\@ym2E
\mathchardef\nshortparallel="3\@ym2F
\mathchardef\nvdash="3\@ym30
\mathchardef\nVdash="3\@ym31
\mathchardef\nvDash="3\@ym32
\mathchardef\nVDash="3\@ym33
\mathchardef\ntrianglerighteq="3\@ym34
\mathchardef\ntrianglelefteq="3\@ym35
\mathchardef\ntriangleleft="3\@ym36
\mathchardef\ntriangleright="3\@ym37
\mathchardef\nleftarrow="3\@ym38
\mathchardef\nrightarrow="3\@ym39
\mathchardef\nLeftarrow="3\@ym3A
\mathchardef\nRightarrow="3\@ym3B
\mathchardef\nLeftrightarrow="3\@ym3C
\mathchardef\nleftrightarrow="3\@ym3D
\mathchardef\divideontimes="2\@ym3E
\mathchardef\varnothing="0\@ym3F
\mathchardef\nexists="0\@ym40
\mathchardef\mho="0\@ym66
\mathchardef\eth="0\@ym67
\mathchardef\eqsim="3\@ym68
\mathchardef\beth="0\@ym69
\mathchardef\gimel="0\@ym6A
\mathchardef\daleth="0\@ym6B
\mathchardef\lessdot="3\@ym6C
\mathchardef\gtrdot="3\@ym6D
\mathchardef\ltimes="2\@ym6E
\mathchardef\rtimes="2\@ym6F
\mathchardef\shortmid="3\@ym70
\mathchardef\shortparallel="3\@ym71
\mathchardef\smallsetminus="2\@ym72
\mathchardef\thicksim="3\@ym73
\mathchardef\thickapprox="3\@ym74
\mathchardef\approxeq="3\@ym75
\mathchardef\succapprox="3\@ym76
\mathchardef\precapprox="3\@ym77
\mathchardef\curvearrowleft="3\@ym78
\mathchardef\curvearrowright="3\@ym79
\mathchardef\digamma="0\@ym7A
\mathchardef\varkappa="0\@ym7B
\mathchardef\hslash="0\@ym7D
\mathchardef\hbar="0\@ym7E
\mathchardef\backepsilon="3\@ym7F

% Blackboard bold (uppercase)

\def\Bbb{\ifmmode\let\next\Bbb@\else
\def\next{\errmessage{Use \string\Bbb\space only in math mode}}\fi\next}
\def\Bbb@#1{{\Bbb@@{#1}}}
\def\Bbb@@#1{\fam\ymfam#1}
\fi

% NUMBER THE DESIGN ELEMENTS

\def\Nulle{0} % null element
\def\Afe{1}   % author affiliation
\def\Hae{2}   % heading A
\def\Hbe{3}   % heading B
\def\Hce{4}   % heading C
\def\Hde{5}   % heading D

% TEMPORARY REGISTERS

\newcount\LastMac       \LastMac=\Nulle

\newskip\half      \half=5.5pt plus 1.5pt minus 2.25pt
\newskip\one       \one=11pt plus 3pt minus 5.5pt
\newskip\onehalf   \onehalf=16.5pt plus 5.5pt minus 8.25pt
\newskip\two       \two=22pt plus 5.5pt minus 11pt

\def\Half{\addvspace{\half}}
\def\One{\addvspace{\one}}
\def\OneHalf{\addvspace{\onehalf}}
\def\Two{\addvspace{\two}}

 % for manually numbered sections

\def\Raggedright{% set lines unjustified
  \rightskip=\z@ plus \hsize\relax
}

\def\Fullout{% set lines justified
  \rightskip=\z@\relax
}

\def\Hang#1#2{% set hanging indentation
  \hangindent=#1%
  \hangafter=#2\relax
}

% Pagestyles

\newif\ifsp@page
\def\pagestyle#1{\csname ps@#1\endcsname}
\def\thispagestyle#1{\global\sp@pagetrue\gdef\sp@type{#1}}

\def\ps@titlepage{%
  \def\@oddhead{\eightpoint\noindent \the\CatchLine
    \ifprod@font\else\qquad Printed\ \today\fi \hfil}%
  \let\@evenhead=\@oddhead
}

\def\ps@headings{%
  \def\@oddhead{\elevenpoint\it\noindent
    \hfill\the\RightHeader\hskip1.5em\rm\folio}%
  \def\@evenhead{\elevenpoint\noindent
    \folio\hskip1.5em\it\the\LeftHeader\hfill}%
}

\def\ps@plate{%
  \def\@oddhead{\eightpoint\noindent\plt@cap\hfil}%
  \def\@evenhead{\eightpoint\noindent\plt@cap\hfil}%
}

% DESIGN ELEMENT DEFINITIONS

% Article opening

\def\title#1{% article title
  \bgroup
    \vbox to 8pt{\vss}%
    \seventeenpoint
    \Raggedright
    \noindent \strut{\bf #1}\par
  \egroup
}

\def\author#1{% article author(s)
  \bgroup
    \ifnum\LastMac=\Afe \OneHalf\else \vskip 21pt\fi
    \fourteenpoint
    \Raggedright
    \noindent \strut #1\par
    \vskip 3pt%
  \egroup
}

\def\affiliation#1{% author(s) affiliation
  \bgroup
    \vskip -4pt%
    \eightpoint
    \Raggedright
    \noindent \strut {\it #1}\par
  \egroup
  \LastMac=\Afe\relax
}

\def\acceptedline#1{% acceptance date
  \bgroup
    \Two
    \eightpoint
    \Raggedright
    \noindent \strut #1\par
  \egroup
}

\long\def\abstract#1{%
  \bgroup
    \vskip 20pt%
    \everypar{\Hang{11pc}{0}}%
    \noindent{\ninebf ABSTRACT}\par
    \tenpoint
    \Fullout
    \noindent #1\par
  \egroup
}

\long\def\keywords#1{% keywords
  \bgroup
    \Half
    \everypar{\Hang{11pc}{0}}%
    \tenpoint
    \Fullout
    \noindent\hbox{\bf Key words:}\ #1\par
  \egroup
}

% The \maketitle macro ensures that the two spanning material appears
% at the top of the first page, and that it has two lines of space
% underneath it. If you forget this in you input, no output will be produced.
% The \BeginOpening (alias \begintopmatter) macro should be called at the
% very start of the input file, so that it is in force when the document
% starts. This ensures that when \maketitle is called that the group is
% closed, and the material gets printed.

\def\maketitle{%
  \EndOpening
  \ifsinglecol \else \MakePage\fi
}

% Page offset

\def\pageoffset#1#2{\hoffset=#1\relax\voffset=#2\relax}

% Headings

\newif\ifAutoNumber \AutoNumberfalse
\newcount\Sec        %  heading auto number counters
\newcount\SecSec
\newcount\SecSecSec

\Sec=\z@

\def\:{\let\@sptoken= } \:  % this makes \@sptoken a space token 
\def\:{\@xifnch} \expandafter\def\: {\futurelet\@tempc\@ifnch}

\def\@ifnextchar#1#2#3{%
  \let\@tempMACe #1%
  \def\@tempMACa{#2}%
  \def\@tempMACb{#3}%
  \futurelet \@tempMACc\@ifnch%
}

\def\@ifnch{%
\ifx \@tempMACc \@sptoken%
  \let\@tempMACd\@xifnch%
\else%
  \ifx \@tempMACc \@tempMACe%
    \let\@tempMACd\@tempMACa%
  \else%
    \let\@tempMACd\@tempMACb%
  \fi%
\fi%
\@tempMACd%
}

\def\@ifstar#1#2{\@ifnextchar *{\def\@tempMACa*{#1}\@tempMACa}{#2}}

\newskip\@tempskipb

\def\addvspace#1{%
  \ifvmode\else \endgraf\fi%
  \ifdim\lastskip=\z@%
    \vskip #1\relax%
  \else%
    \@tempskipb#1\relax\@xaddvskip%
  \fi%
}

\def\@xaddvskip{%
  \ifdim\lastskip<\@tempskipb%
    \vskip-\lastskip%
    \vskip\@tempskipb\relax%
  \else%
    \ifdim\@tempskipb<\z@%
      \ifdim\lastskip<\z@ \else%
        \advance\@tempskipb\lastskip%
        \vskip-\lastskip\vskip\@tempskipb%
      \fi%
    \fi%
  \fi%
}

\newskip\@tmpSKIP

\def\addpen#1{%
  \ifvmode
    \if@nobreak
    \else
      \ifdim\lastskip=\z@
        \penalty#1\relax
      \else
        \@tmpSKIP=\lastskip
        \vskip -\lastskip
        \penalty#1\vskip\@tmpSKIP
      \fi
    \fi
  \fi
}

\newcount\@clubpen   \@clubpen=\clubpenalty
\newif\if@nobreak    \@nobreakfalse

\def\@noafterindent{%
  \global\@nobreaktrue
  \everypar{\if@nobreak
              \global\@nobreakfalse
              \clubpenalty \@M
              {\setbox\z@\lastbox}%
              \LastMac=\Nulle\relax%
            \else
              \clubpenalty \@clubpen
              \everypar{}%
            \fi}
}

\newcount\gds@cbrk   \gds@cbrk=-300

\def\@nohdbrk{\interlinepenalty \@M\relax}

\let\@par=\par
\def\@restorepar{\def\par{\@par}}

\newif\if@endpe   \@endpefalse
 
\def\@doendpe{\@endpetrue \@nobreakfalse \LastMac=\Nulle\relax%
     \def\par{\@restorepar\everypar{}\par\@endpefalse}%
              \everypar{\setbox\z@\lastbox\everypar{}\@endpefalse}%
}

\def\section{\@ifstar{\@ssection}{\@section}}

\def\@section#1{% heading A (\section{....})
  \if@nobreak
    \everypar{}%
    \ifnum\LastMac=\Hae \addvspace{\half}\fi
  \else
    \addpen{\gds@cbrk}%
    \addvspace{\two}%
  \fi
  \bgroup
    \ninepoint\bf
    \Raggedright
    \ifAutoNumber
      \global\advance\Sec \@ne
      \noindent\@nohdbrk\number\Sec\hskip 1pc \uppercase{#1}\par
      \global\SecSec=\z@
    \else
      \noindent\@nohdbrk\uppercase{#1}\par
    \fi
  \egroup
  \nobreak
  \vskip\half
  \nobreak
  \@noafterindent
  \LastMac=\Hae\relax
}

\def\@ssection#1{%  main section heading (\section*{....})
  \if@nobreak
    \everypar{}%
    \ifnum\LastMac=\Hae \addvspace{\half}\fi
  \else
    \addpen{\gds@cbrk}%
    \addvspace{\two}%
  \fi
  \bgroup
    \ninepoint\bf
    \Raggedright
    \noindent\@nohdbrk\uppercase{#1}\par
  \egroup
  \nobreak
  \vskip\half
  \nobreak
  \@noafterindent
  \LastMac=\Hae\relax
}

\def\subsection#1{% heading B
  \if@nobreak
    \everypar{}%
    \ifnum\LastMac=\Hae \addvspace{1pt plus 1pt minus .5pt}\fi
  \else
    \addpen{\gds@cbrk}%
    \addvspace{\onehalf}%
  \fi
  \bgroup
    \ninepoint\bf
    \Raggedright
    \ifAutoNumber
      \global\advance\SecSec \@ne
      \noindent\@nohdbrk\number\Sec.\number\SecSec \hskip 1pc\relax #1\par
      \global\SecSecSec=\z@
    \else
      \noindent\@nohdbrk #1\par
    \fi
  \egroup
  \nobreak
  \vskip\half
  \nobreak
  \@noafterindent
  \LastMac=\Hbe\relax
}

\def\subsubsection#1{% heading C
  \if@nobreak
    \everypar{}%
    \ifnum\LastMac=\Hbe \addvspace{1pt plus 1pt minus .5pt}\fi
  \else
    \addpen{\gds@cbrk}%
    \addvspace{\onehalf}%
  \fi
  \bgroup
    \ninepoint\it
    \Raggedright
    \ifAutoNumber
      \global\advance\SecSecSec \@ne
      \noindent\@nohdbrk\number\Sec.\number\SecSec.\number\SecSecSec
        \hskip 1pc\relax #1\par
    \else
      \noindent\@nohdbrk #1\par
    \fi
  \egroup
  \nobreak
  \vskip\half
  \nobreak
  \@noafterindent
  \LastMac=\Hce\relax
}

\def\paragraph#1{% heading D
  \if@nobreak
    \everypar{}%
  \else
    \addpen{\gds@cbrk}%
    \addvspace{\one}%
  \fi%
  \bgroup%
    \ninepoint\it
    \noindent #1\ \nobreak%
  \egroup
  \LastMac=\Hde\relax
  \ignorespaces
}

% Text

 % provided for backward compatibility

% Lists

\def\beginlist{%
  \par\if@nobreak \else\addvspace{\half}\fi%
  \bgroup%
    \ninepoint
    \let\item=\list@item%
}

\def\list@item{%
  \par\noindent\hskip 1em\relax%
  \ignorespaces%
}

\def\endlist{\par\egroup\addvspace{\half}\@doendpe}

% References

\def\beginrefs{%
  \par
  \bgroup
    \eightpoint
    \Raggedright
    \let\bibitem=\bib@item
}

\def\bib@item{%
  \par\parindent=1.5em\Hang{1.5em}{1}%
  \everypar={\Hang{1.5em}{1}\ignorespaces}%
  \noindent\ignorespaces
}

\def\endrefs{\par\egroup\@doendpe}

% Page heads

\newtoks\CatchLine

\def\@journal{Mon.\ Not.\ R.\ Astron.\ Soc.\ }  % The journal title string
\def\@pubyear{1994}        % Assign a default publication year
\def\@pagerange{000--000}  % Assign a default page-range
\def\@volume{000}          % Assign a default volume number
\def\@microfiche{}         %

\def\pubyear#1{\gdef\@pubyear{#1}\@makecatchline}
\def\pagerange#1{\gdef\@pagerange{#1}\@makecatchline}
\def\volume#1{\gdef\@volume{#1}\@makecatchline}
\def\microfiche#1{\gdef\@microfiche{and Microfiche\ #1}\@makecatchline}

\def\@makecatchline{%
  \global\CatchLine{%
    {\rm \@journal {\bf \@volume},\ \@pagerange\ (\@pubyear)\ \@microfiche}}%
}

\@makecatchline % Assign a catchline, using the above defaults

\newtoks\LeftHeader
\def\shortauthor#1{% left page head
  \global\LeftHeader{#1}%
}

\newtoks\RightHeader
\def\shorttitle#1{% right page head
  \global\RightHeader{#1}%
}

\def\PageHead{% recto/verso running heads
  \begingroup
    \ifsp@page
      \csname ps@\sp@type\endcsname
      \global\sp@pagefalse
    \fi
    \ifodd\pageno
      \let\the@head=\@oddhead
    \else
      \let\the@head=\@evenhead
    \fi
    \vbox to \z@{\vskip-22.5\p@%
      \hbox to \PageWidth{\vbox to8.5\p@{}%
        \the@head
      }%
    \vss}%
  \endgroup
  \nointerlineskip
}

\def\today{%
  \number\day\space
  \ifcase\month\or January\or February\or March\or April\or May\or June\or
    July\or August\or September\or October\or November\or December\fi
  \space\number\year%
}

\def\PageFoot{} % No page footer as default

\def\authorcomment#1{%
  \gdef\PageFoot{%
    \nointerlineskip%
    \vbox to 22pt{\vfil%
      \hbox to \PageWidth{\elevenpoint\noindent \hfil #1 \hfil}}%
  }%
}

% Plate pages

\newif\ifplate@page
\newbox\plt@box

\def\beginplatepage{%
  \let\plate=\plate@head
  \let\caption=\fig@caption
  \global\setbox\plt@box=\vbox\bgroup
  \TEMPDIMEN=\PageWidth % For \fig@caption test
  \hsize=\PageWidth\relax
}

\def\endplatepage{\par\egroup\global\plate@pagetrue}
\def\plate@head#1{\gdef\plt@cap{#1}}

% Letters option

\def\letters{%
  \gdef\folio{\ifnum\pageno<\z@ L\romannumeral-\pageno
    \else L\number\pageno \fi}%
}

% Math setup

\everydisplay{\displaysetup}

\newif\ifeqno
\newif\ifleqno

\def\displaysetup#1$${%
 \displaytest#1\eqno\eqno\displaytest
}

\def\displaytest#1\eqno#2\eqno#3\displaytest{%
 \if!#3!\ldisplaytest#1\leqno\leqno\ldisplaytest
 \else\eqnotrue\leqnofalse\def\eqn{#2}\def\eq{#1}\fi
 \generaldisplay$$}

\def\ldisplaytest#1\leqno#2\leqno#3\ldisplaytest{%
 \def\eq{#1}%
 \if!#3!\eqnofalse\else\eqnotrue\leqnotrue
  \def\eqn{#2}\fi}

\def\generaldisplay{%
\ifeqno \ifleqno 
   \hbox to \hsize{\noindent
     $\displaystyle\eq$\hfil$\displaystyle\eqn$}
  \else
    \hbox to \hsize{\noindent
     $\displaystyle\eq$\hfil$\displaystyle\eqn$}
  \fi
 \else
 \hbox to \hsize{\vbox{\noindent
  $\displaystyle\eq$\hfil}}
 \fi
}

% Finishing notice

\def\@notice{%
  \par\Two%
  \noindent{\b@ls{11pt}\ninerm This paper has been produced using the
    Blackwell Scientific Publications \TeX\ macros.\par}%
}

% redefine \bye to output our identification notice :
\outer\def\bye{\@notice\par\vfill\supereject\end}

% define a sign on :

\def\start@mess{%
  Monthly notices of the RAS journal style (\@typeface)\space
    v\@version,\space \@verdate.%
}

\everyjob{\Warn{\start@mess}}

% Two-column macros

%--------------------------------------------------------%
%                     INITIALISATION                     %
%--------------------------------------------------------%

\newif\if@debug \@debugfalse  %  when false, only warnings displayed

\def\Print#1{\if@debug\immediate\write16{#1}\else \fi}
\def\Warn#1{\immediate\write16{#1}}
\def\wlog#1{}

\newcount\Iteration % temporary loop counter

\def\Single{0} \def\Double{1}                 % ItemSPAN's
\def\Figure{0} \def\Table{1}                  % ItemTYPE's

\def\InStack{0}  % ItemSTATUS
\def\InZoneA{1}
\def\InZoneB{2}
\def\InZoneC{3}

\newcount\TEMPCOUNT % temporary count register
\newdimen\TEMPDIMEN % temporary dimen register
\newbox\TEMPBOX     % temporary box register
\newbox\VOIDBOX     % a box which is permenately void

\newcount\LengthOfStack % number of items currently in stack
\newcount\MaxItems      % maximum number of items allowed in stack
\newcount\StackPointer
\newcount\Point         % used in calculation for generating the
                        % physical address of an item in the stack.
\newcount\NextFigure    % number of next figure to be output
\newcount\NextTable     % number of next table to be output
\newcount\NextItem      % Next item to consider by order in stack

\newcount\StatusStack   % set to point to top of STATUS stack
\newcount\NumStack      % set to point to top of NUMBER stack
\newcount\TypeStack     % set to point to top of TYPE stack
\newcount\SpanStack     % set to point to top of SPAN stack
\newcount\BoxStack      % set to point to top of BOX stack

\newcount\ItemSTATUS    % status of present item
\newcount\ItemNUMBER    % number of present item
\newcount\ItemTYPE      % type of present item
\newcount\ItemSPAN      % span of present item
\newbox\ItemBOX         % box of present item
\newdimen\ItemSIZE      % size of present item
                        % (calculated by GetItemBOX)

\newdimen\PageHeight    % vertical measure of full page
\newdimen\TextLeading   % distance between baselines of body text
\newdimen\Feathering    % amount of interline stretch
\newcount\LinesPerPage  % height of page in text lines
\newdimen\ColumnWidth   % width of 1 column of text
\newdimen\ColumnGap     % size of gap between columns
\newdimen\PageWidth     % = \ColumnWidth * 2 + \ColumnGap
\newdimen\BodgeHeight   % Bodge to align figures and tables with text
\newcount\Leading       % Set to same as \TextLeading above

\newdimen\ZoneBSize  % size of items in ZoneB
\newdimen\TextSize   % size of text in ZoneB
\newbox\ZoneABOX     % contains Zone A material
\newbox\ZoneBBOX     % contains Zone B material
\newbox\ZoneCBOX     % contains Zone C material

\newif\ifFirstSingleItem
\newif\ifFirstZoneA
\newif\ifMakePageInComplete
\newif\ifMoreFigures \MoreFiguresfalse % set true in join stack
\newif\ifMoreTables  \MoreTablesfalse  % set true in join stack

\newif\ifFigInZoneB % used to determine in which zone an item
\newif\ifFigInZoneC % will be placed based on what is in other
\newif\ifTabInZoneB % zones already for a given page.
\newif\ifTabInZoneC

\newif\ifZoneAFullPage

\newbox\MidBOX    % = LeftBOX+gap+RightBOX
\newbox\LeftBOX
\newbox\RightBOX
\newbox\PageBOX   % complete made-up page

\newif\ifLeftCOL  % flags first pass through output routine
\LeftCOLtrue

\newdimen\ZoneBAdjust

\newcount\ItemFits
\def\Yes{1}
\def\No{2}

% Setup file.

\MaxItems=15
\NextFigure=\z@        % used for article opening
\NextTable=\@ne

\BodgeHeight=6pt
\TextLeading=11pt    % baselineskip of body text
\Leading=11
\Feathering=\z@      % amount of interline stretch
\LinesPerPage=61     % number of text lines per full page -1
\topskip=\TextLeading
\ColumnWidth=20pc    % width of text columns
\ColumnGap=2pc       % gap between columns

\newskip\ItemSepamount  % space between floats
\ItemSepamount=\TextLeading plus \TextLeading minus 4pt

\parskip=\z@ plus .1pt
\parindent=18pt
\widowpenalty=\z@
\clubpenalty=10000
\tolerance=1500
\hbadness=1500
\abovedisplayskip=6pt plus 2pt minus 2pt
\belowdisplayskip=6pt plus 2pt minus 2pt
\abovedisplayshortskip=6pt plus 2pt minus 2pt
\belowdisplayshortskip=6pt plus 2pt minus 2pt

\ninepoint % start main text size

%%%\PageHeight=\TextLeading % calculate height of page
%%%\multiply\PageHeight by \LinesPerPage
%%%\advance\PageHeight by \topskip

\PageHeight=682pt
%%%\advance\PageHeight by \topskip

\PageWidth=2\ColumnWidth
\advance\PageWidth by \ColumnGap

\pagestyle{headings}

%--------------------------------------------------------%
%                         STACKS                         %
%--------------------------------------------------------%

% THE ITEM STACK
% The item stack contains contains figures and tables
% in the order in which they appear in the article source
% code.

% allocate stack space

\newcount\DUMMY \StatusStack=\allocationnumber
\newcount\DUMMY \newcount\DUMMY \newcount\DUMMY 
\newcount\DUMMY \newcount\DUMMY \newcount\DUMMY 
\newcount\DUMMY \newcount\DUMMY \newcount\DUMMY
\newcount\DUMMY \newcount\DUMMY \newcount\DUMMY 
\newcount\DUMMY \newcount\DUMMY \newcount\DUMMY

\newcount\DUMMY \NumStack=\allocationnumber
\newcount\DUMMY \newcount\DUMMY \newcount\DUMMY 
\newcount\DUMMY \newcount\DUMMY \newcount\DUMMY 
\newcount\DUMMY \newcount\DUMMY \newcount\DUMMY 
\newcount\DUMMY \newcount\DUMMY \newcount\DUMMY 
\newcount\DUMMY \newcount\DUMMY \newcount\DUMMY

\newcount\DUMMY \TypeStack=\allocationnumber
\newcount\DUMMY \newcount\DUMMY \newcount\DUMMY 
\newcount\DUMMY \newcount\DUMMY \newcount\DUMMY 
\newcount\DUMMY \newcount\DUMMY \newcount\DUMMY 
\newcount\DUMMY \newcount\DUMMY \newcount\DUMMY 
\newcount\DUMMY \newcount\DUMMY \newcount\DUMMY

\newcount\DUMMY \SpanStack=\allocationnumber
\newcount\DUMMY \newcount\DUMMY \newcount\DUMMY 
\newcount\DUMMY \newcount\DUMMY \newcount\DUMMY 
\newcount\DUMMY \newcount\DUMMY \newcount\DUMMY 
\newcount\DUMMY \newcount\DUMMY \newcount\DUMMY 
\newcount\DUMMY \newcount\DUMMY \newcount\DUMMY

\newbox\DUMMY   \BoxStack=\allocationnumber
\newbox\DUMMY   \newbox\DUMMY \newbox\DUMMY 
\newbox\DUMMY   \newbox\DUMMY \newbox\DUMMY 
\newbox\DUMMY   \newbox\DUMMY \newbox\DUMMY 
\newbox\DUMMY   \newbox\DUMMY \newbox\DUMMY 
\newbox\DUMMY   \newbox\DUMMY \newbox\DUMMY

\def\wlog{\immediate\write\m@ne}

% \GetItemSTATUS, \GetItemNUMBER, \GetItemTYPE, \GetItemSPAN,
% \GetItemBox 
% are used to get details of a particular item from the item
% stack. The argument to each of these is the items position
% in the stack (usually \StackPointer)...not the items number.

\def\GetItemAll#1{%
 \GetItemSTATUS{#1}
 \GetItemNUMBER{#1}
 \GetItemTYPE{#1}
 \GetItemSPAN{#1}
 \GetItemBOX{#1}
}

% Note: \LeaveStack uses this routine. Do not destroy \Point
\def\GetItemSTATUS#1{%
 \Point=\StatusStack
 \advance\Point by #1
 \global\ItemSTATUS=\count\Point
}

% Note: \LeaveStack uses this routine. Do not destroy \Point
\def\GetItemNUMBER#1{%
 \Point=\NumStack
 \advance\Point by #1
 \global\ItemNUMBER=\count\Point
}

% Note: \LeaveStack uses this routine. Do not destroy \Point
\def\GetItemTYPE#1{%
 \Point=\TypeStack
 \advance\Point by #1
 \global\ItemTYPE=\count\Point
}

% Note: \LeaveStack uses this routine. Do not destroy \Point
\def\GetItemSPAN#1{%
 \Point\SpanStack
 \advance\Point by #1
 \global\ItemSPAN=\count\Point
}

% Note: \LeaveStack uses this routine. Do not destroy \Point
\def\GetItemBOX#1{%
 \Point=\BoxStack
 \advance\Point by #1
 \global\setbox\ItemBOX=\vbox{\copy\Point}
 \global\ItemSIZE=\ht\ItemBOX
 \global\advance\ItemSIZE by \dp\ItemBOX
 \TEMPCOUNT=\ItemSIZE
 \divide\TEMPCOUNT by \Leading
 \divide\TEMPCOUNT by 65536
 \advance\TEMPCOUNT \@ne
 \ItemSIZE=\TEMPCOUNT pt
 \global\multiply\ItemSIZE by \Leading
}

% item joins stack

\def\JoinStack{%
 \ifnum\LengthOfStack=\MaxItems % stack is full of items
  \Warn{WARNING: Stack is full...some items will be lost!}
 \else
  \Point=\StatusStack
  \advance\Point by \LengthOfStack
  \global\count\Point=\ItemSTATUS
  \Point=\NumStack
  \advance\Point by \LengthOfStack
  \global\count\Point=\ItemNUMBER
  \Point=\TypeStack
  \advance\Point by \LengthOfStack
  \global\count\Point=\ItemTYPE
  \Point\SpanStack
  \advance\Point by \LengthOfStack
  \global\count\Point=\ItemSPAN
  \Point=\BoxStack
  \advance\Point by \LengthOfStack
  \global\setbox\Point=\vbox{\copy\ItemBOX}
  \global\advance\LengthOfStack \@ne
  \ifnum\ItemTYPE=\Figure % used in \MakePage
   \global\MoreFigurestrue
  \else
   \global\MoreTablestrue
  \fi
 \fi
}

% item leaves stack
% #1=physical position of the item to be removed

\def\LeaveStack#1{%
 {\Iteration=#1
 \loop
 \ifnum\Iteration<\LengthOfStack
  \advance\Iteration \@ne
  \GetItemSTATUS{\Iteration}
   \advance\Point by \m@ne
   \global\count\Point=\ItemSTATUS
  \GetItemNUMBER{\Iteration}
   \advance\Point by \m@ne
   \global\count\Point=\ItemNUMBER
  \GetItemTYPE{\Iteration}
   \advance\Point by \m@ne
   \global\count\Point=\ItemTYPE
  \GetItemSPAN{\Iteration}
   \advance\Point by \m@ne
   \global\count\Point=\ItemSPAN
  \GetItemBOX{\Iteration}
   \advance\Point by \m@ne
   \global\setbox\Point=\vbox{\copy\ItemBOX}
 \repeat}
 \global\advance\LengthOfStack by \m@ne
}

% clean stack
% This routine scans through the stack and removes anything
% that does not have STATUS=\InStack.

\newif\ifStackNotClean

\def\CleanStack{%
 \StackNotCleantrue
 {\Iteration=\z@
  \loop
   \ifStackNotClean
    \GetItemSTATUS{\Iteration}
    \ifnum\ItemSTATUS=\InStack
     \advance\Iteration \@ne
     \else
      \LeaveStack{\Iteration}
    \fi
   \ifnum\LengthOfStack<\Iteration
    \StackNotCleanfalse
   \fi
 \repeat}
}

% Find item.
% This macro searches from the top to the bottom of the
% stack for an item of a specified type and number.
% #1=type, #2=number
% If the specified item is found, then \StackPointer is set
% to point to it, else \StackPointer=-1.
% This routine is used to find the next figure or table
% by number.

\def\FindItem#1#2{%
 \global\StackPointer=\m@ne % assume item isn't in stack for now
 {\Iteration=\z@
  \loop
  \ifnum\Iteration<\LengthOfStack
   \GetItemSTATUS{\Iteration}
   \ifnum\ItemSTATUS=\InStack
    \GetItemTYPE{\Iteration}
    \ifnum\ItemTYPE=#1
     \GetItemNUMBER{\Iteration}
     \ifnum\ItemNUMBER=#2
      \global\StackPointer=\Iteration
      \Iteration=\LengthOfStack % terminate loop
     \fi
    \fi
   \fi
  \advance\Iteration \@ne
 \repeat}
}

% Find next type
% This macro searches from the top to the bottom of the stack
% looking for the first item which has STATUS=\InStack.
% If it is a figure then a figure is what will be considered
% next by \MakePage else table.

\def\FindNext{%
 \global\StackPointer=\m@ne % assume stack is empty for now
 {\Iteration=\z@
  \loop
  \ifnum\Iteration<\LengthOfStack
   \GetItemSTATUS{\Iteration}
   \ifnum\ItemSTATUS=\InStack
    \GetItemTYPE{\Iteration}
   \ifnum\ItemTYPE=\Figure
    \ifMoreFigures
      \global\NextItem=\Figure
      \global\StackPointer=\Iteration
      \Iteration=\LengthOfStack % terminate loop
    \fi
   \fi
   \ifnum\ItemTYPE=\Table
    \ifMoreTables
      \global\NextItem=\Table
      \global\StackPointer=\Iteration
      \Iteration=\LengthOfStack % terminate loop
    \fi
   \fi
  \fi
  \advance\Iteration \@ne
 \repeat}
}

% Change status
% Macro to change the status of a specified item in stack.
% #1=item, #2=new status

\def\ChangeStatus#1#2{%
 \Point=\StatusStack
 \advance\Point by #1
 \global\count\Point=#2
}

%--------------------------------------------------------%
%                       MAKEPAGE                         %
%--------------------------------------------------------%

% This macro is called at the start of every new page
% including the first. It scans through the stack picking
% out items which should be placed on this page. It then
% leaves space for the items to be placed later. The routine
% terminates when either there is no room on the page to
% fit the next figure or table, or there are no more items
% in the stack.

\def\Zone{\InZoneA}

\ZoneBAdjust=\z@

\def\MakePage{% allocate space on this page for stack items
 \global\ZoneBSize=\PageHeight
 \global\TextSize=\ZoneBSize
 \global\ZoneAFullPagefalse
 \global\topskip=\TextLeading
 \MakePageInCompletetrue
 \MoreFigurestrue
 \MoreTablestrue
 \FigInZoneBfalse
 \FigInZoneCfalse
 \TabInZoneBfalse
 \TabInZoneCfalse
 \global\FirstSingleItemtrue
 \global\FirstZoneAtrue
 \global\setbox\ZoneABOX=\box\VOIDBOX
 \global\setbox\ZoneBBOX=\box\VOIDBOX
 \global\setbox\ZoneCBOX=\box\VOIDBOX
 \loop
  \ifMakePageInComplete
 \FindNext
 \ifnum\StackPointer=\m@ne
  \NextItem=\m@ne
  \MoreFiguresfalse
  \MoreTablesfalse
 \fi
 \ifnum\NextItem=\Figure
   \FindItem{\Figure}{\NextFigure}
   \ifnum\StackPointer=\m@ne \global\MoreFiguresfalse
   \else
    \GetItemSPAN{\StackPointer}
    \ifnum\ItemSPAN=\Single \def\Zone{\InZoneB}\relax
     \ifFigInZoneC \global\MoreFiguresfalse\fi
    \else
     \def\Zone{\InZoneA}
     \ifFigInZoneB \def\Zone{\InZoneC}\fi
    \fi
   \fi
   \ifMoreFigures\Print{}\FigureItems\fi
 \fi
\ifnum\NextItem=\Table
   \FindItem{\Table}{\NextTable}
   \ifnum\StackPointer=\m@ne \global\MoreTablesfalse
   \else
    \GetItemSPAN{\StackPointer}
    \ifnum\ItemSPAN=\Single\relax
     \ifTabInZoneC \global\MoreTablesfalse\fi
    \else
     \def\Zone{\InZoneA}
     \ifTabInZoneB \def\Zone{\InZoneC}\fi
    \fi
   \fi
   \ifMoreTables\Print{}\TableItems\fi
 \fi
   \MakePageInCompletefalse % assume page is complete
   \ifMoreFigures\MakePageInCompletetrue\fi
   \ifMoreTables\MakePageInCompletetrue\fi
 \repeat
%\Print{TextSize=\the\TextSize}
%\Print{ZoneBSize=\the\ZoneBSize}
 \ifZoneAFullPage
  \global\TextSize=\z@
  \global\ZoneBSize=\z@
  \global\vsize=\z@\relax
  \global\topskip=\z@\relax
  \vbox to \z@{\vss}
  \eject
 \else
 \global\advance\ZoneBSize by -\ZoneBAdjust
 \global\vsize=\ZoneBSize
 \global\hsize=\ColumnWidth
 \global\ZoneBAdjust=\z@
 \ifdim\TextSize<23pt
 \Warn{}
 \Warn{* Making column fall short: TextSize=\the\TextSize *}
 \vskip-\lastskip\eject\fi
 \fi
}

\def\MakeRightCol{% allocate space for the right column of text
 \global\TextSize=\ZoneBSize
 \MakePageInCompletetrue
 \MoreFigurestrue
 \MoreTablestrue
 \global\FirstSingleItemtrue
 \global\setbox\ZoneBBOX=\box\VOIDBOX
 \def\Zone{\InZoneB}
 \loop
  \ifMakePageInComplete
 \FindNext
 \ifnum\StackPointer=\m@ne
  \NextItem=\m@ne
  \MoreFiguresfalse
  \MoreTablesfalse
 \fi
 \ifnum\NextItem=\Figure
   \FindItem{\Figure}{\NextFigure}
   \ifnum\StackPointer=\m@ne \MoreFiguresfalse
   \else
    \GetItemSPAN{\StackPointer}
    \ifnum\ItemSPAN=\Double\relax
     \MoreFiguresfalse\fi
   \fi
   \ifMoreFigures\Print{}\FigureItems\fi
 \fi
 \ifnum\NextItem=\Table
   \FindItem{\Table}{\NextTable}
   \ifnum\StackPointer=\m@ne \MoreTablesfalse
   \else
    \GetItemSPAN{\StackPointer}
    \ifnum\ItemSPAN=\Double\relax
     \MoreTablesfalse\fi
   \fi
   \ifMoreTables\Print{}\TableItems\fi
 \fi
   \MakePageInCompletefalse % assume page is complete
   \ifMoreFigures\MakePageInCompletetrue\fi
   \ifMoreTables\MakePageInCompletetrue\fi
 \repeat
 \ifZoneAFullPage
  \global\TextSize=\z@
  \global\ZoneBSize=\z@
  \global\vsize=\z@\relax
  \global\topskip=\z@\relax
  \vbox to \z@{\vss}
  \eject
 \else
 \global\vsize=\ZoneBSize
 \global\hsize=\ColumnWidth
 \ifdim\TextSize<23pt
 \Warn{}
 \Warn{* Making column fall short: TextSize=\the\TextSize *}
 \vskip-\lastskip\eject\fi
\fi
}

\def\FigureItems{% Stack pointer points to next figure
 \Print{Considering...}
 \ShowItem{\StackPointer}
 \GetItemBOX{\StackPointer} % auto calculates ItemSIZE
 \GetItemSPAN{\StackPointer}
  \CheckFitInZone % check to see if item fits
  \ifnum\ItemFits=\Yes
   \ifnum\ItemSPAN=\Single
     \ChangeStatus{\StackPointer}{\InZoneB} % flag to be output
     \global\FigInZoneBtrue
     \ifFirstSingleItem
      \hbox{}\vskip-\BodgeHeight
     \global\advance\ItemSIZE by \TextLeading
     \fi
     \unvbox\ItemBOX\ItemSep
     \global\FirstSingleItemfalse
     \global\advance\TextSize by -\ItemSIZE% allocate space
     \global\advance\TextSize by -\TextLeading
   \else
    \ifFirstZoneA
     \global\advance\ItemSIZE by \TextLeading
     \global\FirstZoneAfalse\fi
    \global\advance\TextSize by -\ItemSIZE
    \global\advance\TextSize by -\TextLeading
    \global\advance\ZoneBSize by -\ItemSIZE
    \global\advance\ZoneBSize by -\TextLeading
    \ifFigInZoneB\relax
     \else
     \ifdim\TextSize<3\TextLeading
     \global\ZoneAFullPagetrue
     \fi
    \fi
    \ChangeStatus{\StackPointer}{\Zone}
    \ifnum\Zone=\InZoneC \global\FigInZoneCtrue\fi
  \fi
   \Print{TextSize=\the\TextSize}
   \Print{ZoneBSize=\the\ZoneBSize}
  \global\advance\NextFigure \@ne
   \Print{This figure has been placed.}
  \else
   \Print{No space available for this figure...holding over.}
   \Print{}
   \global\MoreFiguresfalse
  \fi
}

\def\TableItems{% Stack pointer points to next table
 \Print{Considering...}
 \ShowItem{\StackPointer}
 \GetItemBOX{\StackPointer} % auto calculates ItemSIZE
 \GetItemSPAN{\StackPointer}
  \CheckFitInZone % check to see of item fits in Zone
  \ifnum\ItemFits=\Yes
   \ifnum\ItemSPAN=\Single
    \ChangeStatus{\StackPointer}{\InZoneB}
     \global\TabInZoneBtrue
     \ifFirstSingleItem
      \hbox{}\vskip-\BodgeHeight
     \global\advance\ItemSIZE by \TextLeading
     \fi
     \unvbox\ItemBOX\ItemSep
     \global\FirstSingleItemfalse
     \global\advance\TextSize by -\ItemSIZE
     \global\advance\TextSize by -\TextLeading
   \else
    \ifFirstZoneA
    \global\advance\ItemSIZE by \TextLeading
    \global\FirstZoneAfalse\fi
    \global\advance\TextSize by -\ItemSIZE
    \global\advance\TextSize by -\TextLeading
    \global\advance\ZoneBSize by -\ItemSIZE
    \global\advance\ZoneBSize by -\TextLeading
    \ifFigInZoneB\relax
     \else
     \ifdim\TextSize<3\TextLeading
     \global\ZoneAFullPagetrue
     \fi
    \fi
    \ChangeStatus{\StackPointer}{\Zone}
    \ifnum\Zone=\InZoneC \global\TabInZoneCtrue\fi
   \fi
%   \Print{TextSize=\the\TextSize}
%   \Print{ZoneBSize=\the\ZoneBSize}
  \global\advance\NextTable \@ne
   \Print{This table has been placed.}
  \else
  \Print{No space available for this table...holding over.}
   \Print{}
   \global\MoreTablesfalse
  \fi
}

% These macros check to see if an item of ItemSIZE will
% fit in a particular zone. If it will, then ItemFits
% will be set true else false.

\def\CheckFitInZone{%
{\advance\TextSize by -\ItemSIZE
 \advance\TextSize by -\TextLeading
 \ifFirstSingleItem
  \advance\TextSize by \TextLeading
 \fi
 \ifnum\Zone=\InZoneA\relax
  \else \advance\TextSize by -\ZoneBAdjust
 \fi
 \ifdim\TextSize<3\TextLeading \global\ItemFits=\No
 \else \global\ItemFits=\Yes\fi}
}

\def\BeginOpening{%
  \thispagestyle{titlepage}%
  \global\setbox\ItemBOX=\vbox\bgroup%
    \hsize=\PageWidth%
    \hrule height \z@
    \ifsinglecol\vskip 6pt\fi % Bodge, to get same pos. as two-column!
}

\let\begintopmatter=\BeginOpening  %  alias for \BeginOpening

\def\EndOpening{%
  \One%  1 line fixed space below opening
  \egroup
  \ifsinglecol
    \box\ItemBOX%
    \vskip\TextLeading plus 2\TextLeading% var. space: min 1, max 3 lines
    \@noafterindent
  \else
    \ItemNUMBER=\z@%
    \ItemTYPE=\Figure
    \ItemSPAN=\Double
    \ItemSTATUS=\InStack
    \JoinStack
  \fi
}

% Figures

\newif\if@here  \@herefalse

\def\no@float{\global\@heretrue}
\let\nofloat=\relax % only enabled for one column

\def\beginfigure{%
  \@ifstar{\global\@dfloattrue \@bfigure}{\global\@dfloatfalse \@bfigure}%
}

\def\@bfigure#1{%
  \par
  \if@dfloat
    \ItemSPAN=\Double
    \TEMPDIMEN=\PageWidth
  \else
    \ItemSPAN=\Single
    \TEMPDIMEN=\ColumnWidth
  \fi
  \ifsinglecol
    \TEMPDIMEN=\PageWidth
  \else
    \ItemSTATUS=\InStack
    \ItemNUMBER=#1%
    \ItemTYPE=\Figure
  \fi
  \bgroup
    \hsize=\TEMPDIMEN
    \global\setbox\ItemBOX=\vbox\bgroup
      \eightpoint\nostb@ls{10pt}%
      \let\caption=\fig@caption
      \ifsinglecol \let\nofloat=\no@float\fi
}

\def\fig@caption#1{%
  \vskip 5.5pt plus 6pt%
  \bgroup % grouping and size change needed for plate pages
    \eightpoint\nostb@ls{10pt}%
    \setbox\TEMPBOX=\hbox{#1}%
    \ifdim\wd\TEMPBOX>\TEMPDIMEN
      \noindent \unhbox\TEMPBOX\par
    \else
      \hbox to \hsize{\hfil\unhbox\TEMPBOX\hfil}%
    \fi
  \egroup
}

\def\endfigure{%
  \par\egroup % end \vbox
  \egroup
  \ifsinglecol
    \if@here \midinsert\global\@herefalse\else \topinsert\fi
      \unvbox\ItemBOX
    \endinsert
  \else
    \JoinStack
    \Print{Processing source for figure \the\ItemNUMBER}%
  \fi
}

% Tables

\newbox\tab@cap@box
\def\tab@caption#1{\global\setbox\tab@cap@box=\hbox{#1\par}}

\newtoks\tab@txt@toks
\long\def\tab@txt#1{\global\tab@txt@toks={#1}\global\table@txttrue}

\newif\iftable@txt  \table@txtfalse
\newif\if@dfloat    \@dfloatfalse

\def\begintable{%
  \@ifstar{\global\@dfloattrue \@btable}{\global\@dfloatfalse \@btable}%
}

\def\@btable#1{%
  \par
  \if@dfloat
    \ItemSPAN=\Double
    \TEMPDIMEN=\PageWidth
  \else
    \ItemSPAN=\Single
    \TEMPDIMEN=\ColumnWidth
  \fi
  \ifsinglecol
    \TEMPDIMEN=\PageWidth
  \else
    \ItemSTATUS=\InStack
    \ItemNUMBER=#1%
    \ItemTYPE=\Table
  \fi
  \bgroup
    \eightpoint\nostb@ls{10pt}%
    \global\setbox\ItemBOX=\vbox\bgroup
      \let\caption=\tab@caption
      \let\tabletext=\tab@txt
      \ifsinglecol \let\nofloat=\no@float\fi
}

\def\endtable{%
  \par\egroup % end \vbox
  \egroup
  \setbox\TEMPBOX=\hbox to \TEMPDIMEN{%
    \hss
    \vbox{%
      \hsize=\wd\ItemBOX
      \ifvoid\tab@cap@box
      \else
        \noindent\unhbox\tab@cap@box
        \vskip 5.5pt plus 6pt%
      \fi
      \box\ItemBOX
      \iftable@txt
        \vskip 10pt%
        \eightpoint\nostb@ls{10pt}%
        \noindent\the\tab@txt@toks
        \global\table@txtfalse
      \fi
    }%
    \hss
  }%
  \ifsinglecol
    \if@here \midinsert\global\@herefalse\else \topinsert\fi
      \box\TEMPBOX
    \endinsert
  \else
    \global\setbox\ItemBOX=\box\TEMPBOX
    \JoinStack
    \Print{Processing source for table \the\ItemNUMBER}%
  \fi
}

\def\UnloadZoneA{%
\FirstZoneAtrue
 \Iteration=\z@
  \loop
   \ifnum\Iteration<\LengthOfStack
    \GetItemSTATUS{\Iteration}
    \ifnum\ItemSTATUS=\InZoneA
     \GetItemBOX{\Iteration}
     \ifFirstZoneA \vbox to \BodgeHeight{\vfil}%
     \FirstZoneAfalse\fi
     \unvbox\ItemBOX\ItemSep
     \LeaveStack{\Iteration}
     \else
     \advance\Iteration \@ne
   \fi
 \repeat
}

\def\UnloadZoneC{%
\Iteration=\z@
  \loop
   \ifnum\Iteration<\LengthOfStack
    \GetItemSTATUS{\Iteration}
    \ifnum\ItemSTATUS=\InZoneC
     \GetItemBOX{\Iteration}
     \ItemSep\unvbox\ItemBOX
     \LeaveStack{\Iteration}
     \else
     \advance\Iteration \@ne
   \fi
 \repeat
}

%--------------------------------------------------------%
%                         DIAGNOSTICS                    %
%--------------------------------------------------------%

\def\ShowItem#1{% Show details of on item entry in stack
  {\GetItemAll{#1}
  \Print{\the#1:
  {TYPE=\ifnum\ItemTYPE=\Figure Figure\else Table\fi}
  {NUMBER=\the\ItemNUMBER}
  {SPAN=\ifnum\ItemSPAN=\Single Single\else Double\fi}
  {SIZE=\the\ItemSIZE}}}
}

\def\ShowStack{% 
 \Print{}
 \Print{LengthOfStack = \the\LengthOfStack}
 \ifnum\LengthOfStack=\z@ \Print{Stack is empty}\fi
 \Iteration=\z@
 \loop
 \ifnum\Iteration<\LengthOfStack
  \ShowItem{\Iteration}
  \advance\Iteration \@ne
 \repeat
}

\def\B#1#2{%
\hbox{\vrule\kern-0.4pt\vbox to #2{%
\hrule width #1\vfill\hrule}\kern-0.4pt\vrule}
}

%-------------------------------------------------------%
%             SINGLE COLUMN OUTPUT ROUTINE              %
%-------------------------------------------------------%

\newif\ifsinglecol   \singlecolfalse

\def\onecolumn{%
  \global\output={\singlecoloutput}%
  \global\hsize=\PageWidth
  \global\vsize=\PageHeight
  \global\ColumnWidth=\hsize
  \global\TextLeading=12pt
  \global\Leading=12
%%%  \global\topskip=\TextLeading%%%
  \global\singlecoltrue
  \global\let\onecolumn=\relax%         stop them using \onecolumn again
  \global\let\footnote=\sing@footnote%  enable footnotes
  \global\let\vfootnote=\sing@vfootnote
  \ninepoint % reset \baselineskip after leading change
  \message{(Single column)}%
}

\def\singlecoloutput{%
  \shipout\vbox{\PageHead\pagebody\PageFoot}%
  \advancepageno
  \ifplate@page
    \shipout\vbox{%
      \sp@pagetrue
      \def\sp@type{plate}%
      \global\plate@pagefalse
      \PageHead\vbox to \PageHeight{\unvbox\plt@box\vfil}\PageFoot%
    }%
    \message{[plate]}%
    \advancepageno
  \fi
  \ifnum\outputpenalty>-\@MM \else\dosupereject\fi%
}

\def\ItemSep{\vskip\ItemSepamount\relax}

\def\ItemSepbreak{\par\ifdim\lastskip<\ItemSepamount
  \removelastskip\penalty-200\ItemSep\fi%
}

% Modify plain's \endinsert so that the mn's spacing is used

\let\@@endinsert=\endinsert % save plain's original \endinsert

\def\endinsert{\egroup % finish the \vbox
  \if@mid \dimen@\ht\z@ \advance\dimen@\dp\z@ \advance\dimen@12\p@
    \advance\dimen@\pagetotal \advance\dimen@-\pageshrink
    \ifdim\dimen@>\pagegoal\@midfalse\p@gefalse\fi\fi
  \if@mid \ItemSep\box\z@\ItemSepbreak
  \else\insert\topins{\penalty100 % floating insertion
    \splittopskip\z@skip
    \splitmaxdepth\maxdimen \floatingpenalty\z@
    \ifp@ge \dimen@\dp\z@
    \vbox to\vsize{\unvbox\z@\kern-\dimen@}% depth is zero
    \else \box\z@\nobreak\ItemSep\fi}\fi\endgroup%
}

% Footnotes (only enabled in single column)

\def\gobbleone#1{}
\def\gobbletwo#1#2{}
\let\footnote=\gobbletwo % Gobble footnote's unless enabled by \onecolumn
\let\vfootnote=\gobbleone

\def\sing@footnote#1{\let\@sf\empty % parameter #2 (the text) is read later
  \ifhmode\edef\@sf{\spacefactor\the\spacefactor}\/\fi
  \hbox{$^{\hbox{\eightpoint #1}}$}\@sf\sing@vfootnote{#1}%
}

\def\sing@vfootnote#1{\insert\footins\bgroup\eightpoint\b@ls{9pt}%
  \interlinepenalty\interfootnotelinepenalty
  \splittopskip\ht\strutbox % top baseline for broken footnotes
  \splitmaxdepth\dp\strutbox \floatingpenalty\@MM
  \leftskip\z@skip \rightskip\z@skip \spaceskip\z@skip \xspaceskip\z@skip
  \noindent $^{\scriptstyle\hbox{#1}}$\hskip 4pt%
    \footstrut\futurelet\next\fo@t%
}

% Kill footnote rule
\def\footnoterule{\kern-3\p@ \hrule height \z@ \kern 3\p@}

\skip\footins=19.5pt plus 12pt minus 1pt
\count\footins=1000
\dimen\footins=\maxdimen

% Landscape

\def\landscape{%
  \global\TEMPDIMEN=\PageWidth
  \global\PageWidth=\PageHeight
  \global\PageHeight=\TEMPDIMEN
  \global\let\landscape=\relax%         stop them using \landscape again.
  \onecolumn
  \message{(landscape)}%
  \raggedbottom
}

%-------------------------------------------------------%
%               TWO COLUMN OUTPUT ROUTINE               %
%-------------------------------------------------------%

\output{%
  \ifLeftCOL
    \global\setbox\LeftBOX=\vbox to \ZoneBSize{\box255\unvbox\ZoneBBOX}%
    \global\LeftCOLfalse
    \MakeRightCol
  \else
    \setbox\RightBOX=\vbox to \ZoneBSize{\box255\unvbox\ZoneBBOX}%
    \setbox\MidBOX=\hbox{\box\LeftBOX\hskip\ColumnGap\box\RightBOX}%
    \setbox\PageBOX=\vbox to \PageHeight{%
      \UnloadZoneA\box\MidBOX\UnloadZoneC}%
    \shipout\vbox{\PageHead\box\PageBOX\PageFoot}%
    \advancepageno
    \ifplate@page
      \shipout\vbox{%
        \sp@pagetrue
        \def\sp@type{plate}%
        \global\plate@pagefalse
        \PageHead\vbox to \PageHeight{\unvbox\plt@box\vfil}\PageFoot%
      }%
      \message{[plate]}%
      \advancepageno
    \fi
    \global\LeftCOLtrue
    \CleanStack
    \MakePage
  \fi
}

% Startup message

\Warn{\start@mess}

 % so articles can see if a format file has been used.

\catcode `\@=12 % @ signs are non-letters

% \dump

% end of mn.tex

%%%%%%%%%%%%%%%%%%%%%%%%%%%%%%%%%%%%%%%%%%%%
% Simple user defined macros               %
%%%%%%%%%%%%%%%%%%%%%%%%%%%%%%%%%%%%%%%%%%%%

\newif\ifprintcomments
 
% uncomment either of the two, to either print or not print comments
 
\printcommentstrue
%\printcommentsfalse

%%%%% Useful pieces of text that occur often

\def\etal{et al.~}

%%%%% Physical Units for use in math mode

\def\Msun{{\rm\,M_\odot}}

%%%%% Physical quantities for use in math mode

\def\MBH{{M_{\rm BH}}}

%%%%% Symbols for use in math mode

\def\deg{^{\circ}}

%  \lta and \gta : produce > and < signs with twiddle underneath

\def\spose#1{\hbox to 0pt{#1\hss}}
\def\lta{\mathrel{\spose{\lower 3pt\hbox{$\sim$}}
    \raise 2.0pt\hbox{$<$}}}
\def\gta{\mathrel{\spose{\lower 3pt\hbox{$\sim$}}
    \raise 2.0pt\hbox{$>$}}}

%%%%% Useful measures of length for use with psfig

\newdimen\hssize
\hssize=8.4truecm
\newdimen\hdsize
\hdsize=17.7truecm

%%%%% Useful macros for this particular paper

\def\df{distribution function }

\def\feelz{$f_{e}(E,L_{z})$ }
\def\foelz{$f_{o}(E,L_{z})$ }

%%%%%%%%%%%%%%%%%%%%%%%%%%%%%%%%%%%%%%%%%%%%%%%%%%%%%%%%%%%%%%%%%%%%%%%

\def\today{\ifcase\month\or
 January\or February\or March\or April\or May\or June\or
 July\or August\or September\or October\or November\or December\fi
 \space\number\day, \number\year}
 
%%%%%%%%%%%%%%%%%%%%%%%%%%%%%%%%%%%%%%%%%%%%%%%%%%%%%%%%%%%%%%%%%%%%%%%
%  Equation numbering
%%%%%%%%%%%%%%%%%%%%%%%%%%%%%%%%%%%%%%%%%%%%%%%%%%%%%%%%%%%%%%%%%%%%%%%

\newcount\eqnumber
\eqnumber=1
\def\chaphead{} 

\def\new{\hbox{(\rm\chaphead\the\eqnumber)}\global\advance\eqnumber by 1}
   % \new macro produces sequentially numbered equations by writing \eqno\new
   % at end of displayed equations
 
\def\first{\hbox{(\rm\chaphead\the\eqnumber a)}\global\advance\eqnumber by 1}
\def\last#1{\advance\eqnumber by -1 
            \hbox{(\rm\chaphead\the\eqnumber#1)}\advance
     \eqnumber by 1}
   % \first and \last are useful for equations in parts.
   %   \first increases eqnumber by 1 and adds "a"
   %   \last{b} doesn't increase by 1 and adds "b"
 
\def\ref#1{\advance\eqnumber by -#1 \chaphead\the\eqnumber
     \advance\eqnumber by #1}
   % to refer to an equation which is 5 equations back, write "equation (\ref5)" 
\def\nref#1{\advance\eqnumber by -#1 \chaphead\the\eqnumber
     \advance\eqnumber by #1}
   % without the parenthesis at the beginning.

\def\eqnam#1{\xdef#1{\chaphead\the\eqnumber}}
   % to name an equation, place command "\eqnam{\Poisson}" before equation, and
   % thereafter "equation \Poisson)" will generate the proper equation number.

%\def\new{\hbox{(\the\eqnumber )}\global\advance\eqnumber by 1}
   % \new macro produces sequentially numbered equations by writing \eqno\new)
   % at end of displayed equations
 
%\def\first{\hbox{(\the\eqnumber a)}\global\advance\eqnumber by 1}
%\def\last#1{\advance\eqnumber by -1 \hbox{(\the\eqnumber#1)}\advance
%     \eqnumber by 1}
   % \first and \last are useful for equations in parts.
   %   \first increases eqnumber by 1 and adds "a"
   %   \last{b} doesn't increase by 1 and adds "b"
 
%\def\ref#1{\advance\eqnumber by -#1 \the\eqnumber
%     \advance\eqnumber by #1}
   % to refer to an equation which is 5 equations back, write "equation \ref5)"
 
%\def\nref#1{\advance\eqnumber by -#1 \the\eqnumber
%     \advance\eqnumber by #1}
   % without the parenthesis at the beginning.

%\def\eqnam#1{\xdef#1{\the\eqnumber}}
   % to name an equation, place command "\eqnam{\Poisson}" before equation, and
   % thereafter "equation \Poisson)" will generate the proper equation number.

%%%%%%%%%%%%%%%%%%%%%%%%%%%%%%%%%%%%%%%%%%%%
% Beginning of paper                       %
%%%%%%%%%%%%%%%%%%%%%%%%%%%%%%%%%%%%%%%%%%%%

% Uncomment if single column format is required
% \onecolumn

% Modify to change offset of the text on the page
%\pageoffset{-0.85truecm}{-1.05truecm}
\pageoffset{-0.85truecm}{0.45truecm}

% Uncomment for author footline
% \authorcomment{}

% Autonumbering of sections, subsections and subsubsections is not used
% \Autonumber

% To be modified at the Journals office
\pagerange{000-000}
\pubyear{1996}
\volume{000}

% Start with material that spans two columns
\begintopmatter

\title{Self-Consistent, Axisymmetric Two-Integral Models of Elliptical 
       Galaxies with Embedded Nuclear Discs}

\author{Frank C.\ van den Bosch and P.\ Tim de Zeeuw}

\affiliation{Leiden Observatory, P.O. Box 9513, 2300 RA Leiden, 
             The Netherlands}

\shortauthor{F.C.\ van den Bosch and P.T.\ de Zeeuw}

\shorttitle{%
Models of Elliptical Galaxies with Nuclear Discs}

% To be filled in at the Journals office
% \acceptedline{}

\abstract{%
Ionized gas discs in the nuclei of ellipticals have proven to be
excellent tools for the determination of the central mass density in
these galaxies.  The recent discovery with the Hubble Space Telescope
of small {\it stellar} discs embedded in the nuclei of a number of
ellipticals and S0s might be of similar importance.  We
construct two-integral axisymmetric models for such systems. The
models consist of a spheroidal bulge with a central density cusp, and
a disc described by a strongly flattened exponential spheroid.  We use
the Hunter \& Qian (1993) method to calculate the even part of the
phase-space distribution function (DF), and specify the odd part by
means of a simple parameterization.  We consider both local stability
against axisymmetric perturbations, as well as global stability
against bar forming modes, and find that our models are stable as long
as the discs are not too flat and/or compact. The margin of stability is
derived as a function of disc scalelength and central surface density.
Its location agrees well with the observed values of these disc parameters.
This suggests that discs build up their mass until they become marginally
stable.

We investigate the photometric as well as the kinematic signatures of
nuclear discs, including their velocity profiles (VPs), and study the
influence of seeing convolution. In particular, we study to what
extent these kinematic signatures can be used to determine the central
density of the galaxy, and to test for the presence of massive black
holes. We consider nuclear discs that are either dynamically coupled
to or decoupled from the host elliptical, including counter-rotating
discs. The latter are models for the counter-rotating cores observed
in a number of galaxies, which are often found to exhibit discy
isophotes in the central region. The counter-rotation is only
detectable when the disc light contributes significantly to the
central velocity profiles. We find that in this case the observed
velocity dispersion will show a central decrease.

The rotation curve of a nuclear disc gives an excellent measure of the
central mass-to-light ratio whenever the VPs clearly reveal the
narrow, rapidly rotating component associated with the nuclear disc.
Steep cusps and seeing convolution both result in central VPs that are
dominated by the bulge light, and these VPs barely show the presence
of the nuclear disc, impeding measurements of the central rotation
velocities of the disc stars. However, if a massive BH is present, the
disc component of the VP can be seen in the wing of the bulge part,
and measurements of its mean rotation provide a clear signature of the
presence of the BH. This signature is insensitive to the uncertainties in 
the velocity anisotropy, which often lead to ambiguity in the interpretation
of a central rise in velocity dispersion as due to a central BH.}
% end of abstract

\keywords{% 
stellar dynamics -- galaxies: kinematics and dynamics --
galaxies:  ellipticals -- galaxies: structure -- galaxies: nuclei -- 
line: profiles} % end of keywords

% finish material that spans two columns
\maketitle  

% Uncomment this for referee mode
% \Referee   

\section{1 Introduction}

Ever since the realization that elliptical galaxies are not smooth
structureless systems of old stars supported by rotation, numerous
studies have revealed the increasingly complex nature of these objects
(de Zeeuw \& Franx 1991). In addition to the bright, pressure
supported triaxial systems with boxy isophotes, there are the discy
ellipticals that are mainly supported by rotation. The central regions
often exhibit complex structures.  Counter-rotating cores, stellar
cusps, and nuclear activity are all phenomena connected with the
nuclei of elliptical galaxies (Kormendy \& Richstone 1995). In
addition, the nuclear regions might conceal massive black holes.

Numerous studies over the last decade have concentrated on the nature
of the discy ellipticals (e.g., Capaccioli, Held, \& Nieto 1987; 
Carter 1987; van den Bergh 1989; Nieto \etal 1991; Rix \& White 1990, 
1992; Scorza \& Bender 1990).  Scorza \& Bender (1995) have shown
that discy ellipticals follow the same trend as observed for spirals
and S0s that discs with smaller scale lengths have higher central
surface brightness (e.g., Kent 1985). This strongly suggests that
discy ellipticals are two-component systems that form a continuing
sequence in the Hubble diagram from S0s to galaxies with smaller
disc-to-bulge ratios (see also Michard 1984; Capaccioli, Caon \&
Rampazzo 1990).

The Hubble Space Telescope (HST) has revealed very small (scale length
$\sim 20$pc), nuclear discs in a number of ellipticals (van den Bosch
\etal 1994; Forbes 1994; Lauer \etal 1995).  Several of these galaxies 
also harbour a much larger kpc-scale disc which may have an inner
cut-off radius at $200-400$pc. Such multiple component discs have been
observed in other galaxies as well (e.g., Seifert 1990; Scorza \&
Bender 1995). For example, Burkhead (1986) and Kormendy (1988) argue
that the huge disc of the Sombrero galaxy (NGC~4594) consists of two
separate components.  Emsellem \etal (1994) even find an indication
for a third disc component in the Sombrero, namely a red nuclear disc
that remains unresolved from the ground.

In this paper we investigate the dynamical properties and observables
of nuclear discs. We construct self-consistent axisymmetric dynamical
models of a spheroidal bulge with an embedded nuclear disc. In
particular we investigate whether the observable dynamical properties
of nuclear discs can be used to discriminate whether or not these
galaxies harbour a nuclear black hole (BH). Proving the presence of a
BH in the nucleus of an elliptical galaxy is complicated by the fact
that the three-dimensional velocity distribution of the stars in the
central region can be strongly anisotropic. For example, the observed
central increase of the velocity dispersion in M~87 can equally well
be fitted by an isotropic model with a central BH as by a radially
anisotropic model without any BH (Young \etal 1978; Sargent \etal
1978; Binney \& Mamon 1982; van der Marel 1994). Discs, on the other
hand, are cold components, strongly dominated by rotation.
Furthermore, they are generally thin, so that the interpretation of
the observed line-of-sight velocity distribution is less complicated
than for the three-dimensional distribution of stars in the
ellipsoidal component. Discs embedded in the nuclei of elliptical
galaxies offer a promising means of detecting nuclear BHs.  The recent
discovery of a disc of ionized gas in the nucleus of M87 (Ford \etal
1994) has indeed greatly strengthened the case for a nuclear BH in
this giant elliptical (Harms \etal 1994). Even more spectacular
evidence for the presence of a nuclear BH is provided by the VLBA
observations of emission of a disc of water masers with a Keplerian
rotation curve at only 0.13 pc from the centre of NGC~4258 (Miyoshi
\etal 1995). However, the interpretation of the dynamics of gas discs
can be complicated, especially in AGNs where outflow, inflow, and
turbulent phenomena are very likely to play a major role (see e.g.,
van den Bosch \& van der Marel 1995; Jaffe \etal 1996). Since the
motion of stars is purely gravitational, {\em stellar} discs do not
suffer from these effects, and their kinematics might offer valuable
tests of the presence of nuclear BHs. Recent absorption line
spectroscopy with the Faint Object Spectrograph aboard the HST of the
nucleus of NGC~3115, one of the galaxies with a bright nuclear,
stellar disc, has indeed provided an excellent case for the presence
of a $2 \times 10^9 \Msun$ BH (Kormendy \etal 1996).

This paper is organized as follows. In Section 2 we describe our
models. We then investigate the photometric observables of the nuclear
discs in Section 3, with attention to their effect on the surface
brightness profiles and isophotes. In Section 4 we construct both the
even and the odd part of the DF. We study the stability of the
resulting models in Section 5. In Section 6 we discuss the kinematic
signatures of nuclear discs and the influence of seeing
convolution. We investigate the influence of a nuclear black hole in
Section 7, and summarize our results in Section 8.

\eqnumber=1
\def\chaphead{\hbox{2.}}

\section{2 Multi-component Models}

We construct galaxy models consisting of a thick disc embedded in an
spheroidal body: the `bulge'. In addition, we allow a black hole to be
present at the centre of the galaxy. The potential of the model can
therefore be written as
\eqnam\totpot
$$\Psi(R,z) = \psi_{\rm bulge}(R,z) + \psi_{\rm disc}(R,z) + 
\psi_{\rm BH}(R,z).                                       \eqno\new $$
We denote the {\it total} potential by $\Psi(R,z)$, and use
$\psi(R,z)$ to indicate the potential of a separate component. We use
the {\it relative} potential $\Psi = - \Phi$ instead of the potential
$\Phi$. Therefore stars at rest in the centre of the potential well
have the {\it maximum} energy.

We define a disc to be {\it nuclear} if its horizontal scale length
$R_{d}$ is smaller than half the core radius $R_{b}$ of the bulge.
This is defined as the inner core- or break radius of the bulge
component which $\sim 2'' - 3''$ at Virgo (Ferrarese \etal 1994), and
should not be confused with the `the Vaucouleurs' effective radius
which is much larger. Although this paper only focuses on such nuclear
discs, the method discussed here is also applicable to larger discs,
and even to S0 galaxies, by simply increasing the ratio $R_{d}/R_{b}$.

Throughout this paper we take the mass-to-light ratio $M/L$ to be
equal to one, so that, e.g., the projected surface brightness is
identical to the projected surface density.

\subsection{2.1 The bulge component}

We describe the spheroidal component by the so-called ($\alpha$,
$\beta$) models, which have a density distribution given by
\eqnam\bulgedens
$$\rho(R,z) = \rho_{0,b} \left({m \over R_b}\right)^{\alpha} 
\left(1 + {m^2 \over R_b^2}\right)^{\beta},                 \eqno\new $$
where 
$$m = \sqrt{R^2 + {z^2 \over q_b^2}}.                       \eqno\new $$
Here $R_b$ is the `break-radius' of the bulge, and $q_b$ is the
flattening of the bulge. For $\alpha = 0$ the central density is
finite and equal to $\rho_{0,b}$. When $\alpha < 0$ the density
profile has a central cusp.  These models have proven to accurately
fit the surface brightness distribution of elliptical galaxies (e.g.,
van der Marel \etal 1994, Qian \etal 1995 (hereafter QZMH), van den
Bosch \& van der Marel 1995). QZMH discuss these models in more
detail.

At large radii the density (\bulgedens) falls off proportional to
$m^{\alpha + 2 \beta}$. We consider only models with $\alpha + 2
\beta =-4$, so that the projected surface density at large
radii falls off as $R^{-3}$. In this case the total mass is finite, and 
given by
\eqnam\bulgemass
$$M = 2 \pi q_b \rho_{0,b} R_b^3 B\bigl( {\textstyle {\alpha \over 2} + 
{3 \over 2},{1 \over 2}}\bigr),\eqno\new $$
where $B$ is the beta-function. These models are similar to the set of
models discussed by Dehnen \& Gerhard (1994), except that their
density has $(1 + m/R_b)^{2\beta}$ as second factor rather than $(1 +
m^2/R_b^2)^{\beta}$.

QZMH showed how the classical double quadrature expression (e.g.,
Chandrasekhar 1969) for the potential of any axisymmetric system in
which the density is stratified on similar concentric spheroids, i.e.,
$\rho = \rho(m^2)$, can always be rewritten as a single quadrature.
For our particular choice of $\alpha + 2 \beta =-4$, however, the
inner integration of the classical formula can be carried out
explicitly, and therefore leads directly to a simple single
quadrature, which is evaluated more efficiently. According to the
classical formula, the potential that corresponds to (\bulgedens) is
given by
\eqnam\classpot
$$\psi(R,z) = \psi_{0,b} - \pi G q_b \int\limits_0^{\infty}
{F(t) {\rm d}\tau \over (\tau + 1) \sqrt{\tau + q_b^2}},\eqno\new $$
where 
\eqnam\psiofm
$$F(t) = \int\limits_{0}^{t^2} \rho(m'^2) {\rm d}m'^2.\eqno\new $$
Substitution of the density (\bulgedens) in the above integral, and
changing to the integration variable $x = {(1 + m^2/R_b^2)\over
m^2/R_b^2}$, yields
\eqnam\respsiofm
$$F(t) = {-\rho_{0,b} R_b^{2(\beta + 2)} \over 1 + \beta} 
{(1 + t^2/R_b^2)^{\beta + 1} \over t^{2(\beta + 1)}},\eqno\new $$
with
\eqnam\mrad
$$t = \sqrt{{R^2\over \tau + 1} + {z^2 \over \tau + q_b^2}}.\eqno\new $$
This result is valid when $\beta < -1$.  Substitution of expression
(\respsiofm) in the integral (\classpot) then gives the potential as
a single quadrature.  The central potential $\psi_{0,b}$ can be 
evaluated explicitly, and is given by
\eqnam\cenbulpot
$$\psi_{0,b} = {-GM \over (1 + \beta) R_b 
B({\alpha \over 2} + {3 \over 2}, {1 \over 2})}
{{\rm arcsin}\sqrt{1 - q_b^2} \over \sqrt{1 - q_b^2}}.\eqno\new $$

\subsection{2.2 The disc component}

Recent work by, e.g., Scorza \& Bender (1995), indicates that the
small discs embedded in ellipticals form a smooth transition from S0s
to galaxies with smaller disc-to-bulge ratios. We therefore assume
that our nuclear discs have a surface brightness distribution that is
similar to the discs of S0s and spirals. Most studies have considered
models with infinitesimally thin discs.  Although such discs provide a
good approximation for some purposes, they are not sufficient for our
needs, because they are maximally cold in the perpendicular ($z$-)
direction.  Real discs have non-zero $z$-motions that contribute to
the line-of-sight velocities when observed at any inclination other
than edge-on.  Furthermore, infinitesimally thin models are unphysical
in that the derivative $\partial\rho/\partial z$ is infinite at $z=0$.
We therefore consider {\it thick} discs. Observations have shown that
the surface brightness of discs is well approximated by an exponential
profile, both in the radial and the vertical direction (e.g., Freeman
1970; Aoki \etal 1991).

%%%%%%%%%%%%%%%%%%%%%%
% Begin of  Figure 1
%%%%%%%%%%%%%%%%%%%%%%
%
%\beginfigure*{1}
%\centerline{\psfig{figure=meri.ps,width=\hdsize}}\smallskip
%\caption{{\bf Figure 1.} Contour plots of potential 
%and density in the central region of the meridional plane of model
%3.1, which has a logarithmic cusp-slope $\alpha=-1.0$, a disc-to-bulge
%ratio $\Delta=0.0303$, a surface brightness ratio $\Lambda=1.0$ at one
%disc scale-length $R_d$, and no central black hole (see Table 1). }
%\endfigure
%
%%%%%%%%%%%%%%%%%%%%%%
% End of  Figure 1
%%%%%%%%%%%%%%%%%%%%%%
%

Few treatments of potential-density pairs of thick discs are available
in the literature. Kuijken \& Gilmore (1989) and subsequently
Cuddeford (1993) presented a very simple method of expanding
infinitesimally thin disc potentials to a thick disc with arbitrary 
vertical density distribution with constant scale height.  Consider an
infinitesimally thin disc, whose surface density is $\rho(R,z) =
f(R)\delta(z)$, and with corresponding potential $\psi(R,z) =
g(R,z)$. Here $f$ and $g$ are arbitrary functions.  One can thicken
this disc by replacing the Kronecker delta function $\delta(z)$ by any
other function $h(z)$ that describes the vertical density
distribution. At each $z'$ parallel to the equatorial plane the
density distribution can again be written with the delta function,
i.e., $\rho(R,z') = f(R)\delta(z-z')$, and for this density
distribution the corresponding potential is simply $\psi(R,z') =
g(R,z-z')$. The potential of the thickened disc can be built up of an
infinite number of such infinitesimally thin discs, each of which has
a corresponding potential, properly weighted by the vertical density
distribution. The potential
\eqnam\gendiscpot
$$\psi(R,z) = \int\limits_{-\infty}^{\infty} g(R,z-z') h(z') 
{\rm d}z'                                                  \eqno\new $$
follows straightforwardly.

Using the scheme outlined above, we could construct the potential
density pair of the double exponential disc, i.e., $\rho(R,z) =
\Sigma_0 \exp(-R/R_d) \exp(-\vert z \vert /z_{0})$.  The potential of
the infinitesimally thin case is already a single quadrature, so that
equation (\gendiscpot) becomes a double integration. For the special
case of a double-exponential disc this can be reduced to a single
quadrature (Kuijken \& Gilmore 1989).  This is an important advantage,
since it speeds up the numerical calculations considerably. However,
the double-exponential has the unphysical characteristic that the
derivative $\partial\rho/\partial z$ is discontinuous at
$z=0$. Furthermore, the slow oscillatory behavior of the Bessel
functions that occur in the integrand makes the numerical evaluation
of the potential and its derivatives tedious.

We therefore decided to use an `exponential spheroid', first
introduced by Kent, Dame \& Fazio (1991), to model the disc. This 
model, like the double-exponential disc, has an exponential surface
brightness along both the radial and the vertical direction.

Consider a sphere whose projected surface brightness is 
\eqnam\sphersb
$$\Sigma(R) = \Sigma_{0} \exp(-R/R_d). \eqno\new $$
The corresponding density distribution follows upon solving the
Abel integral equation
\eqnam\spherdens
$$\Sigma(R) = 2 \int\limits_{R}^{\infty} {\rho r {\rm d}r \over
\sqrt{r^2 - R^2}},\eqno\new $$
which gives
\eqnam\discdens
$$\rho(r) = \rho_{0,d} K_0(r/R_d),\eqno\new $$
where $\rho_{0,d} = \Sigma_0/\pi R_d$, and $K_0$ is a modified Bessel
function.  

We replace the sphere by a spheroid of flattening $q_d$, and introduce
the spheroids $m= \sqrt{R^2 + z^2/q_d^2}$. This leads to axisymmetric
models that project to a surface brightness that is exponential along
any axis, i.e.,
\eqnam\surfdisc
$$\Sigma_i(x,y) = {q_d\over q'_d} \Sigma_0 \exp(-R^{*}/R_d).\eqno\new $$
Here $R^{*} = \sqrt{x^2 + y^2/q'^2_d}$ and $q'_d$ is the projected
axis ratio, which is related to the intrinsic flattening and the
inclination angle $i$ through
\eqnam\projflat
$$q'^2_d = \cos^2 i + q^2_d \sin^2 i.\eqno\new$$
When $q_d \ll 1$ these models represent thick exponential discs.

The potential of the above discs can be evaluated by means of equation
(\classpot).  Substitution of equation (\discdens) in equation
(\psiofm) gives
\eqnam\respsiofm
$$F(t) = -2 \rho_{0,d} R_d t K_1({t\over R_d}),\eqno\new $$
so that the potential of the exponential spheroids is given by
\eqnam\discpot
$$\psi_d(R,z) = {G M_{D} \over \pi R_{d}^{2}} \int\limits_{0}^{\infty}
{t K_1({t \over R_d}) {\rm d}\tau \over (\tau + 1) 
\sqrt{\tau + q_d^2}},\eqno\new $$
where $t$ is again given by equation (\mrad), but with $q_b$ replaced 
by $q_d$.

The potential can be written as a single quadrature with the modified
Bessel function $K_1$ in the integrand. Since these functions decay
exponentially and are positive definite, their numerical evaluation is
much simpler than in the case of the double-exponential disc mentioned
above.  Furthermore, the derivative $\partial\rho/\partial z$ is
continuous at $z=0$.

The central density of the exponential spheroid is infinite, resulting
from the fact that $\lim_{x\to 0} K_0(x) = -{\rm ln}(x)$. However, the
total mass is finite; $M_{D} = 2 \pi^2 q_d \rho_{0,d} R_d^3$.  The
central potential is given by
\eqnam\cendispot
$$\psi_{0,d} = {2 G M_{D} \over \pi R_d} 
{\rm arcsin(\sqrt{1-q_d^2}) \over \sqrt{1-q_d^2}}.\eqno\new $$ 
%

%
%%%%%%%%%%%%%%%%%%%%%
%  Begin of Table 1
%%%%%%%%%%%%%%%%%%%%%
%
\begintable{1}
\caption{{\bf Table 1.} Parameters of the models}
\halign{#\hfil&\quad \hfil#\hfil\quad& \hfil#\hfil\quad&
\hfil#\hfil\quad& \hfil#\hfil\quad& \hfil#\hfil\quad& \hfil#\hfil \cr
Model & $\alpha$ & $\Delta$ & $\Lambda$ & $M_{\rm BH}/M_{\rm bulge}$ &
$f_{o}(E,L_{z})$ & $a$ \cr
1.0 & $0.0$ & $0.0$ & $0.0$ & 0.0 & ST & $+10$ \cr
1.1 & $0.0$ & $0.0146$ & $1.0$ & 0.0 & ST & $+10$ \cr
1.2 & $0.0$ & $0.0146$ & $1.0$ & 0.0 & NR & $+10$ \cr
1.3 & $0.0$ & $0.0146$ & $1.0$ & 0.0 & NR & $-10$ \cr
1.4 & $0.0$ & $0.0146$ & $1.0$ & 0.0 & CR & $+10$ \cr
2.0 & $-0.5$ & $0.0$ & $0.0$ & 0.0 & ST & $+10$ \cr
2.1 & $-0.5$ & $0.0207$ & $1.0$ & 0.0 & ST & $+10$ \cr
3.0 & $-1.0$ & $0.0$ & $0.0$ & 0.0 & ST & $+10$ \cr
3.1 & $-1.0$ & $0.0303$ & $1.0$ & 0.0 & ST & $+10$ \cr
3.2 & $-1.0$ & $0.0$ & $0.0$ & 0.00303 & ST & $+10$ \cr
3.3 & $-1.0$ & $0.0303$ & $1.0$ & 0.00303 & ST & $+10$ \cr
3.4 & $-1.0$ & $0.0$ & $0.0$ & 0.0303 & ST & $+10$ \cr
3.5 & $-1.0$ & $0.0303$ & $1.0$ & 0.0303 & ST & $+10$ \cr
3.6 & $-1.0$ & $0.0303$ & $1.0$ & 0.0 & CR & $+10$ \cr
4.0 & $-1.5$ & $0.0$ & $0.0$ & 0.0 & ST & $+10$ \cr
4.1 & $-1.5$ & $0.0444$ & $1.0$ & 0.0 & ST & $+10$ \cr
}
\tabletext{For each model, the cusp-steepness $\alpha$, the disc-to-bulge 
ratio $\Delta$, the ratio of surface brightness for disc and bulge at $R=R_d$
(edge-on) $\Lambda$, and the ratio of BH mass over bulge mass are shown. In 
addition, the method and the value of the parameter $a$ used to calculate 
the odd part of the DF are given. Here ST stands for the 
`standard' models, NR for `non-rotating bulge' models, and CR for 
`counter-rotating' models (see Section 4.2).}
\endtable
%
%%%%%%%%%%%%%%%%%%%%%
%  End of Table 1
%%%%%%%%%%%%%%%%%%%%%
%
	 
\subsection{2.3 The parameters}
 
In order to constrain the large parameter space of our multi-component
models we fix the values of a number of the parameters.  Since we are
mainly interested in the central parts of the models the only
bulge-parameter that we vary is the cusp steepness $\alpha$.  The
other parameters ($R_b$, $q_b$, and $\rho_{0,b}$) are kept fixed. The
value of $\beta$ is set by our requirement that $\alpha + 2 \beta =
-4$.  We choose a flattening of $0.7$ for the bulge.

For the disc we choose a fixed flattening of $q_d = 0.1$. This is
consistent with the observed flattening of several nuclear discs (van
den Bosch \etal 1994). Instead of parameterizing the disc by its mass
${\rm M_{D}}$ and scalelength $R_{d}$, we use the ratio $R_{d}/R_{b}$
and the disc-to-bulge ratio $\Delta$. We restrict ourselves to models
with $R_d/R_b = 0.2$. This value is chosen to match approximately the
values observed in a number of Virgo ellipticals (Scorza \& van den
Bosch, in preparation). Furthermore, taking $R_d/R_b = 0.2$ ensures that
our choice to restrict ourselves to models with $\alpha + 2 \beta = -4$
does not influence the results. This is due to the fact that at radii 
where the slope of the density distribution of the bulge starts to depend 
on the particular choice of $\alpha + 2 \beta$ (i.e.,  at $R \gta R_{b}$), 
the contribution of the nuclear disk is already negligible.

%%%%%%%%%%%%%%%%%%%%%%
% Begin of  Figure 2
%%%%%%%%%%%%%%%%%%%%%%
%
%\beginfigure*{2}
%\centerline{\psfig{figure=phot.ps,width=\hdsize}}\smallskip
%\caption{{\bf Figure 2.} The ellipticity and ${\rm cos}(4\theta)$ profiles 
%for models with different cusp steepness $\alpha$, for different
%values of the seeing FWHM $\chi$ (which is given in units of the
%scale-length $R_d$ of the nuclear disc), and for different values of
%the inclination angle $i$.}
%\endfigure
%
%%%%%%%%%%%%%%%%%%%%%%
% End of  Figure 2
%%%%%%%%%%%%%%%%%%%%%%
%

We define $\Delta$ as the ratio of the total luminosity of the disc to
the total luminosity of the bulge component. We assume that the
mass-to-light ratios of both components are equal, and then have
\eqnam\dtob
$$\Delta \equiv {M_{\rm disc} \over M_{\rm bulge}} = 
{\pi \over B({\alpha \over 2} + {3 \over 2}, {1 \over 2})}
{q_d \over q_b} {\rho_{0,d} \over \rho_{0,b}}  
\bigl({R_d \over R_b}\bigr)^3 .\eqno\new $$
Figure 1 shows a contour-plot of the potential and density in the
meridional plane of a model with $\alpha = -1.0$, and $\Delta =
0.0303$.  Although the presence of the nuclear disc is hardly visible
from the potential contours, the flattened density contours in the
centre clearly reveal its presence.  

We also define the ratio of the surface brightness of disc and bulge
at $R=R_d$, for an inclination angle of $i=90\deg$,
$$\Lambda \equiv {\Sigma_{i=90}^d(R_d,0) \over \Sigma_{i=90}^b(R_d,0)}. 
                                                              \eqno\new $$
We choose the disc-to-bulge ratio $\Delta$ of our models in such a way
that $\Lambda = 1.0$. In this way, the relative contribution of disc
and bulge to the velocity profiles at $R=R_d$ is equal (for
$i=90\deg$). 

Table 1 summarizes the parameters of the entire set of models.

\eqnumber=1
\def\chaphead{\hbox{3.}}
\section{3 Photometric Observables of Nuclear Discs}

We first investigate the photometric characteristics of nuclear discs
embedded in a spheroidal body.  We project the models defined in
Section 2 at different inclination angles $i$, and subsequently
seeing-convolve them with a Gaussian Point Spread Function (PSF).  The
parameter $\chi \equiv {\rm FWHM}/R_d$ gives the FWHM of the PSF
expressed in units of the horizontal disc scale length $R_d$. We scale
the galaxy such that the FWHM corresponds to $1''$ in all cases, and
take our pixels to be $0.2'' \times 0.2''$. After seeing-convolution
we determine, as a function of radius, the ellipticity and the
amplitude of the ${\rm cos}(4\theta)$-term in the Fourier expansion
that describes the deviation of the isophote from a pure ellipse
(e.g., Lauer 1985, Jedrzejewski 1987).  If this amplitude is positive
it indicates that the isophote is `discy', whereas negative amplitudes
imply `boxy' isophotes.

Figure 2 presents the results. Since we define the FWHM of the PSF to
equal $1''$, changing $\chi$ corresponds to changes in the scale of
the galaxy, i.e., its distance. We plot all parameters out to
$10''$. The presence of the nuclear disc is evident from the central
increase in ellipticity, and the disciness of the isophotes in the
central region. When the bulge harbours a steep cusp the main
photometric signature of the nuclear disc is the disciness of the
isophotes, as can be seen in the panels on the left. The panels in
the middle of Figure 2 show that the nuclear discs become
photometrically undetectable if the FWHM of the PSF exceeds $2-3$
times the disc scale length.  Furthermore, as was already shown by Rix
\& White (1990), the discs become photometrically undetectable for $i
\lta 60\deg$ (panels on the right).

When trying to decompose the surface brightness distribution in a disc
and a bulge component from the photometry alone, a number of
indicators can be used. The radius at which the ${\rm cos} 4\theta$
profile reaches its maximum is strongly correlated with the scale
length $R_d$ of the disc (Scorza \& Bender 1995). The left panels in
Figure 2 illustrate that the central ellipticity profile contains
information on the cusp steepness $\alpha$.  The amplitude of the
central changes in the ellipticity and ${\rm cos} 4\theta$ profiles
are related to the disc-to-bulge ratio (or equivalently, the central
surface brightness of the nuclear disc).  All these indicators also
depend on the inclination angle $i$. Scorza \& Bender (1995) showed
that the ${\rm cos} 6\theta$ parameter can be used to constrain $i$,
under the assumption that the disc is infinitesimally thin. However,
there is a degeneracy between disc thickness and inclination angle,
because our discs are stratified on similar concentric spheroids.
This will lead to some non-uniqueness for the disc and bulge
parameters, but the solution space will be small for sufficiently
inclined galaxies.

\eqnumber=1
\def\chaphead{\hbox{4.}}
\section{4 Construction of the $f(E, L_z)$ models}

Any collisionless system can be fully specified by its distribution
function, which, as stated by the Strong Jeans Theorem, depends on the
phase-space coordinates only through its isolating integrals of
motion. Most stellar orbits in an axisymmetric potential admit three
such isolating integrals, namely the energy $E$, the angular momentum
along the rotation axis $L_{z}$, and some third integral
$I_{3}$. Except for St\"ackel potentials (e.g., de Zeeuw 1985), the
non-classical third integral is generally not known analytically, and
may not exist for all orbits. This is the reason that most
axisymmetric models constructed to date have distribution functions
that depend only on the two classical integrals of motion; $f =
f(E,L_{z})$.  Although such models often predict too much motion on
the major axis due to the fact that the velocity dispersions in the
$R$- and $z$-directions are everywhere equal, $\sigma_{R} =
\sigma_{z}$, they provide reasonable approximations to the
dynamics of some galaxies (van der Marel 1991; QZMH; Dehnen 1995).
here we also limit ourselves to two-integral models. We discuss the 
limitations of this approach in Section 8.

\subsection{4.1 The even part of the distribution function}

The density that corresponds to the combined potential $\Psi(R,z)$ of
bulge, nuclear disc, and BH, depends only on the part of the DF that
is even in $L_z$:
\eqnam\evendf
$$\rho = {4\pi \over R} \int\limits_{0}^{\Psi} {\rm d}E
\int\limits_{0}^{R\sqrt{2(\Psi-E)}} f_{e}(E,L_{z}) {\rm d}L_{z}. \eqno\new$$
For $L_z = 0$ the inverse of this equation is given analytically by 
Eddington's formula
\eqnam\eddington
$$f_{e}(E,0)\! = \! {1 \over \sqrt{8} \pi^2} \biggl[ \int\limits_{0}^{E} \! 
{{\rm d}^2 \rho \over {\rm d}^2 \Psi} {{\rm d} \Psi \over \sqrt{E - \Psi}} +
{1\over \sqrt{E}} {\biggl( \!{{\rm d}\rho \over {\rm d} \Psi} 
                                        \! \biggr)_{\Psi = 0} \biggr]},
                                                                 \eqno\new$$
but the evaluation of \feelz for other values of $L_z$ from a given
$\rho(R,z)$ is not easy.  Lynden-Bell (1962), Hunter (1977), and
Dejonghe (1986) used different transformation methods to evaluate
\feelz for a number of models.  However, since all these methods
require the analytical knowledge of $\rho(\Psi ,R)$, they are not
widely applicable.

Recently more general schemes have become available for the evaluation
of $f_e(E, L_Z)$. Hunter \& Qian (1993, hereafter HQ) found a contour
integral expression, which is the generalization of Eddington's
formula. Other methods for generating \feelz were developed by Dehnen
(1995), Magorrian (1995), and Kuijken (1995). None of these methods
requires that the density can explicitly be written as a function of
$\Psi$, so they allow the evaluation of \feelz for any $\rho(R,z)$.
QZMH and also Dehnen (1995) successfully used these methods to
construct a two-integral model of M32 with a central black hole that
provides an accurate fit to the observed velocity profiles.

We use the HQ-method. The HQ solution for the even part \feelz of the
DF is
\eqnam\hqf
$$ f_{e}(E,L_{z}) = $$
$$ {\; \; \; \; \; \; \;} {1 \over \sqrt{8} \pi^2 i} 
\int\limits_{\Psi_{\infty}}^{[\Psi_{env}(E)+]} {\tilde \rho}_{11} 
\biggl[\xi, {L_{z}^2 \over 2(\xi-E)}\biggr] 
{{\rm d}\xi \over \sqrt{\xi - E}}.\eqno\new$$
Here the integral is along a complex contour in the plane of the
complex potential $\xi$. The two subscripts of ${\tilde \rho}_{11}$
denote the {\it second} partial derivative with respect to the {\it
first} argument ($\xi$). The tilde denotes the complex continuation of
the function $\rho_{11}$, which is given by
\eqnam\hqrho
$$ \rho_{11}(\xi,R^2) = {\rho_{22}(R^2,z^2) \over
[\Psi_2(R^2,z^2)]^2} - {\rho_{2}(R^2,z^2) \Psi_{22}((R^2,z^2) \over 
[\Psi_2(R^2,z^2)]^3}.\eqno\new$$
Each subscript $2$ denotes the partial derivative with respect to
$z^2$. The HQ paper gives further details on the method. The complete
procedure of how to calculate the complex contour integral is
discussed in detail in QZHM.

We split \feelz in two parts; a bulge part $f_{e}^{b}(E,L_{z})$, and a
disc part $f_{e}^{d}(E,L_{z})$. Each of these two terms of \feelz is
calculated by solving equation (\hqf), replacing $\rho \equiv
\rho_{\rm bulge} + \rho_{\rm disc}$ with $\rho_{\rm bulge}$ and 
$\rho_{\rm disc}$ respectively. The potential remains the {\it total}
potential $\Psi \equiv \psi_{\rm bulge} + \psi_{\rm disc}$.

We evaluate $f_{e}(E,L_{z}) = f_{e}^{b}(E,L_{z}) + f_{e}^{d}(E,L_{z})$
on a $40 \times 40$ grid in the region of the $(E,L_{z}^2)$-plane that
corresponds to bound orbits in the potential $\Psi$. This region is
bounded by $L_{z} \geq 0$, $E \leq \Psi_{\infty}$, and by the curve
$L_{z} \leq L_{z,{\rm max}}(E)$. Here $L_{z,{\rm max}}(E)$ is the
maximum allowed value of $L_{z}$ for energy $E$, and corresponds to
the angular momentum of a star with energy $E$ on a circular orbit.
For the 40 grid points that have zero angular momentum we compare the
result of the complex contour integration with the value given by 
Eddington's formula (\eddington), to check that the integral on the
chosen contour converges, and whether any adjustments of the contour
are necessary. We check for self-consistency by calculating
$\rho(R,z)$ from the DF, using equation (\evendf).  Bi-cubic spline
interpolation is used to interpolate between grid points.  The models
are self-consistent to a high degree of accuracy ($\sim 10^{-3}$).

%%%%%%%%%%%%%%%%%%%%%%
% Begin of  Figure 3
%%%%%%%%%%%%%%%%%%%%%%
%
%\beginfigure*{3}
%\centerline{\psfig{figure=dfs2.ps,width=\hdsize}}\smallskip
%\caption{{\bf Figure 3.} Surface plot of $^{10}{\rm Log}[f_{e}(E,L_{z})]$
%as function of $E/\Psi_{0}$ and $\eta^2$, with $\eta=L_z/L_{z,{\rm
%max}}(E)$, for the bulge without cusp and nuclear disc (upper left), a cusped 
%bulge without disc (upper right), non-cusped bulge with disc (lower left), 
%and the cusped bulge with disc (lower right). The ($E$,$\eta$)-grid used is 
%overplotted.}
%\endfigure
%
%%%%%%%%%%%%%%%%%%%%%%
% End of  Figure 3
%%%%%%%%%%%%%%%%%%%%%%

Figure 3 shows ${\rm log}_{10}[f_{e}(E,L_{z})]$ as function of
$E/\Psi_{0}$ and $\eta^2 \equiv [L_{z}/L_{z,{\rm max}}(E)]^2$ for four
different models: model 1.0 (upper left), model 1.1 (lower left),
model 3.0 (upper right), and model 3.1 (lower right). See Table 1 for
the parameters of each of these models. The $40 \times 40$ grid is
also shown.  In order to properly sample the DF, the grid-density
increases with energy and angular momentum. The presence of a central
cusp in the bulge density is evident from a strong rise in the centre
(i.e., at large $E/\Psi_{0}$). The presence of the nuclear disc also
induces such a strong rise. Furthermore, close to the centre the DF is
peaked towards high angular momentum, indicative of the rotational
support of the disc.

\subsection{4.2 The odd part of the \df}

Whereas the even part of the DF generates the density of the system
through equation (\evendf), the odd part of the DF specifies the mean
streaming motion:
\eqnam\odddf
$$ \rho \langle v_{\phi} \rangle = {4\pi \over R^2} \int\limits_{0}^{\Psi} 
{\rm d}E \!\!\! \int\limits_{0}^{R\sqrt{2(\Psi-E)}} \!\!\! f_{o}(E,L_{z}) 
L_{z} {\rm d}L_{z}.\eqno\new$$
Therefore, only if in addition to $\rho(R,z)$, the {\it entire} mean
streaming of the stars is known {\it a priori}, can one obtain \foelz
by inversion of equation (\odddf). Since we are primarily interested
in studying the observable properties of nuclear discs for different
amounts of rotational support, we exploit the freedom to specify
$f_{o}(E,L_{z})$ under the constraint that $f(E,L_{z}) \equiv
f_{e}(E,L_{z}) + f_{o}(E,L_{z})$ be positive.

We consider three different approaches for specifying $f_e(E, L_z)$.
In the first approach we define a function $h_{a}(\eta)$, where $\eta
\equiv L_{z}/L_{z,{\rm max}}(E)$, and simply take
\eqnam\odda
$$f_{o}(E,L_{z}) = h_{a}(\eta) \> f_{e}(E,L_{z}),\eqno\new$$
By doing so we consider the entire system as a {\it one-component}
model, i.e., although the potential is provided by the disc and the
bulge, the two components are dynamically coupled to each other since
we do not treat the streaming motions in them separately.  In order to
limit the number of free parameters we consider a
\foelz that is fully specified by only one parameter. Dejonghe (1986)
showed that there is such a functional form which has the advantage
that it maximizes the entropy of a two-integral axisymmetric
system. We follow van der Marel \etal (1994), and use a modified
version of this parameterization, which is given by
\eqnam\param
$$h_{a}(\eta) \equiv \cases{
{\rm tanh}(a\eta/2) \> / \> {\rm tanh}(a/2) & $(a>0)$\cr 
\eta & $(a=0)$\cr 
(2/a) {\rm arctanh}\big[\eta{\rm tanh}(a/2)\big] & $(a<0)$\cr}.\eqno\new$$
The other advantage of this parameterization is that these models will
automatically be physical (i.e., have positive DF) as long as
$f_{e}(E,L_{z}) > 0$. The free parameter $a$ determines the amount of
rotation, whereby the model of maximum rotation has $a=\infty$. Van
der Marel \etal (1994) and also QZMH used an extended version of this
parameterization in which they included another parameter that
determines the fraction of stars on clockwise, circular orbits in the
equatorial plane. Although this allows the study of more general
models, it invokes yet another free parameter. Therefore, we restrict
ourselves to models in which this fraction is unity, i.e., in which
all the stars on {\it circular} orbits in the equatorial plane have
the same rotation direction.

%
%%%%%%%%%%%%%%%%%%%%%%
% Begin of  Figure 4 
%%%%%%%%%%%%%%%%%%%%%%
%
%\beginfigure{2}
%\centerline{\psfig{figure=maxdtob.ps,width=\hssize}}\smallskip
%\caption{{\bf Figure 4.} The hatched regions indicate the areas
%of non-physical models, i.e., those with negative $f_e(E, L_z)$, in
%the plane of disc-to-bulge ratio $\Delta$ versus disc scale-length
%$R_d$.  We show these areas for models with $\alpha = 0.0$ and $\alpha
%= -0.25$. The forbidden region becomes smaller with decreasing
%$\alpha$ and vanishes completely for $\alpha \leq -0.5$.}
%\endfigure
%
%%%%%%%%%%%%%%%%%%%%%%
% End of  Figure 4
%%%%%%%%%%%%%%%%%%%%%%
%

The second approach allows us to construct models with a non-rotating
bulge and a rotating disc. We calculate $f_{e}^{b}(E,L_{z})$ and
$f_{e}^{d}(E,L_{z})$ separately, and then define the odd part of the
\df of disc and bulge separate from each other. This gives models that
consist of two (dynamically) decoupled components. We consider cases
in which $f_{o}^{b}(E,L_{z}) = 0$ (no rotation of the bulge
component), and
\eqnam\oddb
$$f_{o}^{d}(E,L_{z}) = h_{a}(\eta) \> f_{e}^{d}(E,L_{z}).\eqno\new$$
Here $h_{a}(\eta)$ is defined by equation (\param). 

Finally, in our third approach for specifying \foelz we construct
models with a counter-rotating disc by simply taking
\eqnam\oddca
$$f_{o}^{b}(E,L_{z}) = + h_{a}(\eta) \> f_{e}^{b}(E,L_{z}),\eqno\new$$
and
\eqnam\oddcb
$$f_{o}^{d}(E,L_{z}) = - h_{a}(\eta) \> f_{e}^{d}(E,L_{z}),\eqno\new$$
Throughout the remainder of this paper we will refer to models
constructed with the first method (i.e., with dynamically coupled
components) as `standard models'. The models for which the odd part of
the \df of the bulge is zero as `models with non-rotating bulge', and
the models whose odd part results in counter-rotating discs as
`counter-rotating models'.

\subsection{4.3 The region of physical models}

QZMH showed that two-integral oblate $(\alpha, \beta)$-models with a
nuclear BH are unphysical (i.e., have a region in phase-space where
the DF is negative) when $\alpha > -0.5$. Therefore, it is to be
expected that models with very compact, massive nuclear discs embedded
in an $(\alpha, \beta)$-bulge, with $\alpha > -0.5$ will also be
unphysical.  Since a nuclear disc is a cold component (see also
Section 5.2) located at the centre of the galaxy, most stars in the
centre will be on nearly circular orbits. In the derived DF this
translates into small weights for the more radial orbits close to the
centre. This can clearly be seen from the DF of model 1.1 in Figure
3. The DF is peaked towards large energies (i.e., the centre of the
potential well) and high angular momentum. A local minimum close to
the centre at $L_{z} = 0$ is also visible.

The value of the DF at this local minimum depends on the disc-to-bulge
ratio of the system. The models will become unphysical if $\Delta =
M_{\rm disc}/M_{\rm bulge}$ is too large, since this will bring about
a region in phase-space were \feelz is negative. If $\Delta$ is
increased further the nuclear disc becomes completely dominant over
the bulge. In those cases the DF is positive again over the entire
phase space. So as function of $R_d$ there is an interval of
disc-to-bulge ratios for which the DF has a local minimum with
negative values. Since the angular momentum at this local minimum is
zero, one can test whether a certain $(R_{d}/R_{b},\Delta)$-model is
physical by examining whether $f_{e}(E,L_{z}=0)$ is positive for all
$E < \Psi_{0}$. In order to determine this interval we proceed as
follows. For each model $(R_{d}/R_{b},\Delta)$ we numerically evaluate
$E_{\rm crit}$ for which $f_{e}(E,L_{z}=0)$ is (locally) minimal,
using Brent's method for the minimization. Subsequently, we check
whether $f_{e}(E_{\rm crit},L_{z}=0) < 0$. We repeat this procedure in
a bisective way until the borders of the interval are known to a
certain accuracy.  The values of $f_{e}(E,L_{z}=0)$ are evaluated
using Eddington's formula (\eddington).

The results for the case with $\alpha = 0$ and $\alpha = -0.25$ are
shown in Figure 4, where we plot ${\rm log}_{10}(\Delta)$ as function
of ${\rm log}_{10}(R_{d}/R_{b})$. The hatched areas indicate the
regions where the models are unphysical. Only very small, massive
nuclear discs give rise to unphysical two-integral DFs. As expected,
the region of unphysical models becomes smaller when the cusp
steepness increases (i.e., when the value of $\alpha$ decreases). 

%
%%%%%%%%%%%%%%%%%%%%%%
% Begin of  Figure 5
%%%%%%%%%%%%%%%%%%%%%%
%
%\beginfigure{3}
%\centerline{\psfig{figure=toomre.ps,width=\hssize}}\smallskip
%\caption{{\bf Figure 5.} The Toomre parameter $Q$ multiplied by the
%constant $\gamma$ as function of radius for four different models.
%All models have surface-brightness ratio $\Lambda = 1.0$ at one disc
%scale-length $R_d$, but differ in the value of the cusp steepness:
%$\alpha = 0.0$ (solid line), $\alpha = -0.5$ (short dashed line),
%$\alpha = -1.0$ (dotted line), and $\alpha = -1.5$ (long dashed line).
%The hatched region ($\gamma Q \leq 3.36$) indicates the region where
%infinitesimally thin discs are unstable against axisymmetric
%perturbations.}
%\endfigure
%
%%%%%%%%%%%%%%%%%%%%%%%
% End of Figure 5
%%%%%%%%%%%%%%%%%%%%%%%
%

\eqnumber=1
\def\chaphead{\hbox{5.}}
\section{5 The stability of the nuclear discs}

Our $f(E,L_z)$-models are of little practical value if they are 
unstable, and hence we study their stability in this section.

\subsection{5.1 Local stability}

The best known stability criterion for discs is the Toomre (1964)
criterion.  This criterion ensures that an infinitesimally thin disc
is {\it locally} stable against axisymmetric perturbations as long as
\eqnam\toomre
$$ Q \equiv {\sigma_{R} \kappa \over \gamma G \Sigma}  > 1.\eqno\new$$
Here $\sigma_{R}$ is the radial velocity dispersion, $\Sigma$ is the
surface density of the disc, $\gamma$ is a constant that equals 3.36,
$G$ is the gravitational constant, and $\kappa$ is the epicyclic
frequency.

In an axisymmetric system with distribution function \feelz we have
$\sigma_{R} = \sigma_{z} = \sigma$ everywhere. From the Jeans
equations it follows that (e.g., Hunter 1977)
\eqnam\jeans
$$\sigma^2(R,z) = -{1 \over \rho(R,z)} \int\limits_{z}^{\infty}
\rho(R,z') {\partial \Psi \over \partial z}(R,z') {\rm d}z'.\eqno\new$$
The surface density $\Sigma(R)$ of our disc is given by 
\eqnam\sbint
$$\Sigma(R) = \int\limits_{-\infty}^{\infty} \rho_{\rm disc}(R,z) 
      {\rm d}z.\eqno\new$$
%
%
%%%%%%%%%%%%%%%%%%%%%%
% Begin of  Figure 6
%%%%%%%%%%%%%%%%%%%%%%
%
%\beginfigure{5}
%\centerline{\psfig{figure=stabcusp.ps,width=\hssize}}\smallskip
%\caption{{\bf Figure 6.} The hatched region indicates the region where
%the entire nuclear disc is unstable against axisymmetric perturbations
%(assuming stability requires $\gamma Q \geq 3$) as function of cusp
%steepness ($\alpha$) and disc-to-bulge ratio ($\Delta$). 
%Models with stronger cusps allow larger disc-to-bulge ratios.}
%\endfigure
%
%%%%%%%%%%%%%%%%%%%%%%%
% End of Figure 6
%%%%%%%%%%%%%%%%%%%%%%%
%
{\noindent A problem that arises is that we are considering a thick
disc.  Only a limited amount of work has been done on the stability of
thick discs, but the little work that has been done (e.g., Shu 1968;
Vandervoort 1970) indicates that disc thickness has a stabilizing
effect. This means that the factor $\gamma = 3.36$ derived by Toomre
for the infinitesimally thin case decreases for thick discs.  However,
it is not known what the value of $\gamma$ is for an arbitrary
vertical density profile. Therefore, instead of presenting $Q(R)$, we
show the radial profile of $\gamma\, Q(R)$.}

The results are shown in Figure 5, where we give $\gamma Q$ as
function of radius for four models that differ only in cusp steepness.
The hatched region indicates the regime where infinitesimally thin
discs are unstable against axisymmetric perturbations (i.e., $\gamma Q
\leq 3.36$).  In the case of a thick disc, the upper limit of this
regime will move slightly downwards. The presence of a cusp stabilizes
the disc in the inner region. Only the nuclear disc in the model
without cusp ($\alpha = 0.0$) is mildly unstable in its inner region.
Figure 6 shows the region in the ($\alpha$, $\Delta$)-plane of models
that are stable over the entire radial range (hatched region), where
we have assumed that stability is achieved for $\gamma Q \geq
3.0$. Although this value is uncertain, as we have seen, a 10\% change
will cause only a very small shift in the location of the boundary
curve. It is again evident that a cusp has a strong stabilizing effect
on a nuclear disc.
%
%%%%%%%%%%%%%%%%%%%%%%
% Begin of  Figure 7
%%%%%%%%%%%%%%%%%%%%%%
%
%\beginfigure{4}
%\centerline{\psfig{figure=thick.ps,width=\hssize}}\smallskip
%\caption{{\bf Figure 7.} The effect of the thickness of the nuclear disc
%on its local stability for a model with cusp steepness $\alpha = 0.0$ and
%$\Lambda = 1.0$. The hatched area indicates the unstable region. }
%\endfigure
%
%%%%%%%%%%%%%%%%%%%%%%%
% End of Figure 7
%%%%%%%%%%%%%%%%%%%%%%%
%
  
Although the surface density of the disc falls off strongly, the
velocity dispersion decreases slowly, due to the fact that there is a
large amount of matter (from the bulge-component) beyond several scale
lengths of the disc. Therefore, the outer parts of the disc are stable
in all cases.

Figure 7 shows $\gamma Q$ as function of radius for the cusp-free
model with $\Lambda = 1.0$ (model 0.1), for three different
flattenings of the nuclear disc. Thickening the disc, i.e., increasing
$q_d$, has a stabilizing effect. This can also be understood
intuitively: the thickness of a disc is closely related to the amount
of random motion of the stars in the $z$-direction. Since our models
have DFs that depend on only two integrals of motion, we have
$\sigma_{R} =\sigma_{z}$. Therefore, thickening the disc will increase
the radial velocity dispersion in the disc, and therefore the
parameter $Q$. The dependence of $\kappa$ on $q_d$ is weak.

\subsection{5.2 Global stability}

Besides {\it local} instabilities, discs can also suffer from {\it
global} instabilities. The best known is the bar-instability, which
has proven to play an important role. Ostriker \& Peebles (1973)
calculated that a disc is globally stable to bar-like modes when the
total kinetic energy of rotation $T$ is less than $\sim 14$\% of the
gravitational energy $|W|$. However, this criterion is merely a good
rule of thumb, rather than a proven physical theorem. Athanassoula \&
Sellwood (1986) showed that, even more effectively than a halo
component, random motions in the disc can stabilize it against bar
forming modes. They showed that as long as the mass-weighted value of
$Q$ averaged over the central region exceeds a value of $\sim
2\,-\,2.5$, the entire disc is stable. This is the case for all of the 
models listed in Table 1.

%
%%%%%%%%%%%%%%%%%%%%%%
% Begin of  Figure 8
%%%%%%%%%%%%%%%%%%%%%%
%
%\beginfigure{5}
%\centerline{\psfig{figure=big.ps,width=\hssize}}\smallskip
%\caption{{\bf Figure 8.} The central disc surface brightness 
%(in magn. arcsec$^{-2}$) of discs that are just marginally stable as
%function of $\log_{10}[R_d/R_b]$. Discs above these lines have a
%radial range that is unstable against axisymmetric perturbations.  The
%scale of $\mu_{0}$ is arbitrary.  Note the remarkable similarity
%between this plot and the observed $\mu_0$ {\it vs.} ${\rm log}(R_d)$
%relation of discs (e.g., figure 17 in Scorza \& Bender 1995),
%suggesting that discs build up their mass until they become marginally
%stable.}
%\endfigure
%
%%%%%%%%%%%%%%%%%%%%%%%
% End of Figure 8
%%%%%%%%%%%%%%%%%%%%%%%
%

Toomre (1981) argued that the origin of the bar forming instability is
feedback to the swing amplifier via waves that pass through the
centre.  The occurrence of swing amplification depends on the value of
\eqnam\toomx
$$ X = {\kappa^2 R \over 2 m \pi \Sigma} ,\eqno\new$$
where $m=2$ in the case of bi-symmetric bars. Toomre showed that as
long as $Q \lta 3$ and $X \lta 3$ the gain of the swing-amplifier is
sufficient to form bars. We find that $X$ increases strongly with
radius, but $X \lta 3$ for small $R$. Although this implies that swing
amplification can occur in the very centre of the nuclear discs, one
can ensure stability against bar formation by cutting off the path to
the centre by means of an inner Lindblad resonance. The models with a
cusp ($\alpha < 0$) have $\Omega - {1\over 2}\kappa \rightarrow
+\infty$ for $R \rightarrow 0$, and will therefore have an ILR for any
pattern speed of the bar. An ILR is absent only when $\alpha = 0$ and
high pattern speeds. In that case the swing amplified feed-back loop
is open in the centre ($X \lta 3$), but as there is a large amount of
random motion in the centre (i.e., $Q > 3$) the growth rate for a bar
mode will be insignificant.  Therefore, we expect that the nuclear
discs in all the models listed in Table 1 are stable against bar
formation.

The above results are valid for a bulge with axis ratio of 0.7. N-body
experiments carried out by Dehnen (priv.\ comm.) indicate that
non-rotating and rotating $(\alpha, \beta)$-models with axis ratios of
0.5 or smaller are unstable to bar formation. 

\subsection{5.3 Continuation to larger discs}

We have restricted ourselves to nuclear discs with $R_d/R_b = 0.2$,
but the method outlined above is also applicable to larger discs. We
have investigated the stability of our disc-models for a large range
in $R_d/R_b$.  The general tendency is for larger discs to be more
stable. We have calculated the maximum disc-to-bulge ratio for which,
at given cusp-steepness $\alpha$ and disc scale length $R_d$, $\gamma
Q(R) \geq 3.0$. From equation (\dtob) it follows that $\Sigma_0
\propto \Delta / R^2_d$. Therefore, for given disc-to-bulge ratio and
disc-scale length we can, using an arbitrary scaling, calculate the
central surface brightness $\mu_0$ of the disc (in magnitudes per
arcsec$^2$).

Figure 8 shows the curve of $\mu_0$ as a function of
$\log_{10}[R_d/R_b]$ for discs with maximum disc-to-bulge ratio so
that the entire disc has $\gamma Q \geq 3$. These curves depend on the
cusp steepness $\alpha$, but only for small discs. This Figure is
remarkably similar to the relation between $\mu_0$ and $R_d$ of
observed discs (i.e., compare to Figure 17 of Scorza \& Bender 1995).
This might indicate that discs build up their mass until they become
marginally stable.

\eqnumber=1
\def\chaphead{\hbox{6.}}
\section{6 The kinematics of nuclear discs}

Our models are fully specified by the four parameters $\alpha$,
$\Delta$, $a$, and $\MBH$. In addition, we must choose the method used
to calculate the odd part of the DF. In this section we restrict
ourselves to cases with $\MBH = 0$. The influence of a nuclear BH in
the centre of these models is the topic of Section 7.

\subsection{6.1 Calculation of the velocity profiles}

Once we have determined the complete phase-space density of our
models, we can calculate every observable property. In particular, the
entire VP can be derived by integrating the DF along the line of
sight:
\eqnam\vpa
$${\rm VP}(v_{z'};x',y') = {1 \over \Sigma} \int\!\!\! 
\int\limits_{E>\Psi_{\infty}}\!\!\! \int f(E,L_{z}) {\rm d}v_{x'} 
{\rm d}v_{y'} {\rm d}z'.\eqno\new$$
Here we adopt the Cartesian coordinate system of an observer $(x', y'
,z')$, where the $z'$-axis lies along the line of sight, and $x'$ is
along the apparent major axis of the galaxy projected on the sky.

Upon defining a polar coordinate system $(v_{\perp},\varphi)$ in the
$(v_{x'},v_{y'})$ plane, where $v_{x'} = v_{\perp} {\rm cos}\varphi$,
and $v_{y'} = v_{\perp} {\rm sin}\varphi$, equation (\vpa) can be
written as
\eqnam\vpb
$${\rm VP} (v_{z'};x',y') 
  = {1 \over \Sigma} \int\limits_{z_1}^{z_2} {\rm d}z' \!\!
    \int\limits_{0}^{2\Psi - v_{z'}^2} {\rm d}v_{\perp}^2
    \int\limits_{0}^{\pi} f(E,L_{z}) \> {\rm d}\varphi.\eqno\new$$
The energy is given by
\eqnam\energy
$$ E = \Psi(x',y',z') - {1\over 2}(v_{z'}^2 + v_{\perp}^2),\eqno\new$$
and the angular momentum follows from
\eqnam\angmom
$$ L_{z} = -v_{z'} x' {\rm sin}i$$
$$ \;\;\;\;\;\; + v_{\perp}{\rm cos}\varphi \sqrt{(-y'{\rm cos}i + 
z'{\rm sin}i)^2 + x'^2 {\rm cos}^2 i}.\eqno\new$$
For any given $v_{z'}$ the boundaries $z_1$ and $z_2$ are determined
by solving $\Psi(x',y',z') - {1\over 2} v_{z'}^2 = 0$.

%
%%%%%%%%%%%%%%%%%%%%%%
% Begin of  Figure 9a 
%%%%%%%%%%%%%%%%%%%%%%
%
%\beginfigure*{6}
%\centerline{\psfig{figure=velprof1.ps,width=\hdsize}}\smallskip
%\caption{{\bf Figure 9a.} The velocity profiles of models 1.1 and 1.2
%(see Table 1) at different radii along the major axis.
%The panels at the left correspond to a model where the odd part
%of the \df is determined by the `standard' method (a=10), whereas the panels 
%at the right correspond to a `non-rotating bulge' model (also with a=10). 
%The disc rotates almost with the circular velocity (indicated by the solid 
%triangles).}
%\endfigure
%
%%%%%%%%%%%%%%%%%%%%%%%
% End of Figure 9a
%%%%%%%%%%%%%%%%%%%%%%%
%

For each model we calculate the VPs at a number of $(x',y')$-points.
In all cases discussed here we take an inclination angle of $90\deg$,
i.e., we assume edge-on observation. Figures 9a and 9b compare the VPs
of model 1 ($\alpha = 0.0$) at seven positions along the major axis
$(x'=R,y'=0)$, using four different odd parts of the distribution
function: the first (left panels in Figure 9a) has \foelz specified by
the `standard' method (Section 4.2). The panels on the right in the
same figure correspond to a `non-rotating bulge' model with $a = +10$,
whereas taking $a=-10$ results in the VPs shown in the left panels of
Figure 9b. Finally, in the right panels of Figure 9b, the VPs are
shown that correspond to the case of a counter-rotating disc, i.e.,
where we have reversed the sign of the odd part of the DF of the disc
with respect to that of the bulge.

Even though the {\it dynamics} of bulge and disc are coupled in the
`standard' models, the VPs nevertheless show two very distinct
components.  The solid triangles in Figure 9 indicate the circular
velocity of the models. As can be seen, the disc rotates with almost
this circular velocity. This is true for both the `standard' models,
the `non-rotating bulge' models, and the `counter-rotating' models.
%
%%%%%%%%%%%%%%%%%%%%%%
% Begin of  Figure 9b
%%%%%%%%%%%%%%%%%%%%%%
%
%\beginfigure*{7}
%\centerline{\psfig{figure=velprof2.ps,width=\hdsize}}\smallskip
%\caption{{\bf Figure 9b.} Same as Figure 9a, except now the VPs of
%models 1.3 and 1.4 are shown. Model 1.3 (panels on the left) is again a 
%`non-rotating bulge' model, but now with a=-10. This results in 
%double peaked profiles (see text). The panels on the right (model 1.4) show 
%the VPs of the model with a counter-rotating disc (odd part defined with the
%`counter-rotating' method). 
%The disc rotates with almost the circular velocity.}
%\endfigure
%
%%%%%%%%%%%%%%%%%%%%%%
% End of  Figure 9b
%%%%%%%%%%%%%%%%%%%%%%
%

%
%%%%%%%%%%%%%%%%%%%%%%
% Begin of  Figure 10
%%%%%%%%%%%%%%%%%%%%%%
%
%\beginfigure*{8}
%\centerline{\psfig{figure=realmom.ps,width=\hdsize}}\smallskip
%\caption{{\bf Figure 10.} The four panels show respectively the
%rotation velocity $V_{\rm rot}$, the velocity dispersion $\sigma$,
%the skewness ${\cal S}$, and the kurtosis ${\cal K}$ of the velocity 
%profiles shown in Figures 9a and 9b. The contribution of the disc makes the 
%velocity profiles close to the centre strongly skewed and peaked.}
%\endfigure
%
%%%%%%%%%%%%%%%%%%%%%%%
% End of Figure 10
%%%%%%%%%%%%%%%%%%%%%%%
%

As discussed in Section 4.2, we have constrained the fraction of stars
on clockwise circular orbits in the equatorial plane to be
unity. However, not all stars in the disc are on such circular
orbits. Therefore, upon taking $a\ll 0$ in the odd part of the DF, a
considerable fraction of stars in the disc will be on anti-clockwise
orbits that are not (perfectly) circular. This results in
double-peaked VPs, as can be seen in the left panels of Figure 9b.

\subsection{6.2 Velocity profile analysis}
 
In order to allow for an easy comparison between different models, and
to be able to extract physically meaningful parameters, we need to
find a way of parameterizing the VPs. It has become customary to
expand the VPs in a Gauss--Hermite series (van der Marel \& Franx
1993; Gerhard 1993). Although this approach has proven useful when
parameterizing VPs that deviate slightly from a Gaussian, it is
inappropriate for describing the VPs presented above, since they
strongly deviate from Gaussians. After seeing convolution with a FWHM
that exceeds a few disc scale-lengths the Gauss--Hermite
parameterization {\it can} be used, since the convolution results in
sufficiently smooth VPs. However, for smaller seeing FWHM the
convolved VPs still deviate strongly from a Gaussian.

We therefore opted to use the real moments to quantify the VPs.  The
$n^{\rm th}$ moment of the normalized velocity profile 
${\rm VP}_{0}(v)$ is defined
as
\eqnam\moment
$$ \mu_n \equiv \int\limits_{-\infty}^{\infty} {\rm VP }_{0}(v) \> v^n \>
{\rm d}v.\eqno\new$$
The mean streaming rotation velocity is simply the first moment
$V_{\rm rot} = \mu_1$, the velocity dispersion $\sigma = \sqrt{\mu_2 -
\mu_1^2}$, the skewness ${\cal S} = (\mu_3 - 3 \mu_2 \mu_1 + 2
\mu_1^3) / \sigma^3$, and the kurtosis ${\cal K} = (\mu_4 - 4 \mu_3
\mu_1 + 6 \mu_2 \mu_1^2 - 3 \mu_1^4) / \sigma^4$.

Figure 10 shows these parameters for models 1.1 -- 1.4 whose VPs are
given in Figures 9a and 9b. The presence of the nuclear discs brings
about VPs that deviate strongly from a Gaussian. This is evident from
the large values of both ${\cal S}$ and ${\cal K}$, which are zero and
three respectively for a Gaussian. The disc dominates the kinematics
in the centre, and causes a strong drop of the velocity dispersion
inwards of $\sim 3 R_{d}$. The rotation curve of model 1.4 clearly
reveals a counter-rotating core. This is also evident from the change
of sign of the skewness ${\cal S}$ around $\sim 2.5 R_d$.
%
%%%%%%%%%%%%%%%%%%%%%%
% Begin of  Figure 11 
%%%%%%%%%%%%%%%%%%%%%%
%
%\beginfigure*{9}
%\centerline{\psfig{figure=tot1mom.ps,width=\hdsize}}\smallskip
%\caption{{\bf Figure 11.} The 4 parameters ($V_{\rm rot}$, $\sigma$, 
%${\cal S}$, ${\cal K}$) that characterize the VPs of models with
%different cusp steepness $\alpha$ after seeing and pixel convolution
%with $\chi = 1$.  Solid lines are for the model without nuclear disc,
%whereas dashed lines represent the cases with $\Lambda = 1.0$.}
%\endfigure
%
%%%%%%%%%%%%%%%%%%%%%%%
% End of Figure 11
%%%%%%%%%%%%%%%%%%%%%%%
%

The real moments of the VPs describe the dynamics of the entire
system.  If one is interested in the dynamics of disc and bulge
separately, another parameterization of the VPs is required. Given the
clear two-component character of the VPs, it is convenient to use the
Levenberg-Marquardt method to fit a double Gaussian to the derived
VPs. Under the assumption that the two separate components of the VPs
correspond to the disc and bulge components of the model, and assuming
that both are Gaussian, the double-Gaussian fit measures the
kinematics of disc and bulge separately (Rix \& White 1992).  There is
no {\it a priori} reason why the VPs of both components will be
Gaussian. However, increasing the number of parameters of the fitting
function will result in a large degeneracy of the parameters.
Furthermore, Scorza \& Bender (1995) have shown that including
deviations from a Gaussian only mildly influences the derived
kinematics of bulge and disc.

\subsection{6.3 Observations of nuclear discs}

We now discuss the influence of seeing convolution on the observable
properties of nuclear discs. We mimic long-slit observations where we
take the slit to be aligned along the major axis, the slit-width to be
equal to the FWHM of the seeing, and the pixels to be one-third of the
seeing FWHM. We use the parameter $\chi$ to define the PSF FWHM, in
units of the scale-length $R_d$ of the disc, and subsequently scale
the models so that the FWHM equals $1''$. Therefore varying $\chi$ is
similar to varying the distance of the galaxy from the observer. We
start by constructing a cube that contains the velocity profiles ${\rm
VP}(v_{z'};x',y')$ calculated on a two-dimensional logarithmic grid on
the sky ($x'$, $y'$). We then convolve the velocity profiles for each
$v_{z'}$ with a Gaussian PSF and take the effects of slit-width and
pixel size into account. We use bi-cubic spline interpolations to
interpolate between $(x', y')$ grid points at a given velocity
$v_{z'}$. We use the approach taken by QZMH, who have shown that in
the case of a Gaussian PSF and rectangular pixels, the entire seeing
convolution and pixel averaging can be written as a double
quadrature. After the convolution we parameterize the VPs.  Since the
convolved VPs deviate strongly from a Gaussian for small values of
$\chi$, we use the moments defined in the previous section to describe
the VPs.
%
%%%%%%%%%%%%%%%%%%%%%%
% Begin of  Figure 12
%%%%%%%%%%%%%%%%%%%%%%
%
%\beginfigure*{10}
%\centerline{\psfig{figure=tot2mom.ps,width=\hdsize}}\smallskip
%\caption{{\bf Figure 12.} 
%The 4 parameters ($V_{\rm rot}$, $\sigma$, ${\cal S}$, ${\cal K}$)
%that characterize the VPs of model 1.1 (i.e., $\alpha=-1.0$), after
%seeing and pixel convolution with different values of $\chi$.  Solid
%lines are for the model without nuclear disc, whereas dashed lines
%represent the cases with $\Lambda = 1.0$. }
%\endfigure
%
%%%%%%%%%%%%%%%%%%%%%%%
% End of Figure 12
%%%%%%%%%%%%%%%%%%%%%%%
%

Figure 11 shows the results for four different values of the cusp
steepness $\alpha$. They include the case without nuclear disc (solid
lines) as well as the case with nuclear disc with disc-to-bulge ratio
such that $\Lambda = 1.0$ (dashed lines). The odd part is determined
in the `standard' way with $a=10$. In all cases we take an inclination
angle of $i = 90^{\rm o}$, and convolve the model with a PSF with
$\chi = 1$. The main effect of changing $\alpha$ is that the rotation
curve of the entire galaxy becomes steeper, due to larger central
mass. The kinematic signature of a nuclear disc is therefore most
pronounced in models with steep cusps. The seeing convolution results
in velocity dispersions that are almost equal for the cases with and
without nuclear disc. The strong central decrease of dispersion (as
can be seen in Figure 10) is no longer observable for a seeing with
FWHM equal to the scale-length of the nuclear disc. The kinematic
signatures of the nuclear disc are confined to higher rotation
velocities and stronger deviations of the VPs from a Gaussian.

Figure 12 shows in more detail the effect of the seeing convolution.
Here we plot the moments of the convolved VPs for the model with
$\alpha = -1.0$, again both with and without nuclear disc (models 3.0
and 3.1 respectively), for four different values of $\chi$.  Upon
increasing $\chi$, the VPs of the models with and without nuclear discs
become more and more similar.  For $\chi \gta 3$ the only kinematic
signatures of the nuclear disc that remain are a slightly larger
maximum rotation velocity and central velocity dispersion. However,
the differences are so small that it is unlikely that the presence of
a nuclear disc can be inferred from kinematic observations with a
seeing FWHM exceeding $\sim 3$ times the disc scale-length.  This is
also the maximum $\chi$ beyond which the disc becomes {\it
photometrically} undetectable. Figure 13 summarizes these results for
the velocity dispersion. We give the ratio of central velocity
dispersion for the case with nuclear disc ($\Lambda = 1$) to the case
without nuclear disc ($\Lambda = 0$) as function of $\chi$ for four
different values of cusp steepness. Increasing $\alpha$ results in an
increase of the central mass, and therefore in an increase of the
central rotation velocity. Increasing $\chi$ results in an increase of
the ratio $\sigma_0(\Lambda = 1)/\sigma_0(\Lambda = 0)$ due to seeing
convolution of the rotation curve.

A number of authors have suggested that the presence of a nuclear disc
can mimic the presence of a BH, in that seeing convolution of the
central rotation curve might result in an increase of the central
velocity dispersion.  If the PSF FWHM is so large that the disc is
photometrically undetectable, the large central velocity dispersion
could erroneously be interpreted as due to a BH. However, we are
unable to achieve a ratio of $\sigma_0(\Lambda = 1)/\sigma_0(\Lambda =
0)$ larger than 1.1.  This is due to the fact that the PSF is circular
compared to the highly flattened disc structure. Upon reducing the
slit-width we only find a very mild increase of $\sigma_0(\Lambda =
1)/\sigma_0(\Lambda = 0)$.  The only way of achieving a large enough
ratio so that the nuclear disc might mimic a BH is by strongly
increasing $\alpha$ and the disc-to-bulge ratio $\Delta$. However,
stability arguments prohibit too large values of $\Delta$.
Furthermore, increasing $\Delta$ will make the nuclear disc detectable
in the photometry.

The entire analysis above is based on edge-on projections of the
models. Changing the inclination angle has a number of effects. First
of all, since the disc is highly flattened, the observed rotation of
the disc roughly scales with $\sin i$. Furthermore, decreasing the
inclination angle decreases the surface brightness of the disc
relatively more than that of the bulge component. As a consequence,
decreasing $i$ results in VPs that less and less clearly show a
rapidly rotating component. We projected model 3.1 at a number of
different inclination angles and found the kinematic signatures of the
nuclear disc to disappear for $i \lta 60\deg$.

\subsection{6.4 Counter-rotating discs}

In Section 4.2 we discussed how to define the odd part of the DF in
order to achieve a model in which the nuclear disc counter-rotates
with respect to the bulge. Several galaxies have been observed to
harbour counter-rotating cores. Recent imaging with the Hubble Space
Telescope has shown that in a number of cases there are indications
for the presence of a disc component in the centre (Forbes, Franx, \&
Illingworth 1995, Carollo \etal 1996).  Therefore, counter-rotating
discs seem a natural explanation for the observed kinematics.  Here we
investigate the kinematic signatures of such counter-rotating nuclear
discs.

We find that for $\chi \lta 2$ the VPs are double-peaked, clearly
revealing the counter-rotation. In these cases one observes also a
strong central decrease of the velocity dispersion. This is due to the
fact that the dynamically cold disc component dominates the central
VPs, and to the fact that, at larger radii, the second moment of the
VP becomes large due to the increasing separation of the disc- and
bulge-components of the VPs.  It is remarkable that in most galaxies
with a counter-rotating core, the central velocity dispersion
strongly {\em increases} towards the centre, e.g., IC 1459 (van der
Marel \& Franx 1993), NGC 3608 (Jedrzejewski \& Schechter 1989). More
extensive modeling of these systems, possibly with a nuclear BH, is
clearly required.

For cases with $\chi \gta 2$, the only kinematic signature of the
counter-rotating disc are strongly skewed VPs. However, the
interpretation of these as being due to a counter-rotating disc is
ambiguous.
%
%%%%%%%%%%%%%%%%%%%%%%
% Begin of  Figure 13
%%%%%%%%%%%%%%%%%%%%%%
%
%\beginfigure{11}
%\centerline{\psfig{figure=sigrat.ps,width=\hssize}}\smallskip
%\caption{{\bf Figure 13.} The ratio $\sigma_0(\Lambda = 1) / 
%\sigma_0(\Lambda = 0)$ as a function of $\chi$ for different values of the 
%cusp steepness $\alpha$. This ratio shows the effect of seeing
%convolution of the rotation curve of the nuclear disc on the central
%dispersion.  The fact that this ratio is larger for steeper cusps is
%due to the fact that the amount of rotation of the nuclear disc is
%governed by the mass in the centre of the galaxy.}
%\endfigure
%
%%%%%%%%%%%%%%%%%%%%%%%
% End of Figure 13
%%%%%%%%%%%%%%%%%%%%%%%
%

%
%%%%%%%%%%%%%%%%%%%%%%
% Begin of  Figure 14
%%%%%%%%%%%%%%%%%%%%%%
%
%\beginfigure{12}
%\centerline{\psfig{figure=convvcirc.ps,width=\hssize}}\smallskip
%\caption{{\bf Figure 14.} The parameter $\omega_d = v_{\rm disc}/v_{\rm circ}$
%(where $v_{\rm disc}$ is derived from the double Gaussian fit to the VPs) as 
%function of the scaling parameter $\chi$ for model 3.1 ($\alpha = -1.0$). 
%The results
%are shown for three different radii. Clearly, the measured rotation at
%$\sim 2 R_d$ is a good measure of the circular velocity and therefore of
%the mass inside that radius. For sufficiently small values of $\chi$,
%$v_{\rm disc}$ at smaller radii provides a good measure of the central
%density of the galaxy.}
%\endfigure
%
%%%%%%%%%%%%%%%%%%%%%%%
% End of Figure 14
%%%%%%%%%%%%%%%%%%%%%%%
%

\subsection{6.5 Deriving the central mass density}

The mean streaming of the disc stars follows from the odd part of the
DF by means of equation (\odddf), and substituting $\rho_d$ and
$f_o^d$ for $\rho$ and $f_o$ respectively.  This mean streaming can be
compared to the circular velocity, which is given by
\eqnam\vcirc
$$v^2_{\rm circ}(R) = -R {d\Psi \over dR} = 4 \pi G q \int\limits_{0}^{R}
{\rho(m^2) m^2 dm \over \sqrt{R^2 - m^2(1 - q^2)}}.\eqno\new$$
As shown in Section 6.1, the disc rotates with nearly the circular
velocity (${\bar v}_{\phi}$ of the disc stars $\gta 0.95 v_{\rm circ}$). 
Fitting a double Gaussian to the observed VPs reveals a hot, mildly
rotating component, and a rapidly rotating, cold component. The mean rotation 
of the latter component is an excellent measure of the mean streaming
of the nuclear disc, and therefore of the circular velocity. In turn, this 
can be used to constrain the central mass density.

We investigated the influence of seeing convolution on the rotation
velocity of the disc as measured by fitting a double Gaussian to the
VPs. The results are shown in Figure 14 for model 3.1 ($\alpha = -1.0,
\Lambda = 1.0$). We plot $\omega_d \equiv v_{\rm disc}/v_{\rm circ}$ 
as function of the scaling parameter $\chi$ for three different radii:
$R = 0.5$, $1.0$, and $2.0$ times the scale length of the nuclear
disc. Seeing convolution decreases $\omega_d$ considerably, especially
at $R \lta 2 R_d$. For PSFs whose FWHM exceeds $2 R_d$ (i.e., $\chi >
2$), the VPs no longer reveal two separate components, and the double
Gaussian fit becomes meaningless.  The fact that $\omega_d \sim 0.8$
at $R= 0.5 R_d$ for $\chi = 0$ (i.e., for the unconvolved VPs), is due
to the fact that at small radii the cusped bulge dominates the surface
brightness, and therefore these central VPs do not clearly reveal two
components, making the interpretation of the double Gaussian fit in
terms of a disc- and bulge-component somewhat ambiguous. This is not
the case for a non-cusped bulge ($\alpha = 0.0$), where, as shown in
Figure 9, the disc dominates the central VPs, and has $\omega_d
\approx 0.95$.
%
%%%%%%%%%%%%%%%%%%%%%%
% Begin of  Figure 15
%%%%%%%%%%%%%%%%%%%%%%
%
%\beginfigure*{13}
%\centerline{\psfig{figure=bhvps.ps,width=\hdsize}}\smallskip
%\caption{{\bf Figure 15.} The 4 parameters ($V_{\rm rot}$, $\sigma$, $\cal S$,
%$\cal K$) that characterize the seeing convolved VPs of models with 
%$\alpha = -1.0$ but with different $\Lambda$ and $M_{\rm BH}$. The panels in 
%the two columns on the left show the results for the models {\it without} 
%nuclear disc ($\Lambda = 0$) for three different values of the BH mass: 
%$M_{\rm BH}/M_{\rm disc} = 0$ (solid lines), $M_{\rm BH}/M_{\rm disc} = 
%0.1$ (dashed lines), and $M_{\rm BH}/M_{\rm disc} = 1.0$ (dotted lines). 
%The panels in the two columns on the right show the same results, but now
%for models {\it with} nuclear disc ($\Lambda = 1.0$).}
%\endfigure
%
%%%%%%%%%%%%%%%%%%%%%%%
% End of Figure 15
%%%%%%%%%%%%%%%%%%%%%%%

\section{7 The influence of a nuclear Black Hole}

Many, if not all, galaxies might harbour massive BHs in their
nuclei. In order to study their kinematic signature in the presence of
a nuclear disc, we have constructed several models with central BH
(see Table 1).  A massive BH in the nucleus of a galaxy strongly
influences the central dynamics. Early studies based on spherical
models showed that adiabatic growth of the BH results in a power-law
cusp in the stellar density profile with a logarithmic slope of
central surface brightness in the range from $-1/2$ to $-5/4$.
Furthermore, hydrostatic equilibrium requires the RMS velocity of the
stars surrounding the BH to have a $r^{-1/2}$-cusp (Bahcall \& Wolff
1976; Young 1980; Quinlan et al.\ 1995). 

%
%%%%%%%%%%%%%%%%%%%%%%
% Begin of  Figure 16
%%%%%%%%%%%%%%%%%%%%%%
%
%\beginfigure*{14}
%\centerline{\psfig{figure=bhvcirc.ps,width=\hdsize}}\smallskip
%\caption{{\bf Figure 16.} The parameter $\omega_d = v_{\rm disc}/v_{\rm circ}$ 
%as a function of radius (in arcsec) for three different scalings $\chi$.
%The results are plotted for three different models, all having $\alpha
%= -1.0$ and $\Lambda = 1.0$, but differing only in the mass of the BH:
%$M_{\rm BH} = 0$ (solid lines), $M_{\rm bh} = 0.1 M_{\rm disc}$
%(dashed lines), and $M_{\rm bh} = M_{\rm disc}$ (dotted lines).}
%\endfigure
%
%%%%%%%%%%%%%%%%%%%%%%%
% End of Figure 16
%%%%%%%%%%%%%%%%%%%%%%%

QZMH showed that oblate ($\alpha$, $\beta$)-models with a BH are
physical (i.e., have non-negative DF) when $\alpha \leq -0.5$.  Under
this condition a nuclear disc can be added, while maintaining a
positive DF. At sufficiently small radii the potential is dominated by
the BH, and an explicit expression for $\rho(\Psi,R)$ can be
obtained. Therefore, an asymptotic expression for $f_e(E,L_z)$ can be
calculated in terms of elementary functions (see Appendix A).

We have added BHs of both $M_{\rm BH} = 0.1 M_{\rm disc}$ and $M_{\rm
BH} = M_{\rm disc}$ to the model with $\alpha = -1.0$, both for the
model with and without nuclear disc (see Table 1 for the parameters).
The real moments of the VPs after seeing convolution are shown in
Figure 15. With this parameterization of the VPs the signature of the
BH with mass $0.00303 M_{\rm bulge}$ is limited to a small increase of
the velocity dispersion provided that $\chi \lta 0.5$.  The signature
of a BH that is ten times more massive, is much more pronounced.  Not
only is a strong central increase of velocity dispersion visible (even
for $\chi = 1$), but in addition, the mean rotation is considerably
larger than in the case without BH. There is no major difference of
BH-signature between the cases with and without nuclear disc, as
judged from the four velocity profile moments presented here. This is
due to the fact that the nuclear disc only adds a small fraction of
the light to the VPs, especially at small radii, where the cusp
dominates.

We have also fitted double Gaussians to the seeing convolved VPs.
Figure 16 shows the parameter $\omega_d$ as function of radius in
arcseconds for three different values of $\chi$ for models 3.1 (solid
lines), 3.3 (dashed lines), and 3.5 (dotted lines).  The main result
is that for the model with $M_{\rm BH} = M_{\rm disc}$ (model 3.5) the
rotation velocity of the nuclear disc as measured from the double
Gaussian fit remains over 90\% of the circular velocity down to $0.5
R_d$ as long as $\chi \lta 1$. The fact that one can actually measure
this is due to the fact that the circular velocity in the centre is so
large that the small part of the VP that is due to the nuclear disc,
clearly shows up in the wing of the bulge part. In the case of the
less massive BH (model 3.3), the effect of the BH on the circular
velocity is so small that the disc part of the VP remains hidden in
the dominating bulge part, so that the double Gauss fit is rather
ambiguous, and can no longer be interpreted as due to a disc- and
bulge-component.

Since the circular velocity at $0.25 R_d$ for the case with $M_{\rm
BH} = M_{\rm disc}$ is 4.2 times larger than for the case without BH,
measuring $v_{\rm circ}$ so close to the centre provides an excellent
measure of the central mass-density and will immediately reveal the
presence of the BH, without the ambiguity with respect to velocity
anisotropies that hampers the interpretation of an observed, central
increase of velocity dispersion. Therefore, nuclear discs can be used
to put unambiguous constraints on the presence of nuclear BHs in these
galaxies.

\section{8 Summary and Discussion}

We have constructed axisymmetric, two-integral models of elliptical
galaxies harbouring small, nuclear discs. We used the HQ method to
calculate the even part of the phase-space distribution function (DF)
for models that consist of an ($\alpha$, $\beta$)-spheroid (the bulge)
which contains a highly flattened exponential spheroid (the disc). In
addition, we considered the effects of a massive nuclear BH. The bulge
has a central power-law density cusp ($\rho \propto r^{\alpha}$). The
disc is thick, and has an exponential surface brightness profile.
Although we concentrated on nuclear discs, the models presented here
can equally well be used to describe, e.g., S0 galaxies, by simply
increasing the horizontal scale length of the disc.

For models with only a moderate cusp (i.e., $\alpha > -0.5$), there 
is a region in the plane of disc-to-bulge ratio $\Delta$ versus disc 
scale-length $R_d$ where the even part of the DF is not strictly positive. 
For the cusped models investigated here ($\alpha = -0.5, -1.0, -1.5$) 
no such region in ($\Delta$, $R_d$) parameter space exists, so all these
models are physical.  We have considered different odd parts of the
DF, which lead to models with a large variety of streaming motions.
Upon defining the odd part of disc and bulge separately, we obtain
disc- and bulge-components that are dynamically decoupled. As an
example of such systems, we have constructed models with
counter-rotating nuclear discs.

Our models are locally stable against axisymmetric perturbations
according to the Toomre criterion.  The nuclear discs are stable as
long as they are not too flattened or compact.  The presence of a
central cusp strongly increases the stability of the nuclear disc.
Our two-integral nuclear discs are also stable against bar
formation. Steeper cusps of the bulge ensure better stability.  The
central surface brightness of the disc with disc-to-bulge ratio such
that the disc is only just locally stable over its entire extension
depends on the value of the disc scale-length $R_d$.  The relation is
remarkably similar to the observed $\mu_0$-$R_d$ relation for a large
sample of spirals, S0s, and discy ellipticals. If stability arguments
are indeed responsible for this relation, this suggests that discs
build up their mass until they become marginally stable.

We have investigated the effects of seeing convolution on the
kinematic observables of nuclear discs. We considered PSFs with FWHM
in the range of $0.5 R_d$ up to $5 R_d$. If $R_d = 0.2''$, as observed
in a number of Virgo ellipticals, this corresponds to the range from
$0.1''$ (i.e., HST resolution) up to $1''$ (i.e., typical ground-based
resolution). For $\chi \equiv {\rm FWHM}/R_d \gta 3$ the nuclear disc
becomes kinematically undetectable. This is similar to the value of
$\chi$ for which the disc becomes photometrically undetectable. One
does, however, observe a slightly larger central velocity dispersion
than would be the case without nuclear disc. This is due to seeing
convolution of the rotation curve of the nuclear disc. It has been
argued by a number of authors that this effect might mimic the
presence of a central BH. However, we find that the effect is rather
small, not exceeding 10\%, and conclude that nuclear discs can {\it
not} mimic the presence of a massive BH.

In the case of counter-rotating nuclear discs the counter-rotation is
only unambiguously detectable from the VPs if the seeing FWHM is
smaller than $\sim 2 R_d$ (for $\Lambda = 1$). For such small seeing
FWHM, the central line-of-sight velocity profile is dominated by the
disc, and therefore one observes in addition to the counter-rotation a
central {\em decrease} in velocity dispersion. It is surprising that
in most cases where counter-rotation is observed, one finds an
additional strong, central {\em increase} in velocity
dispersion. Although we note that this might be an indication for a
nuclear BH, in which case the high central dispersion is due to seeing
convolution of the Keplerian rotation curve, further dynamical
modeling is required to confirm this.

When fitting a double Gaussian to the VPs one extracts information of
the kinematics of both components. Although the assumption
that both components have Gaussian VPs is clearly simplified, it gives a
first order estimate of the rotation velocities of the disc. The disc
rotates with almost the circular velocity ($v_{\rm disc} \approx
0.95\; v_{\rm circ}$). The relative contribution of disc- and bulge
light to the VP depends strongly on radius, and on the parameters of
the model. All constructed models have disc-to-bulge ratio chosen so
that at $R=R_d$ the disc contributes 50\% to the light of the
VP (i.e., $\Lambda = 1$). Therefore, the relative disc contribution to 
the VPs at $R < R_d$ decreases strongly with increasing cusp steepness 
of the bulge. For shallow cusps, the disc still contributes a significant 
amount of light to the central VPs and a central decrease in velocity 
dispersion is observed. For more steeply cusped bulges, the disc 
becomes only detectable at somewhat larger radii.

The fact that the nuclear disc rotates with almost the circular
velocity has important implications. Measuring the mean rotation of
the nuclear disc allows an accurate determination of the central
density (or central mass-to-light ratio) of its host galaxy. We have
tried to "measure" the rotation of the nuclear disc by fitting a
double Gaussian to the seeing convolved VPs. We compared these results
to the mean rotation $v_{\rm disc}$ as calculated from equation
(\odddf). For $R \gta 2 R_d$ and $\chi \lta 1$ we found the mean of
the rapidly rotating Gaussian to be in good agreement (within a few
percent) with $v_{\rm disc}$. There VPs can therefore be used to put
strong constraints on the central mass-to-light ratio of the host
galaxy.  At smaller radii the cusp dominates and the seeing
convolution has important influences. Therefore, these central VPs do
not clearly show two distinct components and the double Gaussian fit
is somewhat arbitrary.  The fit can no longer be interpreted as
representing the disc- and bulge component.

We have also included a massive BH in the centre of these models to
investigate whether nuclear discs can be used to test for the presence
of such BHs in the same way as ionized gas discs can. In both the
cases with and without nuclear disc a strong central increase in
velocity dispersion is observed (for a BH that is massive enough).
The observed rotation velocities, as measured from the first moment of
the VP, increases.  We have fitted double Gaussians to the VPs, to see
if we can measure the rotation velocity of the nuclear disc. Since the
disc rotates with almost the circular velocity, such measurements
provide an accurate estimate of the central mass density of the
galaxy. We find that for a BH whose mass equals that of the nuclear
disc, the circular velocity becomes so large close to the centre, that
even the VPs at only $0.5\;R_d$ from the centre clearly reveal two
components. The mean velocity of the rapid rotating component is an
excellent indicator of the circular velocity. Since at these small
radii, $v_{\rm circ}$ is considerably larger than for the case without
BH, the observed rotation velocity of the nuclear disc provides strong
evidence for the presence of the BH. Most importantly, interpretation
of these rotation velocities is far less ambiguous than interpretation
of a central increase of velocity dispersion.

Although the stellar dynamics of the disc and the bulge is likely to
be influenced by a third integral of motion, we suspect that the
results discussed here, based upon two-integral models, are not
strongly influenced by this oversimplification. It is well known that
when two-integral models require the presence of a massive BH to fit a
central increase of velocity dispersion, one always has the
possibility that a three-integral model without BH might equally well
fit the data. However, if one observes a nuclear disc whose rotation
velocity exceeds the circular velocity that would correspond to a
constant mass-to-light ratio model, invoking a third integral of
motion can not alter the conclusion that there has to be a central
increase in mass-to-light ratio.

The high spatial-resolution spectra obtainable with HST can be used to
accurately measure the rotation curve of the nuclear discs, giving an
excellent measure of the central density of the galaxy.  Although the
spectral resolution of the Faint Object Spectrograph (FOS) aboard the
HST may be too poor to resolve the VPs in enough detail to allow for
$v_{\rm disc}$ to be measured, observations with the Space Telescope
Imaging Spectrograph (STIS), to be installed during the next servicing
mission, will allow the VPs to be measured with high enough both
spatial, and spectral resolution to put stringent constraints on the
presence of possible massive BHs in the nuclei of these galaxies.

\section*{Acknowledgements}

It is a pleasure to thank Richard Arnold, Marcella Carollo, Eric
Emsellem, and Eddie Qian for useful discussions. The authors are
indebted to Jerry Sellwood for helpful advice concerning the stability
of the nuclear discs. FCvdB was supported by the Netherlands
Foundation for Astronomical Research (ASTRON), \#782-373-055, with
financial aid from the Netherlands Organization for Scientific
Research (NWO).

\looseness=-1

\section*{APPENDIX: Asymptotic approximation of even part of DF close to BH}

At sufficiently small radii the density of the composite disc/bulge system
can be approximated by
\eqnam\apa
$$\rho(R,z) = \rho_{0,b} \left({m_b \over R_b}\right)^{\alpha} 
 \!- \rho_{0,d} \; {\rm ln}\left({m_d \over R_d}\right), \eqno({\rm A1})$$
where $m_b$ and $m_d$ are the constant density spheroids of bulge and disc
respectively. As long as $\alpha > -2$ and $\alpha + 2 \beta < -2$, the
central potential can be approximated by
\eqnam\apb
$$\Psi(R,z) = {G M_{\rm BH} \over \sqrt{R^2 + z^2}} + 
\Psi_0^{*}.\eqno({\rm A2}) $$
Here $\Psi_0^{*}$ is the finite, central stellar potential given by
the sum of $\psi_{0,b}$ (equation \cenbulpot) and $\psi_{0,d}$
(equation \cendispot). Solving $z^2$ from equation (\apb) and
substitution in equation (A1) yields an explicit expression of
$\rho(\Psi,R)$.  Using the alternative form of the contour integral
given by equation (3.3) in HQ, and integrating along a circular
contour parameterized by angular parameter $\theta$ (i.e., $\Psi = E +
E e^{i \theta}$, $-\pi \leq \theta \leq
\pi$), an analytic solution for the approximating $f_e(E,L_z) = 
f_e^b(E,L_z) + f_e^d(E,L_z)$ is found.

For $\alpha = -1$ and $\alpha + 2 \beta = -4$ we find
$$f_e^b(E,L_z) = {\rho_{0,b} q_b \over \sqrt{8} \pi^2 \Phi_b^{3/2}}
\left( E - \Psi_0^* \over \Phi_b\right)^{-1/2} 
{1 + \xi_b \over (1 - \xi_b)^2},\eqno({\rm A3}) $$
where $\xi_b = (1 - q_b^2) \eta^2$, and $\Phi_b = G M_{\rm BH} / R_b$
(see also Dehnen and Gerhard 1994, and QZMH).  Similarly, for the disc
part
$$\eqalign{f_e^d(E,L_z) &= 
{\rho_{0,d} \over 2 \sqrt{8} \pi^2 \Phi_d^{3/2}} 
\left({E - \Psi_0^* \over \Phi_d}\right)^{-3/2} \;\; \times \cr
&\biggl[ g(\xi_d) - {\rm ln}\left({E - \Psi_0^* \over \Phi_d}\right) - 
{\rm ln}q_d - 2 {\rm ln}2 - 3 \biggr], \cr}\eqno({\rm A4}) $$
where
$$\eqalign{g(\xi_d) = 
&{3 - 2 \xi_d \over 1 - \xi_d} + \cr
&\left[1 - {3 - 4\xi_d \over (1 - \xi_d)^2}\right]
\sqrt{1 - \xi_d \over \xi_d} {\rm arctan}\sqrt{\xi_d \over 1 - \xi_d}. 
\cr}\eqno({\rm A5})$$
Here $\xi_d = (1 - q_d^2) \eta^2$, and $\Phi_d = G M_{\rm BH} / R_d$.
For the radial, $\eta = 0$ orbits $g(\xi_d) = 1$.

\end